\documentclass[a4paper,11pt]{article}
\pdfoutput=1
\usepackage{jheppub}
\usepackage[T1]{fontenc}
\usepackage{graphicx}
\usepackage{amsmath}
\usepackage{xspace}
\usepackage{subfig}
\usepackage[table]{xcolor}
\usepackage{mathtools}
\usepackage{braket}
\usepackage{multirow}
\usepackage{slashed}

\title{Constraining the SMEFT Extended with Sterile Neutrinos at FCC-ee}

\author[a,b]{Patrick D. Bolton,}
\emailAdd{patrick.bolton@ijs.si}

\author[c]{Frank F. Deppisch,} 
\emailAdd{f.deppisch@ucl.ac.uk}

\author[d]{Suchita Kulkarni,} 
\emailAdd{suchita.kulkarni@uni-graz.at}

\author[c]{Chayan Majumdar,} 
\emailAdd{c.majumdar@ucl.ac.uk}

\author[c,e]{Wenna Pei,} 
\emailAdd{wenna.pei.20@ucl.ac.uk}

\affiliation[a]{SISSA, International School for Advanced Studies, \\
INFN, Sezione di Trieste, Via Bonomea 265, I-34136 Trieste, Italy}

\affiliation[b]{Jožef Stefan Institute, Jamova 39, 1000 Ljubljana, Slovenia}

\affiliation[c]{University College London, Gower Street, London WC1E 6BT, UK}

\affiliation[d]{Institute of Physics, NAWI Graz, University of Graz, Universitätsplatz 5, A-8010 Graz, Austria}

\affiliation[e]{Institute of Physics, ELTE Eötvös Loránd University, Pázmány Péter sétány 1/A, H-1117\newline Budapest, Hungary}

\abstract{We investigate how extensions of the Standard Model (SM) involving heavy neutral leptons (HNLs) can be probed at FCC-ee, the proposed high-energy circular $e^+e^-$ collider. Using the effective field theory (EFT) approach, we determine the impact of new interactions on the production and decay of HNLs at FCC-ee. In particular, we consider $d\leq 7$ $\nu$SMEFT operators which induce vector, scalar and tensor four-fermion and effective charged- and neutral-current interactions of HNLs, that may also mix with the active neutrinos of the SM. We consider sensitivities to the active-sterile mixing and EFT Wilson coefficients from monophoton searches and displaced vertex decay signatures. In both analyses, we consider the scenarios where HNLs are Majorana or Dirac fermions. We translate the upper bounds on the Wilson coefficients to lower limits on the scale of new physics.}
\makeatletter
\gdef\@fpheader{\phantom{a}}
\makeatother

\begin{document}

\maketitle
\flushbottom

%%%%%%%%%%%%%%%%%%%%%%%%%%%%%%%%%%%%%%%%
\section{Introduction} 
\label{sec:intro}
%%%%%%%%%%%%%%%%%%%%%%%%%%%%%%%%%%%%%%%%

Since the direct observation of atmospheric and solar neutrino oscillations over a quarter of a century ago~\cite{Super-Kamiokande:1998kpq, SNO:2002tuh}, it is well-known that neutrinos are massive fermions. However, the Standard Model (SM) of particle physics in its current form does not include right-handed (RH) neutrinos, $N$, forbidding renormalisable Yukawa terms that generate neutrino masses after electroweak (EW) symmetry breaking. Physics beyond the SM is therefore necessary to explain the non-zero neutrino masses.

Adding at least two RH neutrinos is the most straightforward extension of the SM to explain the neutrino oscillation data. No gauge symmetry forbids a Majorana mass term for the RH neutrinos if one allows the violation of total lepton number, $L$, which is conserved accidentally in the SM. In the standard type-I seesaw mechanism~\cite{Minkowski:1977sc,Gell-Mann:1979vob,Yanagida:1979as,Mohapatra:1979ia,Schechter:1980gr}, one obtains three light Majorana active neutrinos and an arbitrary number of sterile neutrinos or heavy neutral leptons (HNLs). While the minimal type-I seesaw mechanism has some other appealing features, such as the viable generation of the baryon asymmetry of the universe via high-scale thermal leptogenesis~\cite{Fukugita:1986hr,Covi:1996wh,Buchmuller:2004nz,Davidson:2008bu}, it is difficult to probe the sterile states at collider and neutrinoless double beta ($0\nu\beta\beta$) decay experiments. For HNLs kinematically accessible at colliders, the mixing between active and sterile states implied by the neutrino oscillation data would be too small to produce an observable number of HNLs.

This has spurred interest in other low-scale seesaw mechanisms such as the inverse seesaw~\cite{Mohapatra:1986aw,Nandi:1985uh,Mohapatra:1986bd,Gonzalez-Garcia:1988okv}. There, the light neutrino masses are proportional to small approximately lepton number violating Majorana masses among the RH neutrinos. Consequently, pairs of heavy Majorana states form nearly degenerate pseudo-Dirac states. An attractive feature of this scenario is that the mixing between the active neutrinos and the pseudo-Dirac states is decoupled from the light neutrino masses and can in principle be large. These states can be produced in collider and beam dump experiments~\cite{Gorbunov:2007ak, Atre:2009rg, Das:2012ze, Deppisch:2015qwa, Cai:2017mow, Das:2018usr, Bondarenko:2018ptm, Ballett:2019bgd, Bolton:2019pcu, Coloma:2020lgy, Abdullahi:2022jlv}, and also generate the baryon asymmetry via the resonant leptogenesis mechanism~\cite{Pilaftsis:1997jf, Akhmedov:1998qx, Pilaftsis:2003gt, Asaka:2005pn, Pilaftsis:2005rv, Dev:2017wwc, Drewes:2017zyw, Klaric:2021cpi, Drewes:2021nqr, Sandner:2023tcg}. In the lepton number conserving limit, the contribution of this mechanism to the light neutrino masses vanishes and the pseudo-Dirac pairs become exactly Dirac fermions. The standard type-I and inverse seesaw mechanisms can be seen as different limits of the RH neutrino parameter space, with intermediate scenarios also being possible~\cite{Bolton:2019pcu, Bolton:2022tds}.

Given the large number of SM extensions involving RH neutrinos, a phenomenological study of HNLs can benefit from a model-independent effective field theory (EFT) approach. The EFT valid for the unbroken phase of the SM (SMEFT) has been studied in detail, with a complete basis of non-redundant operators and their renormalisation group (RG) running examined in~\cite{Buchmuller:1985jz, Grzadkowski:2010es, Jenkins:2013zja, Jenkins:2013wua, Alonso:2013hga, Lehman:2014jma, Liao:2016hru, Brivio:2017vri, Murphy:2020rsh, Liao:2020jmn, Li:2020gnx, Harlander:2023psl, Zhang:2023kvw}. Likewise, the EFT of the SM broken phase (LEFT) has been explored in~\cite{Jenkins:2017jig, Jenkins:2017dyc, Liao:2020zyx, Murphy:2020cly, Li:2020tsi, Hamoudou:2022tdn}. In the EFT framework, models giving rise to neutrino masses are those that generate the $d = 5$ Weinberg operator and other $\Delta L = \pm 2$ operators~\cite{Weinberg:1979sa, Babu:2001ex, deGouvea:2007qla, Gargalionis:2020xvt, Fridell:2024pmw}. If the Majorana masses of RH neutrinos are at or below the energy scale of interest, $N$ must also be included as a light degree of freedom in the EFT. This has motivated studies of the so-called $\nu$SMEFT and $\nu$LEFT in the unbroken and broken phases of the SM, respectively~\cite{delAguila:2008ir, Aparici:2009fh, Bhattacharya:2015vja, Liao:2016qyd, Li:2021tsq, Datta:2020ocb, Bischer:2019ttk}, which extend the usual operator bases to include RH neutrinos. Operators in the $\nu$SMEFT can result in phenomenology distinct from the active-sterile mixing.

\begin{figure}[t!]
\centering
\includegraphics[width=0.22\textwidth]{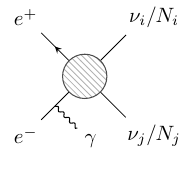}
\includegraphics[width=0.31\textwidth]{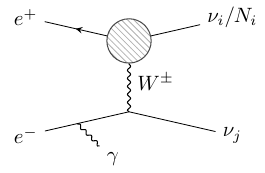}
\includegraphics[width=0.32\textwidth]{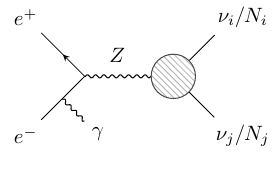}
\caption{Single and pair production of HNLs at FCC-ee via the EFT operators considered in this work: four-fermion operators (left) and effective $W^\pm$ (centre) and $Z$ (right) interactions. The active-sterile mixing $V_{e N}$ induces the $W^\pm$ and $Z$ diagrams, while $V_{\mu N}$ and $V_{\tau N}$ induce the $Z$ diagram only.}
\label{fig:feynman-diagrams}
\end{figure}

In this work, we consider the phenomenology of HNLs in the presence of $\nu$SMEFT operators at the Future Circular $e^+ e^-$ Collider (FCC-ee)~\cite{FCC:2018evy}. Like the proposed Circular Electron-Positron Collider (CEPC)~\cite{CEPCStudyGroup:2018ghi}, FCC-ee will significantly improve on measurements made at the Large Electron Positron (LEP) collider. Operating at the $Z$ pole and three higher centre-of-mass energies, FCC-ee and/or CEPC will enable precision tests of physics at the EW scale, which may be sensitive to the effects of TeV-scale new physics. Global constraints on SMEFT operators impacting EW precision observables (EWPOs) have been placed using LEP and SLD data~\cite{Han:2004az, Berthier:2015oma, Falkowski:2015krw, Efrati:2015eaa, Falkowski:2017pss, Ellis:2018gqa, Ellis:2020unq, Aoude:2020dwv, Ethier:2021bye, Corbett:2021eux, deBlas:2022ofj, Bartocci:2023nvp, Allwicher:2023shc}, and forecasted for the future $e^+e^-$ colliders~\cite{Allwicher:2023shc, Celada:2024mcf, Ge:2024pfn, Greljo:2024ytg}. 

In this study, we instead target two general final states to constrain the $\nu$SMEFT at FCC-ee; monophoton plus missing energy (mono-$\gamma$ plus $\slashed{E}$) and displaced vertex (DV) signatures. We first note that FCC-ee will be able to improve on LEP measurements of the invisible $Z$ decay width $\Gamma_{\text{inv}}$ and therefore the number of active neutrinos $N_\nu$~\cite{Dolgov:1972sp, Gaemers:1978fe, Ma:1978zm, Montagna:1995wp, L3:1992rlp, ALEPH:2005ab} via direct measurements at the $Z$ pole~\cite{AlcarazMaestre:2021ssl} and the radiative return process $e^+e^- \to Z \gamma$, leading to a mono-$\gamma$ plus $\slashed{E}$ signature at higher centre-of-mass energies~\cite{FCC:2018byv}.
The measured value of $\Gamma_{\text{inv}}$ can also be interpreted as a constraint on the presence of heavy sterile neutrinos from non-unitarity~\cite{Jarlskog:1990kt, Carena:2003aj, Akhmedov:2013hec, deGouvea:2015euy, Escrihuela:2019mot, Blennow:2023mqx}. Furthermore, direct measurements of $\Gamma_{\text{inv}}$ and the mono-$\gamma$ plus $\slashed{E}$ signature can be used to constrain new light degrees of freedom, including viable dark matter candidates~\cite{Fox:2011fx,Kundu:2021cmo}. The light new states must be effectively stable for such signatures to be relevant. However, if they do decay and are long-lived, FCC-ee will also be highly constraining via searches for displaced vertices~\cite{Blondel:2014bra, Antusch:2016ejd, Antusch:2016vyf, Blondel:2022qqo, Ajmal:2024kwi}.

Here, we aim to leverage the large luminosity and clean environment of FCC-ee to assess the ability of mono-$\gamma$ plus $\slashed{E}$ and DV signatures to constrain the active-sterile mixing $V_{\alpha N}$ and the more general $\nu$SMEFT landscape. For the latter, we consider $d \leq 7$ operators which may be generated at tree-level in ultra-violet (UV) complete extensions of the SM. 
At an $e^+e^-$ collider, the active-sterile mixing $V_{\alpha N}$ can permit the single production of HNLs, $e^+ e^- \to \nu N$, via $t$-channel $W^\pm$ exchange (Fig.~\ref{fig:feynman-diagrams}, centre) and $s$-channel $Z$ exchange (Fig.~\ref{fig:feynman-diagrams}, right). The considered $\nu$SMEFT operators induce all of the diagrams in Fig.~\ref{fig:feynman-diagrams}, with pair production of HNLs, $e^+ e^- \to N N$, also possible.

\begin{figure}[t!]
\centering
\includegraphics[width=0.28\textwidth]{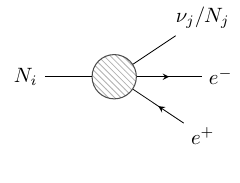}
\includegraphics[width=0.32\textwidth]{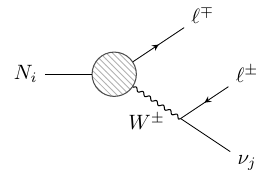}
\includegraphics[width=0.32\textwidth]{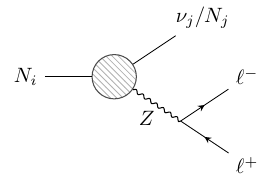}
\caption{A selection of decays of HNLs via the same EFT operators in Fig.~\ref{fig:feynman-diagrams}. The active-sterile mixing $V_{\alpha N}$ also induces the $W^\pm$ and $Z$ diagrams.}
\label{fig:feynman-diagrams-decay}
\end{figure}

If the HNLs do not decay inside the FCC-ee detector, the processes $e^+ e^- \to \nu N\gamma /NN\gamma$ with an initial state photon can give the sought-after mono-$\gamma$ plus $\slashed{E}$ signature. However, the active-sterile mixing $V_{\alpha N}$ and/or $\nu$SMEFT operators that trigger the production of HNLs can also cause their decay, as shown in Fig.~\ref{fig:feynman-diagrams-decay}. Given these decay modes, the HNLs must be sufficiently long-lived to appear as missing energy. Decreasing the coupling responsible for the decay, however, comes at the cost of suppressing the production cross section. Two viable scenarios are if the HNLs are sufficiently light (with masses below the GeV scale) or are pair produced, so that decays are kinematically forbidden. Alternatively, a pair of HNLs with a small mass splitting (such as in the pseudo-Dirac scenario) can be effectively stable over the detector length. Clearly, there is also ample room in the parameter space for the HNLs to decay over a macroscopic distance in the detector, and the DV signature will be more constraining. It is therefore interesting to ask how the constraints depend not only on the overall HNL mass scale, but on the HNL mass splittings.

A number of works in the literature have already examined the phenomenology of $\nu$SMEFT operators at colliders~\cite{delAguila:2008ir, Aparici:2009fh, Duarte:2016caz, Caputo:2017pit, Magill:2018jla, Alcaide:2019pnf, Butterworth:2019iff, Biekotter:2020tbd, Barducci:2020icf, DeVries:2020jbs, Cottin:2021lzz, Beltran:2021hpq, Beltran:2022ast, Mitra:2022nri, Barducci:2022hll, Delgado:2022fea, Duarte:2023tdw, Zapata:2023wsz, Barducci:2023hzo, Fernandez-Martinez:2023phj, Beltran:2024twr, Chun:2024mus, Biswas:2024gtr, Mitra:2024ebr, Beltran:2025ilg, Duarte:2025zrg}. Signatures related to the direct production of HNLs at LEP ($e^+e^- \to \nu N/NN$) and LHC ($pp \to \ell N/\nu N/NN$) include prompt and displaced decays, same-sign leptons and missing energy accompanied by photons, leptons or jets. HNLs can also be produced from rare decays of top quarks ($t \to q+\ell N/\nu N/NN$), tau leptons ($\tau \to \ell +\nu N/NN,M N$), light and heavy mesons ($M\to (M'+)\ell N/\nu N/NN$) at colliders, contributing to missing energy or decaying downstream at a proposed far detector experiment. Other studies of the specific operators we consider in this work are Refs.~\cite{Biekotter:2020tbd, Mitra:2022nri, Fernandez-Martinez:2023phj} and Ref.~\cite{Barducci:2022hll}, which use the mono-$\gamma$ plus $\slashed{E}$ and DV signatures at LEP/FCC-ee, respectively. We intend to add to these analyses by exploring the effects of $d = 7$ operators in the $\nu$SMEFT, the Dirac or Majorana nature of the HNLs, and different mass splittings between a pair of HNLs.

This paper is organised as follows. In Sec.~\ref{sec:model}, we outline the $d\leq 7$ $\nu$SMEFT operators considered in this work, examining their impact, in addition to the active-sterile mixing, on the production and decay of HNLs at FCC-ee (more details are given in Appendices~\ref{app:2to2_xsec} and~\ref{app:decays}, respectively). We briefly explore which simple UV models can produce these operators at tree-level in App.~\ref{app:matching}. In Sec.~\ref{sec:active_sterile_mixing}, we then explore the sensitivity of mono-$\gamma$ plus $\slashed{E}$ searches at FCC-ee to the electron-flavour mixing strength, $V_{e N}$, for a single Dirac HNL, comparing to current constraints and previously estimated sensitivities from DV searches at FCC-ee~\cite{Blondel:2014bra, Antusch:2016vyf, Blondel:2022qqo, Ajmal:2024kwi}. In Sec.~\ref{sec:ff_monophoton}, we estimate the sensitivity of the mono-$\gamma$ plus $\slashed{E}$ signature to the EFT operators of interest, assuming vanishingly small active-sterile mixing and a pair of Majorana or Dirac HNLs. The sensitivities for different mass splittings between the HNL pair are examined. Next, in Sec.~\ref{sec:displacedvertex}, we perform a sensitivity analysis for the same operators using DV searches at FCC-ee. Finally, in Sec.~\ref{sec:discussion}, we translate the maximum reach of FCC-ee via mono-$\gamma$ plus $\slashed{E}$ and DV searches to the scale of new physics of the $\nu$SMEFT operators. We compare and contrast to other experimental probes of these operators. We summarise our findings in Sec.~\ref{sec:conclusions}.

%%%%%%%%%%%%%%%%%%%%%%%%%%%%%%%%%%%%%%%%
\section{Heavy Neutral Leptons in the $\nu$SMEFT}
\label{sec:model}
%%%%%%%%%%%%%%%%%%%%%%%%%%%%%%%%%%%%%%%%

The $\nu$SMEFT consists of a basis of $SU(3)_c \times SU(2)_L \times U(1)_Y$ invariant operators built from SM degrees of freedom and the RH neutrino field $N$. The effective Lagrangian for this theory can be written as
\begin{align}
\label{eq:vSMEFT}
\mathcal{L} = \mathcal{L}_{\text{SM}} + \bar{N}i\slashed{\partial}N - \bigg[\bar{L}Y_\nu N\tilde{H} + \frac{1}{2}\bar{N}^c M N + \text{h.c.}\bigg] + \sum_i C_i^{(d)} Q_i^{(d)}\,,
\end{align}
where $C_i^{(d)} = 1/\Lambda^{d-4}$ are Wilson coefficients (WCs) and $N = \mathcal{C}\bar{N}^T$, with $\mathcal{C}$ the charge conjugation matrix. A complete basis of independent operators in the $\nu$SMEFT, taking into account redundancies from Fierz identities, integration by parts and equations of motion, can be found in Refs.~\cite{Aparici:2009fh,Bhattacharya:2015vja,Liao:2016qyd,Li:2021tsq}. 

%%%%%%%%%%%%%%%%%%%%%%%%%%%%%%%%%%%%%%%%
\subsection{Active-Sterile Mixing}
\label{sec:active_sterile_mixing_model}
%%%%%%%%%%%%%%%%%%%%%%%%%%%%%%%%%%%%%%%%

In Eq.~\eqref{eq:vSMEFT}, the terms in the square brackets are the renormalisable Yukawa coupling and RH neutrino Majorana mass term, respectively. After EW symmetry breaking, these terms induce mixing between the left-handed (active) and RH (sterile) neutrino fields as
\begin{align}
\label{eq:mixing_Majorana}
\begin{pmatrix}
\nu \\
N^c
\end{pmatrix} = 
P_L\begin{pmatrix}
1 - \frac{1}{2}\Theta\Theta^\dagger & \Theta \\
- \Theta^\dagger & 1 - \frac{1}{2}\Theta^\dagger\Theta
\end{pmatrix}\begin{pmatrix}
\nu' \\
N'
\end{pmatrix} \,; \quad \Theta_{\alpha i} \equiv V_{\alpha N_i} = \frac{v(Y_\nu)_{\alpha i}}{\sqrt{2}M_i} \,,
\end{align}
where we have assumed that $\Theta$ is small, permitting an expansion up to $\mathcal{O}(\Theta^2)$ terms (leading to non-unitarity effects in lepton mixing), and we have taken $M$ to be diagonal without loss of generality. We note that the mixing in Eq.~\eqref{eq:mixing_Majorana} receives further modifications from the higher-dimensional operators in Eq.~\eqref{eq:vSMEFT}. A further rotation $\nu_{\alpha}' = U_{\alpha i}\nu_i'$, with $U$ the Pontecorvo–Maki–Nakagawa–Sakata (PMNS) matrix, is required to diagonalise the $3\times 3$ light neutrino mass matrix. The resulting massive states are Majorana fermions ($N' = N^{\prime c}$) and the active-sterile mixing $V_{\alpha N_i}$ couples the heavy neutral leptons (HNLs) $N_i'$ to the SM via the charged and neutral currents. In the type-I seesaw scenario, the typical expectation for the active-sterile mixing is $|V_{\alpha N_i}| \sim \sqrt{m_\nu/m_{N_i}}$ for the light neutrino mass scale $m_\nu$ and HNL mass $m_{N_i} \approx M_i$. The phenomenology of Majorana HNLs with non-zero active-sterile mixing has been reviewed in Refs.~\cite{Atre:2009rg, Bolton:2019pcu, Fernandez-Martinez:2023phj}.

It is also interesting to consider the lepton number conserving ($\Delta L = 0$) limit of Eq.~\eqref{eq:vSMEFT}. To do this, it is convenient to introduce the SM gauge-singlet fermion field $S$ and assign the lepton numbers $L(\nu) = L(N) = L(S) = +1$. For simplicity, we assume an equal number of $N$ and $S$ fields. In the $\Delta L = 0$ limit, the effective Lagrangian becomes
\begin{align}
\label{eq:vSMEFT_S}
\mathcal{L} = \mathcal{L}_{\text{SM}} + \bar{N}i\slashed{\partial}N + \bar{S}i\slashed{\partial}S - \bigg[\bar{L}Y_\nu N\tilde{H} + \bar{S} M' N + \text{h.c.} \bigg] + \sum_i C_i^{(d)} Q_i^{(d)}\,,
\end{align}
where the sum is over $\Delta L = 0$ operators constructed from SM degrees of freedom plus $N$ and $S$. Active-sterile mixing is again induced in the broken phase, up to $\mathcal{O}(\Theta^2)$, as 
\begin{align}
\label{eq:mixing_Dirac}
\begin{pmatrix}
\nu \\
N^c \\
S
\end{pmatrix} = 
P_L\begin{pmatrix}
1 - \frac{1}{2}\Theta\Theta^\dagger & 0 & \Theta \\
0 & 1 & 0 \\
- \Theta^\dagger & 0 & 1 - \frac{1}{2}\Theta^\dagger\Theta
\end{pmatrix}\begin{pmatrix}
\nu' \\
N^{\prime c} \\
N'
\end{pmatrix} \,; \quad \Theta_{\alpha i} \equiv V_{\alpha N_i} = \frac{v(Y_\nu)_{\alpha i}}{\sqrt{2}M_i'} \,.
\end{align}
The physical states are three massless Weyl fermions ($\nu'$) and massive Dirac fermions ($N'$). Modifications to this picture are required to generate the observed neutrino masses. For example, small $\Delta L = \pm 2$ terms can be added such as the $d = 5$ Weinberg operator $Q_5 = (\bar{L}\tilde{H})(\tilde{H}^T L^c)$ or the Majorana mass $-\frac{1}{2}\mu \bar{S} S^c$. The heavy states are then expected to be pseudo-Dirac fermions, equivalent to a pair of Majorana fermions with a small mass splitting. Alternatively, 
the light neutrinos may themselves be Dirac fermions if additional RH neutrino fields $N$ are present. Models where such a situation arises often involve additional discrete symmetries to forbid $\Delta L = \pm 2$ terms~\cite{Ma:2014qra,Valle:2016kyz,CentellesChulia:2016rms}. The phenomenology of (pseudo-)Dirac HNLs with active-sterile mixing is similar to the Majorana HNL case, apart from the (suppression) absence of lepton number violating ($\Delta L = \pm 2$) signatures~\cite{Anamiati:2016uxp,Drewes:2019byd,Abada:2022wvh}.

%%%%%%%%%%%%%%%%%%%%%%%%%%%%%%%%%%%%%%%%
\subsection{EFT Operators}
\label{subsec:EFT}
%%%%%%%%%%%%%%%%%%%%%%%%%%%%%%%%%%%%%%%%

%
\begin{table}[t!]
\centering
\renewcommand{\arraystretch}{1.25}
\setlength\tabcolsep{2.7pt}
\begin{tabular}{c|c|c|c}
\hline 

\multicolumn{2}{c|}{$\psi^4$} & \multicolumn{2}{c}{$\psi^2 H^2 D$} \\ \hline

$Q_{ll}$ & $(\bar{L} \gamma_\mu L)(\bar{L} \gamma^\mu L)$ & $Q^{(1)}_{Hl}$ & $(\bar{L} \gamma_\mu L)(H^{\dagger} i \overleftrightarrow{D}^\mu H)$ \\ 

$Q_{le}$ & $(\bar{L} \gamma_\mu L)(\bar{e} \gamma^\mu e)$ & $Q^{(3)}_{Hl}$ & $(\bar{L} \gamma_\mu \tau^I L)(H^{\dagger} i \overleftrightarrow{D}^{I\mu} H)$ \\

$Q_{lNle}$ & $\epsilon_{ij}(\bar{L}^i N)(\bar{L}^j e)$ & $Q_{HN}$ & $(\bar{N} \gamma_\mu N)(H^{\dagger} i \overleftrightarrow{D}^\mu H)$ \\

$Q_{lN}$ & $(\bar{L} \gamma_\mu L)(\bar{N} \gamma^\mu N)$ & $Q_{HNe}$ & $(\bar{N} \gamma_\mu e)(\tilde{H}^{\dagger} i D^\mu H)$ \\

$Q_{eN}$ & $(\bar{e} \gamma_\mu e)(\bar{N} \gamma^\mu N)$ &  &  \\
\hline
\end{tabular} \\

\vspace{1em}

\begin{tabular}{c|c|c|c}
\hline 
\multicolumn{2}{c|}{$\psi^4 H$} & \multicolumn{2}{c}{$\psi^2 H^3 D$} \\ \hline

$Q_{llleH}$ & $\epsilon_{ij}\epsilon_{mn}(\bar{e} L^i)(\bar{L}^{jc} L^m)H^n$ & $Q_{Nl1}$ & $\epsilon_{ij}(\bar{N}^c\gamma_\mu L^i)(iD^\mu H^j)(H^\dagger H)$ \\

$Q_{lNlH}$ & $\epsilon_{ij}(\bar{L} \gamma_\mu L)(\bar{N}^c \gamma^\mu L^i)H^j$ & $Q_{Nl2}$ & $\epsilon_{ij}(\bar{N}^c\gamma_\mu L^i)H^j(H^\dagger i \overleftrightarrow{D}^\mu H)$ \\

$Q_{eNlH}$ & $\epsilon_{ij}(\bar{e}\gamma_\mu e)(\bar{N}^c \gamma^\mu L^i)H^j$ & $Q_{leHD}$ & $\epsilon_{ij}\epsilon_{mn}(\bar{L}^{ic}\gamma_\mu e)H^j H^m D^\mu H^n$ \\

$Q_{lNeH}$ & $(\bar{L} N)(\bar{N}^c e)H$ &  &  \\

$Q_{elNH}$ & $H^\dagger(\bar{e} L)(\bar{N}^c N)$ &  &  \\

\hline
\end{tabular}
\caption{($\nu$)SMEFT operators at $d = 6$ (above) and $d = 7$ (below) which contribute to the four-fermion operators in Eq.~\eqref{eq:L_fourfermion} and effective $W^\pm$ and $Z$ interactions in Eq.~\eqref{eq:L_gauge}. The tree-level matching conditions are given in App.~\ref{app:matching}.}
\label{tab:vSMEFT-operators}
\end{table}

In this work, we consider the $\nu$SMEFT operators of dimension $d\leq 7$ listed in Table~\ref{tab:vSMEFT-operators}. At FCC-ee, these operators modify the SM process $e^+ e^- \to \nu \nu$, contribute to the single and pair production of HNLs, $e^+ e^- \to \nu N/N N$, and induce HNL decays. These operators can be generated at tree-level in simple SM extensions, as reviewed in App.~\ref{app:matching}.

At the energies relevant for FCC-ee, the $SU(2)_L \times U(1)_Y$ gauge group is spontaneously broken by the non-zero Higgs VEV, but the $W^\pm$, $Z$ and Higgs fields remain as dynamical degrees of freedom. Thus, we expand the Higgs doublet in the $\nu$SMEFT operators around $v$ and rotate the EW gauge fields to the mass basis. The WCs of the resulting operators are in principle determined by the size of the $\nu$SMEFT WCs at $\Lambda$, run down to the EW scale via the appropriate RG equations. 

In Table~\ref{tab:vSMEFT-operators}, the $\nu$SMEFT operators of type $\psi^4 H^n$ result in the following vector, scalar and tensor four-fermion operators,
\begin{align}
\label{eq:L_fourfermion}
\mathcal{L} &\supset \bigg[\frac{1}{2}C_{\nu e}^{V,LX}(\bar{\nu}
\gamma_\mu \nu) + C_{\nu Ne}^{V,RX}(\bar{\nu}^c \gamma_\mu N) + \frac{1}{2}C_{Ne}^{V,RX}(\bar{N}
\gamma_\mu N)\bigg](\bar{e} \gamma^\mu P_X e) \nonumber \\
& \hspace{1em} + \bigg[\frac{1}{2}C_{\nu e}^{S,LX}(\bar{\nu}^c \nu) + C_{\nu Ne}^{S,RX}(\bar{\nu}
N) + \frac{1}{2}C_{Ne}^{S,RX}(\bar{N}^c N)\bigg](\bar{e} P_X e)
\nonumber \\
& \hspace{1em} + \bigg[\frac{1}{2}C_{\nu e}^{T,LX}(\bar{\nu}^c \sigma_{\mu\nu} \nu) + C_{\nu Ne}^{T,RX}(\bar{\nu}
\sigma_{\mu\nu} N) + \frac{1}{2}C_{Ne}^{T,RX}(\bar{N}^c \sigma_{\mu\nu} N)\bigg](\bar{e} \sigma^{\mu\nu} P_X e) + \text{h.c.} \,,
\end{align}
for $X \in \{L,R\}$, where $P_X$ is the chirality projection operator and we omit the weak basis indices $p, r, s, t$ for simplicity. Furthermore, the operators of type $\psi^2 H^n D^2$ generate the effective $W^\pm$ and $Z$ interactions,
\begin{align}
\label{eq:L_gauge}
\mathcal{L} &\supset - \frac{g}{\sqrt{2}}\Big[W_\nu^L (\bar{\nu} \gamma_\mu e) + W_\nu^R (\bar{\nu}^c \gamma_\mu e) + W_N^L (\bar{N}^c \gamma_\mu e) +W_N^R (\bar{N} \gamma_\mu e)\Big]W^{+\mu} \nonumber \\
& \hspace{1em}  - \frac{g}{c_w} \bigg[\frac{1}{2}Z_{\nu}^L  (\bar{\nu} \gamma_\mu \nu) + Z_{\nu N}^R (\bar{\nu}^c \gamma_\mu N) + \frac{1}{2}Z_{N}^R (\bar{N} \gamma_\mu N)\bigg]Z^\mu  + \text{h.c.} \,,
\end{align}
where $g$ is the $SU(2)_L$ gauge coupling, $c_w = \cos \theta_w$ with $\theta_w$ the weak mixing angle, and we again omit the weak basis indices $p, r$. The tree-level matching conditions between the WCs in Eqs.~\eqref{eq:L_fourfermion} and~\eqref{eq:L_gauge} and the ($\nu$)SMEFT WCs are given in Tables~\ref{tab:matching_2} and~\ref{tab:matching_3} of App.~\ref{app:matching}, respectively. The interactions above are relevant in the Majorana HNL scenario, i.e. Eq.~\eqref{eq:vSMEFT}, because we allow $\Delta L = \pm 2$ operators, identified as the operators containing the charge conjugate fields $\nu^c$ and $N^c$. The $\Delta L = 0$ and $\Delta L = \pm 2$ operators are generated by the $d = 6$ and $d = 7$ operators in Table.~\ref{tab:vSMEFT-operators}, respectively.

In Table~\ref{tab:vSMEFT-operators}, we do not include all $d \leq 7$ operators relevant for the production and decay of HNLs at FCC-ee. Firstly, there are operators of type $\psi^2 H^n$, e.g. $Q_N = (\bar{N}^c N)(H^\dagger H)$, which modify the extended neutrino mass matrix but also couple the Higgs boson to HNLs. However, the process $e^+e^- \to h \to \nu N$/$NN$ is suppressed by the small electron Yukawa coupling. Other ways to constrain these operators include $e^+ e^- \to Zh/\nu_e \bar{\nu}_e h$ followed by $h \to \nu N/NN$, which has been studied in Ref.~\cite{Barducci:2020icf}. There are also dipole operators of type $\psi^2 X H^n$, e.g. $Q_{NNB} = (\bar{N}^c \sigma_{\mu\nu} N) B^{\mu\nu}$, which vanishes for a single RH neutrino. Some studies have already examined the impact of these operators at FCC-ee, for example Refs.~\cite{Barducci:2020icf, Barducci:2022hll, Mitra:2022nri, Chun:2024mus}. For $Q_{NNB}$, a sensitivity of $\Lambda \gtrsim 4 \times 10^{3}$~TeV is found from the mono-$\gamma$ plus $\slashed{E}$ signal at the $Z$ pole. As $Q_{NNB}$ can only be generated at first-loop order (or higher), including a loop factor yields the equivalent mass scale $M \sim \Lambda /16\pi^2 \sim 25$~TeV, which is comparable to sensitivities of the operators considered in this analysis. Thus, the operators considered by us can be relevant in generic UV completions, and be of comparable importance. The dipole operators also display interesting phenomenology in other contexts~\cite{Shoemaker:2018vii, Brdar:2020quo, Ismail:2021dyp, Bolton:2021pey}. For $d \geq 7$, we note that there are derivative operators of type $\psi^2 D^2 H^n$ and $\psi^4 D H^n$ which can also be relevant. Finally, there are $\nu$SMEFT operators which cannot contribute to the production of HNLs at FCC-ee, but can enable their decay, such as $Q_{lNqd} = (\bar{L}N)\epsilon(\bar{Q}d_R)$ and other operators involving quarks. A full analysis of all operators is beyond the scope of this work. Nevertheless, such an analysis could benefit from the distinction of different HNL scenarios, as done in this work.

\begin{table}[t!]
\centering
\renewcommand{\arraystretch}{1.25}
\setlength\tabcolsep{2.7pt}
\begin{tabular}{c|c|c|c}
\hline 

\multicolumn{2}{c|}{$\psi^4$} & \multicolumn{2}{c}{$\psi^2 H^2 D$} \\ \hline

$Q_{lS}$ & $(\bar{L} \gamma_\mu L)(\bar{S} \gamma^\mu S)$ & $Q_{HS}$ & $(\bar{S} \gamma_\mu S)(H^{\dagger} i \overleftrightarrow{D}^\mu H)$ \\

$Q_{eS}$ & $(\bar{e} \gamma_\mu e)(\bar{S} \gamma^\mu S)$ &  &  \\

\hline
\end{tabular} \\
\vspace{1em}
\begin{tabular}{c|c|c|c}
\hline 
\multicolumn{2}{c|}{$\psi^4 H$} & \multicolumn{2}{c}{$\psi^2 H^3 D$} \\\hline

$Q_{lSlH}$ & $\epsilon_{ij}(\bar{L} \gamma_\mu L)(\bar{S} \gamma^\mu L^i)H^j$ & $Q_{Sl1}$ & $\epsilon_{ij}(\bar{S} \gamma_\mu L^i)(iD^\mu H^j)(H^\dagger H)$ \\

$Q_{eSlH}$ & $\epsilon_{ij}(\bar{e}\gamma_\mu e)(\bar{S} \gamma^\mu L^i)H^j$ & $Q_{Sl2}$ & $\epsilon_{ij}(\bar{S} \gamma_\mu L^i)H^j(H^\dagger i \overleftrightarrow{D}^\mu H)$ \\

$Q_{lSNeH}$ & $(\bar{L} S^c)(\bar{N}^c e)H$ &  &  \\

$Q_{elSNH}$ & $H^\dagger(\bar{e} L)(\bar{S} N)$ &  &  \\

\hline
\end{tabular}
\caption{Additional $\Delta L = 0$ $\nu$SMEFT operators at $d = 6$ (above) and $d = 7$ (below) when the gauge singlet field $S$ is present.}
\label{tab:vSMEFT-operators-2}
\end{table}

The Lagrangians in Eqs.~\eqref{eq:L_fourfermion} and~\eqref{eq:L_gauge} are applicable to the Majorana HNL scenario. For the Dirac HNL scenario, as described in the previous section, we add the field $S$ and forbid $\Delta L = \pm 2$ operators. As such, we need to consider the additional $\Delta L = 0$ operators in the $\nu$SMEFT shown in Tab.~\ref{tab:vSMEFT-operators-2}. At the energies probed at FCC-ee, these lead to the additional four-fermion interactions, 
\begin{align}
\label{eq:L_fourfermion_S}
\mathcal{L} &\supset \bigg[C_{\nu S e}^{V,LX}(\bar{\nu}
\gamma_\mu S) + \frac{1}{2}C_{Se}^{V,LX}(\bar{S}
\gamma_\mu S)\bigg](\bar{e} \gamma^\mu P_X e) \nonumber \\
& \hspace{1em} + C_{S Ne}^{S,RX}(\bar{S}
N)(\bar{e} P_X e) + C_{S Ne}^{S,RX}(\bar{S}
\sigma_{\mu\nu} N)(\bar{e} \sigma^{\mu\nu} P_X e) + \text{h.c.} \,,
\end{align}
and effective $W^\pm$ and $Z$ interactions
\begin{align}
\label{eq:L_gauge_S}
\mathcal{L} &\supset - \frac{g}{\sqrt{2}}W_S^L (\bar{S} \gamma_\mu e) W^{+\mu} - \frac{g}{c_w} \bigg[Z_{\nu S}^L (\bar{\nu} \gamma_\mu S) + \frac{1}{2}Z_{S}^L (\bar{S} \gamma_\mu S)\bigg]Z^\mu + \text{h.c.} \,,
\end{align}
in addition to the $\Delta L = 0$ interactions in Eqs.~\eqref{eq:L_fourfermion} and~\eqref{eq:L_gauge}. Again, the tree-level matching conditions between the WCs in Eqs.~\eqref{eq:L_fourfermion_S} and~\eqref{eq:L_gauge_S} and the $\nu$SMEFT WCs are provided in Tables~\ref{tab:matching_2} and~\ref{tab:matching_3} of App.~\ref{app:matching}, respectively.

The interactions above are written in the weak basis. For phenomenology in the broken phase, we must rotate to the mass basis according to Eqs.~\eqref{eq:mixing_Majorana} and~\eqref{eq:mixing_Dirac} in the Majorana and Dirac cases, respectively. However, to simplify the analyses of Secs.~\ref{sec:ff_monophoton} and~\ref{subsec:ff_DV}, we will take one EFT coefficient to be non-zero at a time with vanishing active-sterile mixing, $\Theta = 0$. Thus, the rotations in Eqs.~\eqref{eq:mixing_Majorana} and~\eqref{eq:mixing_Dirac} are trivial. Nevertheless, we can write the four-fermion operators in the mass basis as
\begin{align}
\label{eq:L_fourfermion_mass}
\mathcal{L} &\supset \nonumber \bigg(\frac{1}{2}\bigg) 
\Big[C_{\nu e}^{i,XY}(\bar{\nu}\,\Gamma_i P_X \nu) (\bar{e} \,\Gamma_i P_Y e) + 
C_{\nu N e}^{i,XY}(\bar{\nu}\,\Gamma_i P_X N) (\bar{e} \,\Gamma_i P_Y e) \nonumber \\
&\hspace{4em} + 
C_{N\nu e}^{i,XY}(\bar{N}\,\Gamma_i P_X \nu) (\bar{e} \,\Gamma_i P_Y e) + 
C_{N e}^{i,XY}(\bar{N}\,\Gamma_i P_X N) (\bar{e}\,\Gamma_i P_Y e)\Big]\,,
\end{align}
where $\Gamma_i \in \{\gamma_\mu, 1, \sigma_{\mu\nu}\}$ for $i \in \{V, S, T\}$ and for simplicity we rewrite the mass eigenstate fields as $\nu' \to \nu$ and $N'\to N$. The effective $W^\pm$ and $Z$ interactions are given by
\begin{align}
\label{eq:L_gauge_mass}
\mathcal{L} &\supset \nonumber - \frac{g}{\sqrt{2}}\Big[W_\nu^X(\bar{\nu} \gamma_\mu P_X e) + W_N^X(\bar{N} \gamma_\mu P_X e)\Big]W^{+\mu} + \text{h.c.}\\
& \hspace{1em}- \bigg(\frac{1}{2}\bigg)\frac{g}{c_w}\Big[Z_{\nu}^{X}(\bar{\nu}\gamma_\mu P_X \nu)  + 
Z_{\nu N}^{X}(\bar{\nu}\gamma_\mu P_X N) \nonumber \\
&\hspace{6.3em} + 
Z_{N\nu}^{X}(\bar{N}\gamma_\mu P_X \nu)  + 
Z_{N}^{X}(\bar{N}\gamma_\mu P_X N)\Big]Z^\mu \,.
\end{align}
The Lagrangians above are applicable in both the Majorana and Dirac scenarios, i.e. $N$ can represent a Majorana or Dirac HNL. However, in the Majorana case, the factor of $1/2$ in parentheses is present. In the Dirac case, the following WCs vanish,
\begin{gather}
C_{\nu e}^{V,RX} = C_{\nu Ne}^{V,RX} = C_{\nu e}^{S,XY}= C_{\nu Ne}^{S,LX} = C_{\nu e}^{T,XX} = C_{\nu Ne}^{T,LL} = W_\nu^R = Z_\nu^R = Z_{\nu N}^R = 0 \,,
\end{gather}
because we take the limit where the three light neutrinos are massless Weyl fermions with LH components only. In the Dirac and Majorana scenarios, there are further relations between the WCs in Eqs.~\eqref{eq:L_fourfermion_mass} and \eqref{eq:L_gauge_mass}, given in App.~\ref{app:matching}.

Before moving the main analysis, we note that $W_\nu^L$ and $Z_\nu^L$ contain the SM charged- and neutral-current interactions involving neutrinos. They can be split up into contributions from the SM and heavy new physics as
\begin{align}
\label{eq:SM_couplings}
W_\nu^L = W_{\nu}^L\big|_{\text{SM}} + \delta W_\nu^L \,, \quad Z_\nu^L = Z_{\nu}^L\big|_{\text{SM}} + \delta Z_\nu^L \,,
\end{align}
where $W_{\nu}^L\big|_{\text{SM}} = \mathbb{I}$ and $Z_\nu^L\big|_{\text{SM}} = g_L^\nu \mathbb{I}$, with $g_L^\nu = 1/2$. With only the SM interactions, rotating the active and sterile neutrino fields to the mass basis according to Eq.~\eqref{eq:mixing_Majorana} or~\eqref{eq:mixing_Dirac} yields, up to $\mathcal{O}(\Theta^2)$,
\begin{align}
\label{eq:L_mix}
\mathcal{L} &\supset \nonumber - \frac{g}{\sqrt{2}}\bigg[\bigg(1 - \frac{1}{2}\Theta\Theta^\dagger\bigg)(\bar{\nu} \gamma_\mu P_L e) + \Theta^\dagger(\bar{N} \gamma_\mu P_L e)\bigg]W^{+\mu} + \text{h.c.}\\
& \hspace{1em}- \frac{g}{c_w}g_L^\nu\Big[\big(1 - \Theta\Theta^\dagger\big)(\bar{\nu}\gamma_\mu P_L \nu)  + 
\Theta(\bar{\nu}\gamma_\mu P_L N) \nonumber \\
&\hspace{5.3em} + 
\Theta^\dagger(\bar{N}\gamma_\mu P_L \nu)  + 
\Theta^\dagger \Theta(\bar{N}\gamma_\mu P_L N)\Big]Z^\mu \,,
\end{align}
in both the Majorana and Dirac HNL scenarios, where the light neutrino fields $\nu$ are those before diagonalisation via the PMNS matrix. Eq.~\eqref{eq:L_mix} shows that if the HNLs are heavy, i.e. not kinematically accessible at FCC-ee, their impact is still felt in the non-unitarity of the light neutrino mixing. This is equivalent to generating the operators $Q_{Hl}^{(1)}$ and $Q_{Hl}^{(3)}$, with WCs
\begin{align}
C_{Hl}^{(1)} = -C_{Hl}^{(3)} = \frac{Y_\nu Y_\nu^\dagger}{4M^2} =  \frac{\Theta\Theta^\dagger}{2v^2} \,,
\end{align}
after integrating the HNLs out of the theory~\cite{Broncano:2002rw,Elgaard-Clausen:2017xkq}.

%%%%%%%%%%%%%%%%%%%%%%%%%%%%%%%%%%%%%%%%
\section{Monophoton Constraints at FCC-ee}
\label{sec:monophoton}
%%%%%%%%%%%%%%%%%%%%%%%%%%%%%%%%%%%%%%%%

Here, we establish the sensitivity of the mono-$\gamma$ plus $\slashed{E}$ search at FCC-ee. To keep the analysis straightforward, we consider the presence of one HNL interaction at a time. In Sec.~\ref{sec:active_sterile_mixing}, we first examine the scenario where all EFT WCs are zero, $C_i = 0$, and the active-sterile mixing is significant for Dirac HNLs, $V_{\alpha N} \neq 0$. In Sec.~\ref{sec:ff_monophoton}, we then explore the opposite limit, where the active-sterile mixing is negligible and the EFT WCs are important for either Majorana or Dirac HNLs. 

We consider two of the proposed centre-of-mass energies at FCC-ee, $\sqrt{s} = 91.2$~GeV (Tera-$Z$) and $\sqrt{s} = 240$~GeV ($Zh$) with the forecasted integrated luminosities $\mathcal{L} = 100~\rm{ab}^{-1}$ and $\mathcal{L} = 5~\rm{ab}^{-1}$, respectively. The two other possible centre-of-mass energies, $\sqrt{s} = 161$~GeV ($W^+W^-$) and $\sqrt{s} = 350/365$~GeV ($t\bar{t}$) can also be considered, but are likely to provide intermediate or weaker sensitivities with respect to $\sqrt{s} = 91.2$~GeV and $240$~GeV due to the smaller luminosity in the $t\bar{t}$ scenario~\cite{Bernardi:2022hny} and a reduced HNL mass reach for the intermediate $W^+ W^-$ centre-of-mass energy.

%%%%%%%%%%%%%%%%%%%%%%%%%%%%%%%%%%%%%%%%
\subsection{Active-Sterile Mixing}
\label{sec:active_sterile_mixing}
%%%%%%%%%%%%%%%%%%%%%%%%%%%%%%%%%%%%%%%%

At FCC-ee, the active-sterile mixing $V_{\alpha N}$ can lead to $e^+e^- \to \nu N (\gamma)$ via the centre and right diagrams of Fig.~\ref{fig:feynman-diagrams}, with each circle denoting an insertion of $V_{\alpha N}$. The $t$-channel $W^\pm$ exchange is only present for $\alpha = e$, while the $s$-channel $Z$ exchange is present for $\alpha = e, \mu, \tau$. The general $2\to 2$ cross section for $\ell_\alpha^+ \ell_\beta^- \to \nu_{i} N_j$ (Majorana) and $\ell_\alpha^+ \ell_\beta^- \to \nu_i \bar{N}_{j} + \bar{\nu}_i N_j$ (Dirac) is given in Eq.~\eqref{eq:tot_Maj_cs_mixing}, which can be calculated from the Lagrangian in Eq.~\eqref{eq:L_mix}. The corresponding $2\to 3$ cross sections with a final-state photon may be approximated by applying Eq.~\eqref{eq:xsec_monophoton} to Eq.~\eqref{eq:tot_Maj_cs_mixing}. To observe an exclusive monophoton plus $\slashed{E}$ signal, we require that the HNL does not decay inside the detector. For $V_{\alpha N} \neq 0$, $e^+e^- \to NN(\gamma)$ can also proceed via the $Z$ exchange diagram, but the cross section is suppressed by two additional powers of $V_{\alpha N}$ and is thus neglected. 

We also note that in general $V_{\alpha N}$ modifies (leads to a decrease in the cross section for) the SM process $e^+e^- \to \nu \nu (\gamma)$ via the non-unitary modifications in Eq.~\eqref{eq:L_mix}. In the limit of large HNL masses, this becomes the only impact of the active-sterile mixing. Being equivalent to the WCs $C_{Hl}^{(1)} = -C_{Hl}^{(3)}$, we indirectly constrain the active-sterile in the $\nu$SMEFT analysis, discussed in more detail in Sec.~\ref{sec:discussion}. The non-unitarity also affects the massless HNL limit; when the HNL masses are negligible with respect to the centre-of-mass energy, the decrease in the cross section for $e^+e^- \to \nu \nu(\gamma)$ exactly cancels the increase from $e^+e^- \to \nu N(\gamma)/NN(\gamma)$. This is simply due to the unitarity of the extended mixing matrix. The active-sterile mixing thus leads to no deviation from the SM prediction for $e^+e^- \to \nu \nu(\gamma)$, and any constraints on $V_{\alpha N}$ must vanish. However, non-unitarity also induces shifts in input parameters to the cross section, also discussed in Sec.~\ref{sec:discussion}. For the purposes of this section, we neglect the effect of non-unitarity in the light neutrino mixing and input scheme, treating the SM prediction for $e^+e^- \to \nu \nu\gamma$ as a background to $e^+e^- \to \nu N\gamma$.

To simulate the $e^+e^- \to \nu N \gamma$ process in \texttt{MadGraph5\_aMC@NLO}, we use the UFO output of the \texttt{Feynrules} model file \texttt{SM HeavyN Dirac CKM Masses LO}~\cite{Pascoli:2018heg}. The model file adds three generations of Dirac HNLs, $N_{1,2,3}$, with $N \equiv N_1$ taken to be the lightest. For now we only consider a Dirac HNL; the difference in the sensitivities for Dirac and Majorana HNLs is similar to that the EFT operators, so we defer that comparison to Sec.~\ref{sec:ff_monophoton}. For simplicity, we consider only the electron-flavour mixing, with the mono-$\gamma$ plus $\slashed{E}$ signature constraining the $(m_N, |V_{eN}|^2)$ parameter space. Similar constraints can be placed on the other flavour mixing strengths, $|V_{\mu N}|^2$ and $|V_{\tau N}|^2$, with small differences arising from the smaller cross sections in those scenarios (only $Z$ exchange contributing).

%%%%%%%%%%%%%%%%%%%%%%%%%%%%%%%%%%%%%%%%
\subsubsection{Sensitivity Estimate}
\label{sec:mixing_sim}
%%%%%%%%%%%%%%%%%%%%%%%%%%%%%%%%%%%%%%%%

Using \texttt{MadGraph5\_aMC@NLO}, we simulate the signal process $e^+ e^- \to \nu_e \bar{N}\gamma+ \bar{\nu}_e N\gamma$ alongside the irreducible SM background process $e^+ e^- \rightarrow \sum \nu \bar{\nu} \gamma$, each with $N_{\text{tot}} = 5\times 10^4$ events, requiring  $p_T^\gamma > 1 $~GeV at generator level. In the case of the signal process, the simulation is carried out for HNL masses up to the kinematic threshold $m_N \leq \sqrt{s}$.

\begin{figure}[t!]
\centering
\includegraphics[width=0.49\textwidth]{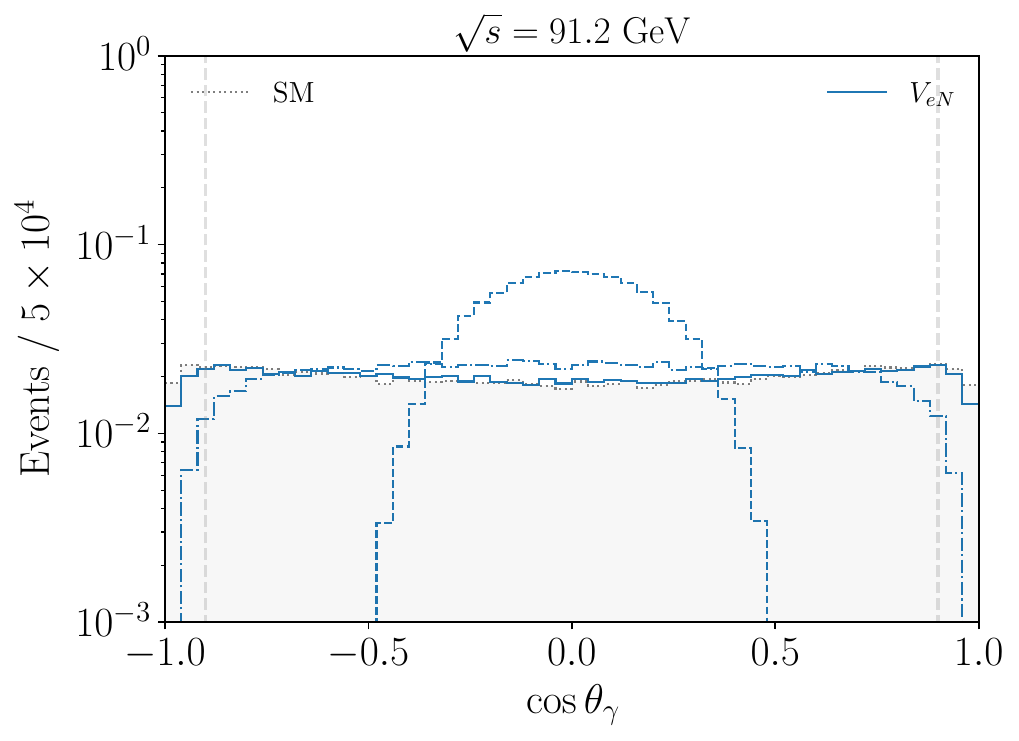}
\includegraphics[width=0.49\textwidth]{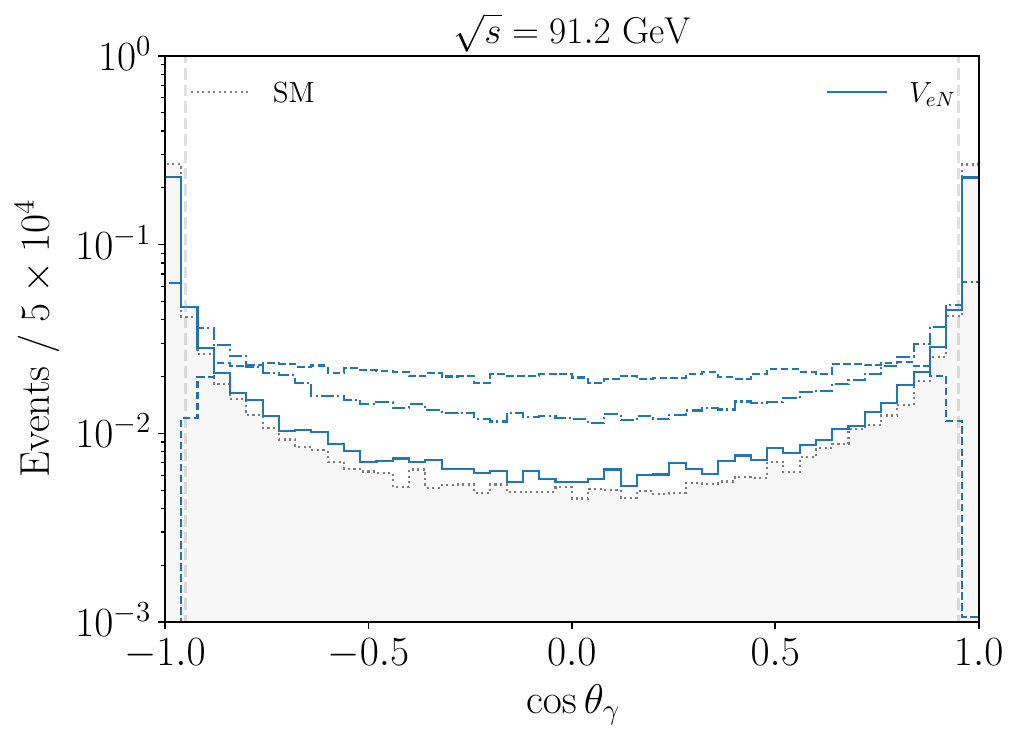}
\includegraphics[width=0.49\textwidth]{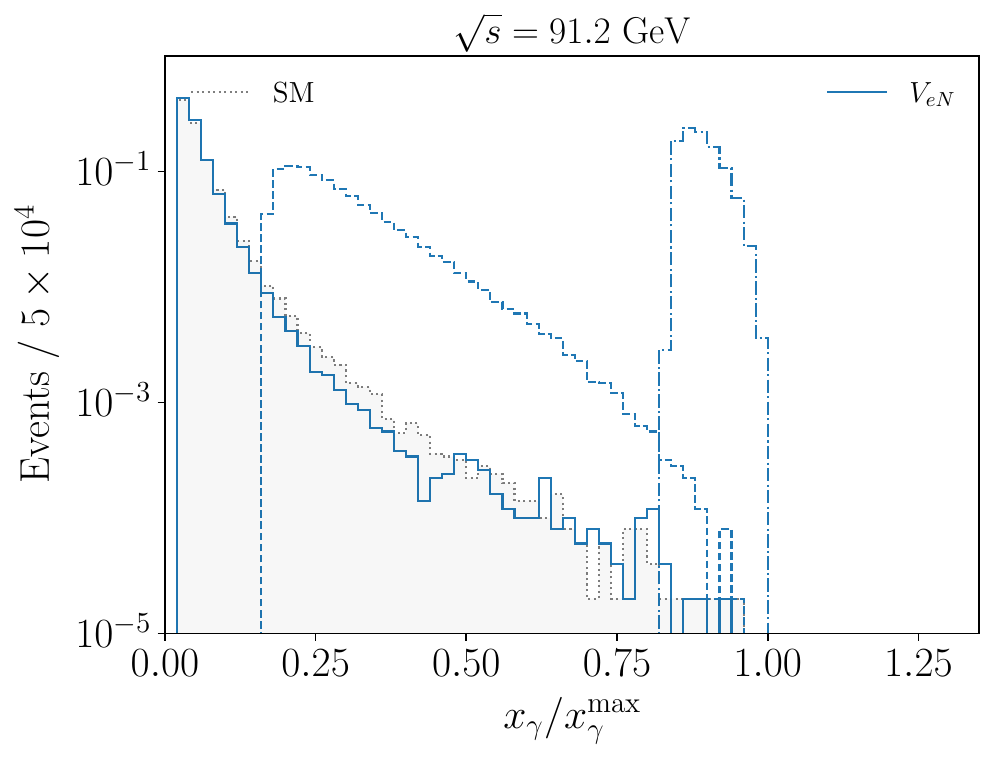}
\includegraphics[width=0.49\textwidth]{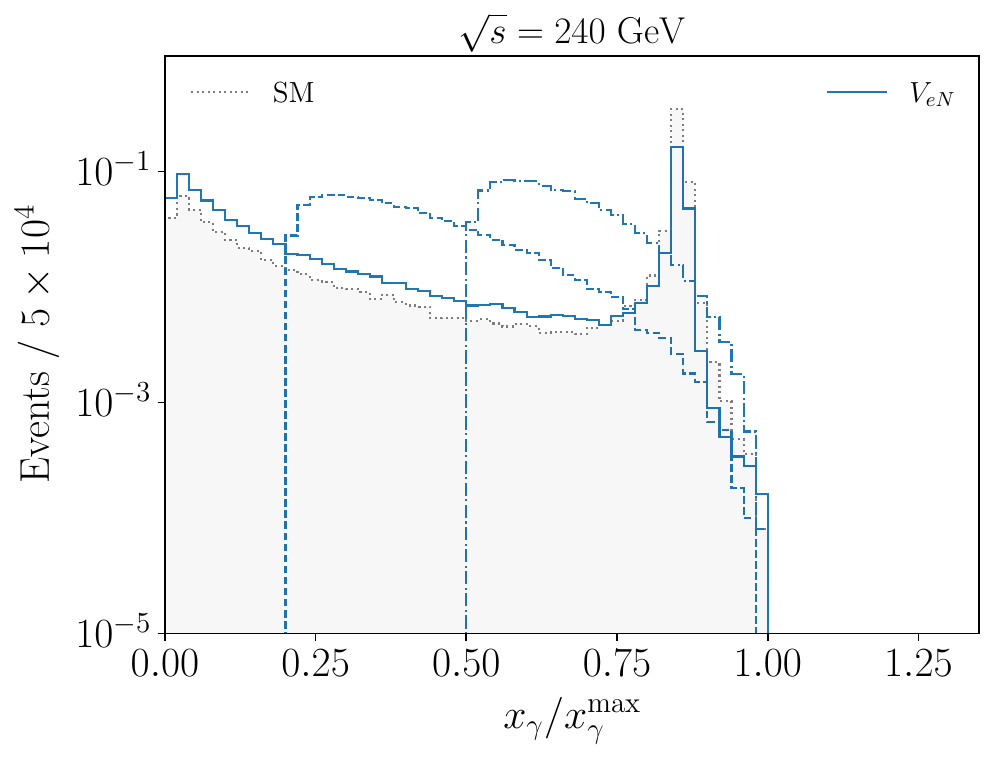}
\caption{Normalised binned distributions in the outgoing photon angle $\cos\theta_\gamma$ (above) and energy $x_\gamma = 2E_\gamma/\sqrt{s}$ (below) for mono-$\gamma$ processes induced by the electron-flavour mixing $V_{eN}$ and SM background in the Dirac HNL scenario. Distributions are shown for $\sqrt{s} = 91.2$~GeV (left) and $\sqrt{s} = 240$~GeV (right). Solid lines indicate the scenario with $m_N = 10$~GeV. We also show the distributions for $m_N$ close to the kinematic threshold using dashed and dot-dashed lines.}
\label{fig:kinematics_active_sterile_monophoton}
\end{figure}

For the simulated signal and backgrounds at $\sqrt{s} = 91.2$~GeV and~$240$~GeV, we apply the kinematic acceptance cuts shown in Tab.~\ref{tab:active_sterile_cuts} to the outgoing photon angle $\theta_\gamma$ and energy $E_\gamma$. These cuts are designed to maximise the signal-to-background ratio given the different signal and background distributions in $\cos\theta_\gamma$ and $E_\gamma$ (or equivalently, $x_\gamma = 2E_\gamma /\sqrt{s}$), shown in Fig.~\ref{fig:kinematics_active_sterile_monophoton}. In the third column of this table, we show the ratios of signal-to-background ratios before and after implementing the kinematic cuts for a benchmark scenario with $m_N = 10$~GeV and $|V_{eN}| = 10^{-3}$. Due to the similarity between the signal and background on the $Z$ resonance, no significant improvement is achieved. The distribution in $x_\gamma$ in units of the maximum possible photon energy,
\begin{align}
x_\gamma^{\rm{max}} = \frac{2E_\gamma^{\rm{max}}}{\sqrt{s}} = 1- \frac{m_N^2}{s} \,.
\end{align}
For $m_N = 10$~GeV (solid, blue), the distributions are compared to the SM background (grey dashed, shaded). The distributions for $m_N$ close to the kinematic threshold are also shown.

\begin{table}[t!]
\centering
\renewcommand{\arraystretch}{1.25}
\setlength\tabcolsep{2.7pt}
\begin{tabular}{c|c|c}
\hline
$\sqrt{s}$~[GeV]  & Cuts & $\left(\frac{S}{B}\right)_\text{cuts}/\left(\frac{S}{B}\right)$ \\\hline
91.2 & $|\cos\theta_\gamma| < 0.9$, $E_\gamma < 4$ GeV & 1.03 \\
240 & $|\cos\theta_\gamma| < 0.95$, $E_\gamma < 90$ GeV & 1.50 \\\hline
\end{tabular}
\caption{Kinematic cuts used to maximise the signal-to-background ratio for each $\sqrt{s}$ in the active-sterile mixing sensitivity analysis. The improvement in the signal-to-background ratios after cuts is shown in the last column for the benchmark scenario $m_N = 10$~GeV and $|V_{eN}| = 10^{-3}$.}
\label{tab:active_sterile_cuts}
\end{table}
Along with these cuts, we note that the HNL can decay to SM particles via the active-sterile mixing with diagrams such as those in Fig.~\ref{fig:feynman-diagrams-decay}, which show the contribution of $V_{\alpha N}$ to the leptonic decays $N \to \nu \ell^-\ell^+$. Other leptonic and semi-leptonic decays channels are open, such as $N \to \nu \nu \bar{\nu}$, $N \to \nu q \bar{q}$ and $N \to \ell^- u \bar{d}$, with the quarks hadronising to form single pseudoscalar and vector mesons for HNL masses below the QCD scale, and forming multi-hadron final states/jets above. The total Dirac HNL width can be found from the general decay rates in App.~\ref{app:decays} by rotating the SM charged and neutral-current interactions to the mass basis. After this, we find the same total width $\Gamma_{N}$ for the Dirac HNL as explored in~\cite{Atre:2009rg, Bondarenko:2018ptm, Coloma:2020lgy, Feng:2024zfe}. The total decay width in the Majorana HNL scenario is twice as large.

We first consider mono-$\gamma$ plus $\slashed{E}$ signal, and do not consider HNL decays, effectively treating it as a stable particle. However, for the parameter space considered in this analysis, HNLs can indeed decay inside the detector, which results in a displaced vertex (DV) signature. Therefore, the mono-$\gamma$ analysis can be considered as an inclusive analysis if the HNL decays are not considered, while it can be considered as an exclusive analysis if mono-$\gamma$ has no DV signature associated with it, i.e, when HNLs decay outside the detector volume. The HNL decay probability therefore needs to be appropriately accounted for.

The HNL decay probability is a function of detector geometry and boost. Based on the preliminary FCC-ee proposal~\cite{FCC:2018byv}, we approximate the detector geometry and consider it as a spherical detector of radius $L = 5$~m. The probability of the HNL decaying within the interval $\lbrace L_1, L_2 \rbrace$ can be written as,
\begin{align}
\label{eq:prob_decay}
\mathcal{P}_{\text{in}}(L_1, L_2, \sqrt{s}, m_N, V_{\alpha N}) = \int db \, f(\sqrt{s}, m_N, b)\left[e^{-L_1/b\tau_N}-e^{-L_2/b\tau_N}\right] \,,
\end{align}
where $b \equiv \beta\gamma$ is the boost factor of the HNL with the probability distribution $f(\sqrt{s}, m_N, b)$ and $\tau_N = 1/\Gamma_N$ is the proper lifetime of the HNL. As we consider a $2\to 3$ scattering process, the boost distribution does not have a simple analytical form like the equivalent $2\to 2$ process without the photon, where $f(\sqrt{s}, m_N, b) = \delta(b - b')$, with $b' = (s - m_N^2)/(2m_N\sqrt{s})$. We follow a simple approach to overcome this difficulty; on an event-by-event basis for mono-$\gamma$ events that pass the kinematic cuts in Tab.~\ref{tab:active_sterile_cuts}, we take the boost factor of the HNL directly from the MadGraph simulation. For each event $i$, the probability of the HNL decaying outside the FCC-ee fiducial volume is obtained as $\mathcal{P}_{\text{out}}^{i} \equiv 1 - \mathcal{P}_{\text{in}}^{i}$, where $\mathcal{P}_{\text{in}}^{i}$ is found by setting $f(\sqrt{s}, m_N, b) = \delta(b - b_i)$ in Eq.~\eqref{eq:prob_decay}. Then, the overall geometric acceptance can be approximated as $\mathcal{P}_\text{out} \approx (b_\text{max} - b_{\text{min}})\sum_i \mathcal{P}_\text{out}^i/(\epsilon_k N_{\text{tot}})$, where $\epsilon_k N_{\text{tot}}$ is the number of events surviving the kinematic cuts. Through the boost distribution $f$ and the proper lifetime $\tau_N$, the estimated probability $\mathcal{P}_{\mathrm{out}}$ depends the HNL mass $m_N$ and the active-sterile mixing strength $V_{\alpha N}$.

The method above defines sensitivity for the exclusive mono-$\gamma$ plus $\slashed{E}$ signature at FCC-ee. Complementary to this, the inclusive mono-$\gamma$ search relaxes the condition of HNL stability in the fiducial volume. Clearly, for the inclusive search one cannot identify the photon energy with the missing energy due to the presence of additional potentially prompt visible final states. This also implies that the inclusive analysis may have additional backgrounds, which we do not account for. Our sensitivity estimate is still applicable if the HNLs can decay to additional exotic invisible final states, thereby suppressing the branching ratios for visible decays in the detector. For the inclusive search, we set $\mathcal{P}_{\text{out}} = 1$. We generally assume that the HNLs decay outside the detector or have lifetimes long enough to be considered stable. An example scenario with a potentially dominant invisible decay mode is discussed in Ref.~\cite{Deppisch:2024izn} where the HNL couples to a light axion-like pseudoscalar particle $a$, leading to the two-body decay $N\to a\nu$. Satisfying experimental and cosmological constraints, this decay mode can be dominant for HNL masses $1~\text{GeV} \lesssim m_N \lesssim 100$~GeV in our region of interest with branching ratios to invisible states suppressed by factors of $10^{-6}$ ($m_N \approx 1$~GeV) to $10^{-2}$ ($m_N \approx 100$~GeV). While the presence of an extra light exotic state $a$ would result in additional operators within an EFT framework, the operators discussed here will still be present.

After applying the kinematic cuts and estimating the geometric acceptance, the total number of surviving mono-$\gamma$ plus $\slashed{E}$ signal events is given by
\begin{align}
\label{eq:sig_events}
S = \mathcal{L} \times \sigma \times \mathcal{P}_{\text{out}}\times \epsilon_k \,,
\end{align}
which depends on the HNL mass and active-sterile mixing through the cross section for the signal process, $\sigma$, and $\mathcal{P}_{\text{out}}$. The number of background events, $B$, is similarly found by multiplying the SM cross section by the integrated luminosity and the associated kinematic efficiency. To determine the sensitivity to the active-sterile mixing, $V_{eN}$, we compute the median significance for a counting experiment of known background~\cite{Cowan:2010js},
\begin{align}
\label{eq:simp_sigsens}
\mathcal{S} = \sqrt{2\bigg((S + B)\ln\bigg(1 + \frac{S}{B}\bigg) - S\bigg)} \approx \frac{S}{\sqrt{B}} \,,
\end{align}
where in the second equality, we assume that for relevant parameter space the cross section for $e^+ e^- \rightarrow \sum \nu \bar{\nu}\gamma$ is much larger than that for $e^+ e^- \to \sum_{i}\nu_i \bar{N}\gamma+ \bar{\nu}_i N\gamma$, such that $S \ll B$. For each HNL mass point up to $m_N \leq \sqrt{s}$, the expected significance is computed for different values of the active-sterile mixing strength; bounds at 90\%~CL are then placed by determining the value(s) of $|V_{eN}|^2$ corresponding to $\mathcal{S} = 1.28$, which delimit an excluded region of the parameter space with $\mathcal{S} > 1.28$.

%%%%%%%%%%%%%%%%%%%%%%%%%%%%%%%%%%%%%%%%
\subsubsection{Results}
\label{sec:mixing_res}
%%%%%%%%%%%%%%%%%%%%%%%%%%%%%%%%%%%%%%%%

%
\begin{figure}[t!]
\centering    
\includegraphics[width=0.7\textwidth]{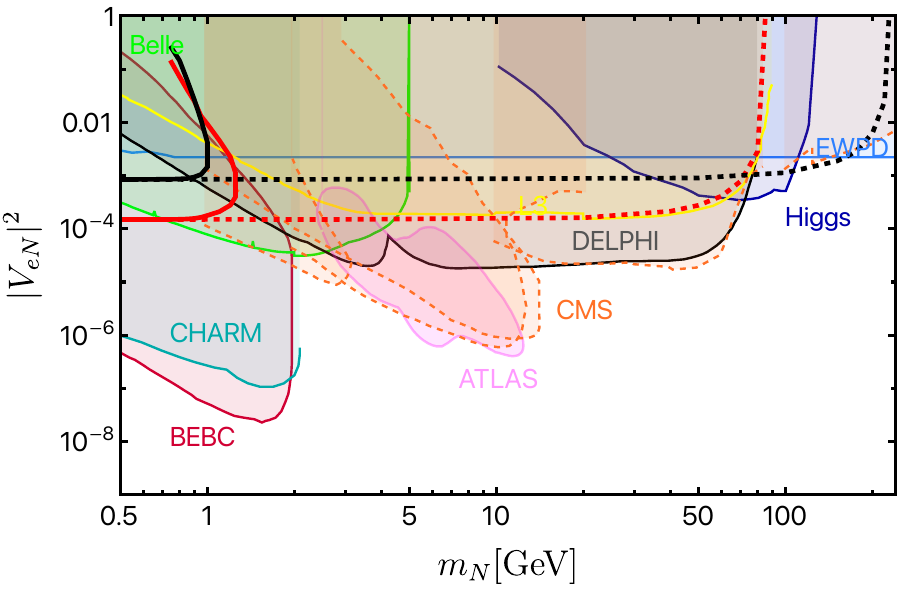}
\caption{Sensitivity of mono-$\gamma$ plus $\slashed{E}$ searches at FCC-ee to the electron-flavour mixing strength as a function of the HNL mass at $90\%$ CL, for $\sqrt{s} = 91.2$~GeV (red) and $\sqrt{s}
= 240$~GeV (black). Shown are the results of the exclusive signal analysis (solid), taking into account the probability of the HNL decaying outside the detector of length $L = 5$~m, and inclusive signal analysis (dashed), where this requirement is relaxed. The shaded regions correspond to the currently excluded regions of the parameter space.}
\label{fig:mixing_exclusions}
\end{figure}

Using the procedure outlined above, we present in Fig.~\ref{fig:mixing_exclusions} the estimated sensitivity of the mono-$\gamma$ plus $\slashed{E}$ search at FCC-ee to the electron-flavour mixing strength, $|V_{eN}|^2$, for HNL masses $m_N$ between $500$~MeV and $240$~GeV. The sensitivities are shown for the exclusive (solid) and inclusive searches (dashed) for $\sqrt{s} = 91.2$~GeV (red) and $\sqrt{s} = 240$~GeV (black), for a detector length of $L=5$~m. The shaded regions indicate the excluded regions of the parameter space from complementary beam dump, prompt and displaced vertex collider searches; BEBC~\cite{Barouki:2022bkt}, CHARM~\cite{CHARM:1985nku,CHARMII:1994jjr}, DELPHI~\cite{DELPHI:1996qcc}, ATLAS~\cite{ATLAS:2022atq}, CMS~\cite{CMS:2024bni}, Higgs decays~\cite{Das:2017zjc} and electroweak precision data (EWPD)~\cite{deBlas:2013gla}. 

We immediately observe that the $\sqrt{s} = 91.2$~GeV run is more sensitive by an order of magnitude with respect to $\sqrt{s} = 240$~GeV. This is expected from the enhancement of the cross section at the $Z$ pole and the $\sim 20$ times larger integrated luminosity. For the exclusive mono-$\gamma$ plus $\slashed{E}$ search, we see that the reach of FCC-ee is severely impacted for HNL masses above $m_N = 1$~GeV, where the HNLs are very unlikely to decay outside the detector and appear as missing energy. For each HNL mass, competing effects yield two $|V_{eN}|^2$ solutions for the condition $\mathcal{S} = 1.28$; for large values of the active-sterile mixing strength, both the cross section and the decay rate are large, leading to a small probability to decay outside the detector and a reduced value of $\mathcal{S}$. Conversely, for small values of the active-sterile mixing strength, the probability of decaying outside the fiducial volume is enhanced, but the cross section and therefore $\mathcal{S}$ is suppressed. For the inclusive signal, no such suppression from the geometric acceptance applies, removing the lower limits on $|V_{eN}|^2$ and extending the bounds up to the kinematic thresholds.

Overall, the reaches of the exclusive and inclusive searches are around $|V_{eN}|^2\sim 10^{-4}$ and $|V_{eN}|^2\sim 10^{-3}$ for $\sqrt{s} = 91.2$~GeV and $\sqrt{s} = 240$~GeV, respectively. In the parameter space depicted in Fig.~\ref{fig:mixing_exclusions}, the sensitivities therefore lie almost entirely within the region excluded by existing searches, unlike the proposed displaced vertex searches for HNLs, which can probe $|V_{eN}|^2 \sim 10^{-11}$ for $m_N \sim 30 - 60$~GeV~\cite{Blondel:2014bra,Antusch:2016vyf,Blondel:2022qqo,Ajmal:2024kwi}. Only the unconstrained region around $m_N \sim 80 - 130$~GeV, below the reach of EWPD constraints, can be excluded by the inclusive mono-$\gamma$ search at $\sqrt{s} = 240$~GeV. However, we note that the exclusive and inclusive search sensitivities both extend to lower values of the HNL mass. For $m_N \sim 10~\text{eV} - 2.5~\text{MeV}$, bounds from kink searches in $\beta$ decay spectra, excluding $|V_{eN}|^2 \gtrsim 10^{-4} - 10^{-3}$, are generally less stringent than the sensitivities presented here for the FCC-ee mono-$\gamma$ plus $\slashed{E}$ search at $\sqrt{s} = 91.2$~GeV.

%%%%%%%%%%%%%%%%%%%%%%%%%%%%%%%%%%%%%%%%
\subsection{EFT Operators}
\label{sec:ff_monophoton}
%%%%%%%%%%%%%%%%%%%%%%%%%%%%%%%%%%%%%%%%

Given the wide range of existing constraints on the active-sterile mixing strength, it is now interesting to explore the mono-$\gamma$ plus $\slashed{E}$ constraints on the WCs considered in Sec.~\ref{sec:model}, which are generally less constrained. 

At FCC-ee, the four-fermion operators and effective $W^\pm/Z$ interactions can lead to the single and pair production of HNLs, $e^+e^-\to \nu N (\gamma)$ and $e^+e^-\to N N (\gamma)$, respectively, via the diagrams in Fig.~\ref{fig:feynman-diagrams}. Heavy new physics may also contribute to operators involving two active neutrino fields, modifying the SM process $e^+e^-\to \nu\nu (\gamma)$. For Majorana and Dirac HNLs, the general $2\to 2$ scattering cross sections via the EFT operators are in Eqs.~\eqref{eq:tot_Maj_cs} and~\eqref{eq:tot_Dirac_cs}, respectively. The cross sections with an additional final-state photon can be approximated through the use of Eq.~\eqref{eq:xsec_monophoton}. For the analysis, we simulate the $2\to 3$ process in \texttt{MadGraph5\_aMC@NLO}, implementing the relevant operators in \texttt{Feynrules} and obtaining the necessary UFO input. The technical difficulty in \texttt{MadGraph5\_aMC@NLO} related to four-fermion operators containing two Majorana fermions is solved using the method described in App.~\ref{app:2to2_xsec}.

In this analysis, we turn on a single EFT operator coefficient $C_i$ in Sec.~\ref{subsec:EFT} at a time. The operators are taken to be in the mass basis, with the diagonalisation of the extended neutrino mass matrix resulting in two Majorana or Dirac HNLs, $N_1$ and $N_2$, with $m_{N_1} < m_{N_2}$. We consider:
\begin{itemize}
\item Diagonal WCs of the four-fermion and effective $Z$ interactions with two HNLs:
\begin{align}
\label{eq:diagonal}
C_i \in \Big\{C_{\underset{iiee}{Ne}}^{V,RR}\,,~
C_{\underset{iiee}{Ne}}^{S,RR}\,,~
C_{\underset{iiee}{Ne}}^{T,RR}\,,~
\frac{2}{v^2}[Z_{N}^R]_{ii}\Big\}\,,
\end{align}
with $i = 2$, which induce the pair production of HNLs via $e^+e^- \to N_2 N_2 \gamma$ (Majorana) or $e^+e^- \to N_2 \bar{N}_2 \gamma$ (Dirac). As the HNLs cannot decay via these WCs, only the mono-$\gamma$ plus $\slashed{E}$ signal can constrain these operators at FCC-ee. 

\item Off-diagonal WCs of four-fermion and effective $Z$ interactions with two HNLs:
\begin{align}
\label{eq:off-diagonal}
C_i \in \Big\{C_{\underset{ijee}{Ne}}^{V,RR}\,,~ 
C_{\underset{ijee}{Ne}}^{S,RR}\,,~
C_{\underset{ijee}{Ne}}^{T,RR}\,,~
\frac{2}{v^2}[Z_{N}^R]_{ij}\Big\}\,,
\end{align}
with $i = 1$ and $j = 2$, leading to HNL pair production via $e^+e^- \to N_1 N_2\gamma$ (Majorana) or $e^+e^- \to N_1 \bar{N}_2 \gamma + \bar{N}_1 N_2\gamma$ (Dirac). We consider three different mass splitting ratios, $\delta \equiv (m_{N_2} - m_{N_1})/m_{N_2}$, between the HNLs: $\delta = 0.01$, $0.1$, and $1$. Because $N_2$ can decay to $N_1$ via these WCs, the DV signature can also provide complementary constraints, as explored in Sec.~\ref{sec:displacedvertex}.

\item WCs of the four-fermion and effective $Z$ interactions involving a light neutrino and HNL, and effective $W^\pm$ interactions involving an HNL;
\begin{align}
\label{eq:one-light-one-heavy}
C_i \in \Big\{C_{\underset{\alpha jee}{\nu Ne}}^{V,RR}\,,~ 
C_{\underset{\alpha jee}{\nu Ne}}^{S,RR}\,,~
C_{\underset{\alpha jee}{\nu Ne}}^{T,RR}\,,~
\frac{2}{v^2}[W_N^R]_{je}\,,~
\frac{2}{v^2}[W_N^L]_{je}\,,~
\frac{2}{v^2}[Z_{\nu N}^R]_{\alpha j}\Big\}\,,
\end{align}
with $\alpha = e,\mu,\tau$ and $j = 2$, which lead to the single production of HNLs via the processes $e^+e^- \to \sum_{i}\nu_i N_2\gamma$ (Majorana) and $e^+e^- \to \sum_{i}\nu_i \bar{N}_2\gamma + \bar{\nu}_i N_2\gamma$ (Dirac). While the WCs $C_{\nu Ne}^{V,RR}$ and $Z_{\nu N}^{R}$ are not present in the Dirac case, we take them in the following as shorthand for $C_{\nu Ne}^{V,LR}$ and $Z_{\nu N}^{L}$, which are non-vanishing. For the four-fermion and effective $Z$ interaction WCs in Eq.~\eqref{eq:one-light-one-heavy}, the mono-$\gamma$ plus $\slashed{E}$ constraints can be obtained from those for the off-diagonal WCs in Eq.~\eqref{eq:off-diagonal}, with $\delta = 1$.

\item WCs of the four-fermion and effective $Z$ interactions with two light neutrinos and the effective $W^\pm$ interactions involving a light neutrino;
\begin{align}
\label{eq:two-light}
C_i \in \Big\{C_{\underset{\alpha\beta ee}{\nu e}}^{V,LL}\,,~
C_{\underset{\alpha\beta ee}{\nu e}}^{V,LR}\,,~
C_{\underset{\alpha\beta ee}{\nu e}}^{S,LL},~C_{\underset{\alpha\beta ee}{\nu e}}^{T,LL}\,,~\frac{2}{v^2}[W_\nu^R]_{\alpha e}\,,~
\frac{2}{v^2}[\delta W_\nu^L]_{\alpha e}\,,~
\frac{2}{v^2}[\delta Z_{\nu}^L]_{\alpha\beta}\Big\}\,,
\end{align}
with $\alpha,\beta = e,\mu,\tau$, modifying the SM process $e^+e^- \to \sum \nu \bar{\nu} \gamma$. We note that the WCs $C_{\nu e}^{S,LL}$, $C_{\nu e}^{T,LL}$ and $W_\nu^R$ are not present in the Dirac case. In the Majorana case, these operators do not interfere with the SM; thus, constraints on them can be obtained from the limits on the diagonal and off-diagonal WCs in Eqs.~\eqref{eq:diagonal} and~\eqref{eq:off-diagonal}, respectively, in the limit $m_{N_2} \to 0$. New physics contributions to $C_{\nu e}^{V,LL}$, $C_{\nu e}^{V,LR}$, $\delta W_\nu^L$ and $\delta Z_\nu^L$, meanwhile, interfere with the SM, depending on the flavour of the fields involved. Interference with the SM can be the leading effect of these operators, and must be taken into account when deriving limits on the associated $\nu$SMEFT WCs in Sec.~\ref{sec:discussion}.

\end{itemize}
%

%%%%%%%%%%%%%%%%%%%%%%%%%%%%%%%%%%%%%%%%
\subsubsection{Sensitivity Estimate}
\label{sec:EFT_sim}
%%%%%%%%%%%%%%%%%%%%%%%%%%%%%%%%%%%%%%%%

The sensitivity analysis for the EFT operators now proceeds similarly to the active-sterile mixing analysis in Sec.~\ref{sec:mixing_sim}, with the simulation performed in \texttt{MadGraph5\_aMC@NLO}. Firstly, for the four-fermion operators, we simulate the signal processes $e^+ e^- \to N_2 \bar{N}_2 \gamma$ and $e^+ e^- \to N_1 \bar{N}_2 \gamma + \bar{N}_1 N_2 \gamma$ for the diagonal and off-diagonal WCs, respectively. As described in App.~\ref{app:2to2_xsec}, Majorana HNLs are treated as Dirac HNLs in the simulation, with the definition of WCs in the \texttt{FeynRules} model file ensuring the correct behaviour of the Majorana four-fermion operators. For the effective $W^\pm$ and $Z$ interactions, the HNLs can instead be treated as Majorana fermions in \texttt{MadGraph5\_aMC@NLO}. For the former, the processes $e^+e^- \to \nu_e N_2 \gamma + \bar{\nu}_e N_2 \gamma$ and $e^+e^- \to \nu_e \bar{N}_2 \gamma + \bar{\nu}_e N_2 \gamma$ are simulated in the Majorana and Dirac cases, respectively. For the latter, we simulate $e^+e^- \to N_2 N_2 \gamma$ (Majorana) and $e^+e^- \to N_2 \bar{N}_2 \gamma$ (Dirac) for the diagonal WCs and $e^+e^- \to N_1 N_2 \gamma$ (Majorana) and $e^+e^- \to N_1 \bar{N}_2 \gamma + \bar{N}_1 N_2 \gamma$ (Dirac) for the off-diagonal WCs. For all signal scenarios, $N_{\text{tot}} = 5 \times 10^4$ events are generated. For the irreducible SM background, we use the same simulated events as in Sec.~\ref{sec:mixing_sim}.

\begin{table}[t!]
\centering
\renewcommand{\arraystretch}{1.25}
\setlength\tabcolsep{2.7pt}
\begin{tabular}{c|c|c|c|c|c|c}
\hline
\multirow{2}{*}{$\sqrt{s}$~[GeV]}  & \multirow{2}{*}{Cuts} & \multicolumn{5}{c}{$\left(\frac{S}{B}\right)_\text{cuts} / \left(\frac{S}{B}\right)$} \\ \cline{3-7}
  &  & $C_{Ne}^{V,RR}$ & $C_{Ne}^{S,RR}$ & $C_{Ne}^{T,RR}$ & $W_{N}^{R}$ & $Z_{N}^{R}$  \\ \hline
91.2 & $|\cos\theta_\gamma| < 0.4$,  $|\cos\theta_\gamma| > 0.8$ & 1.21 & 1.20 & 1.20 & 1.22  & 1.01 \\
240 & $|\cos\theta_\gamma| <0.95$, $E_{\gamma}<40$~GeV & 2.30  & 2.30 & 2.23 & 2.00 & 0.05 \\\hline

\end{tabular}
\caption{Universal kinematic cuts for maximising the signal-to-background ratio for each $\sqrt{s}$ in the EFT operator sensitivity analysis, in both the Majorana and Dirac HNL scenarios. The improvement in the signal-to-background ratios after cuts is shown in the last column, for benchmark scenarios involving a Dirac HNL with $m_{N_2} = 10$~GeV.}
\label{tab:universal_cuts}
\end{table}

\begin{figure}[t!]
\centering
\includegraphics[width=0.49\textwidth]{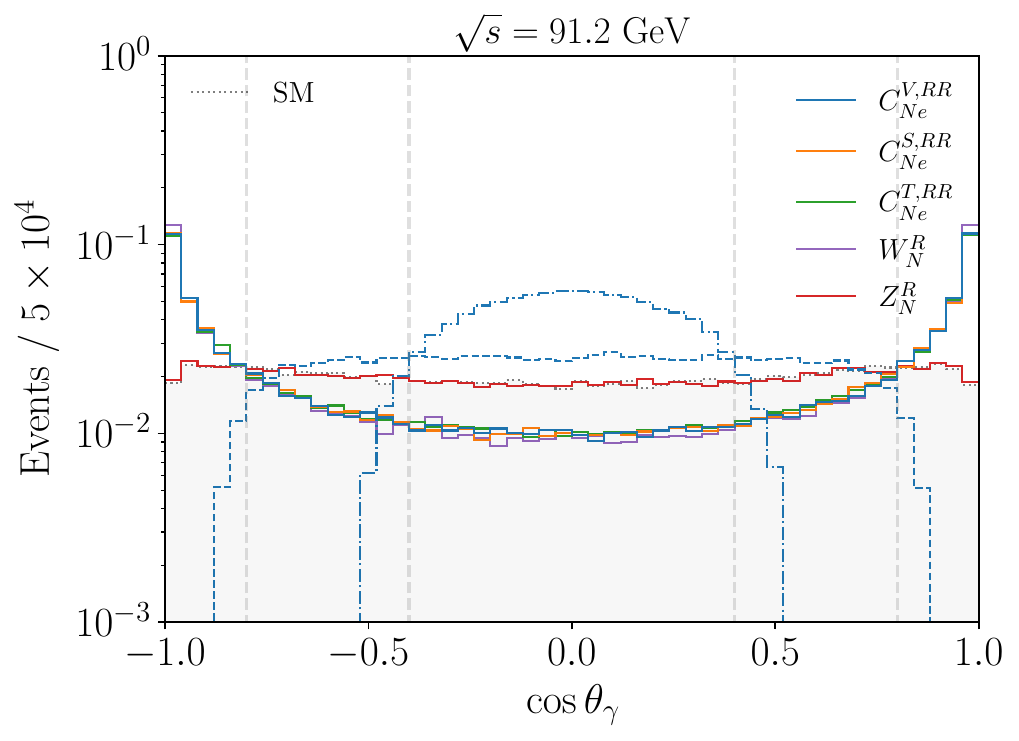}
\includegraphics[width=0.49\textwidth]{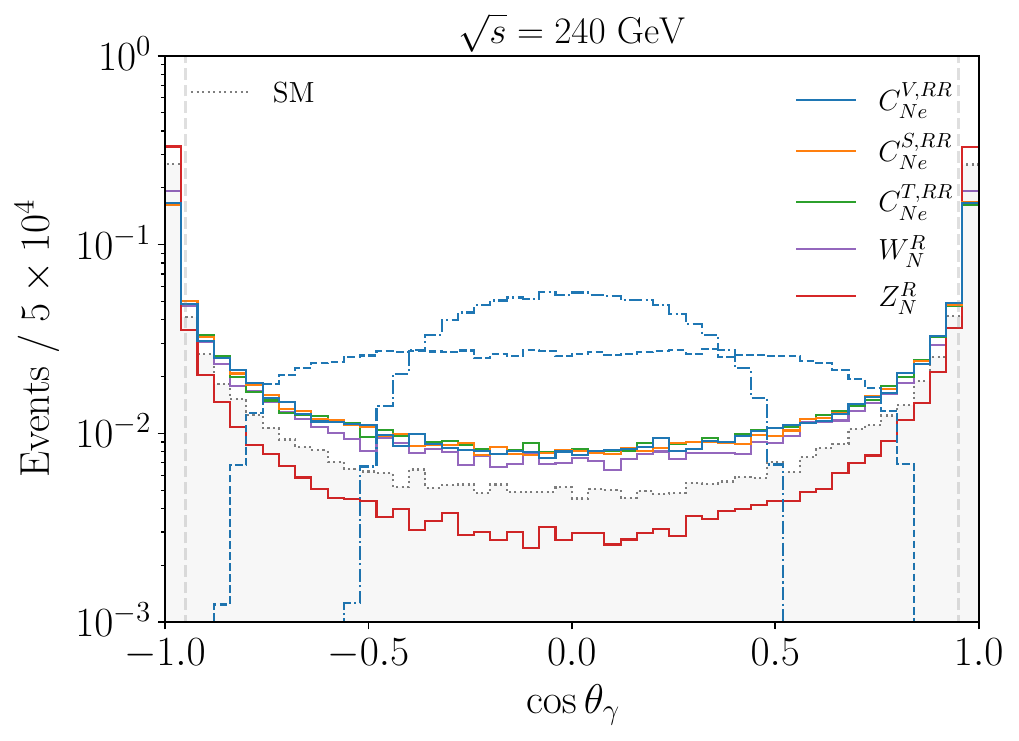}
\includegraphics[width=0.49\textwidth]{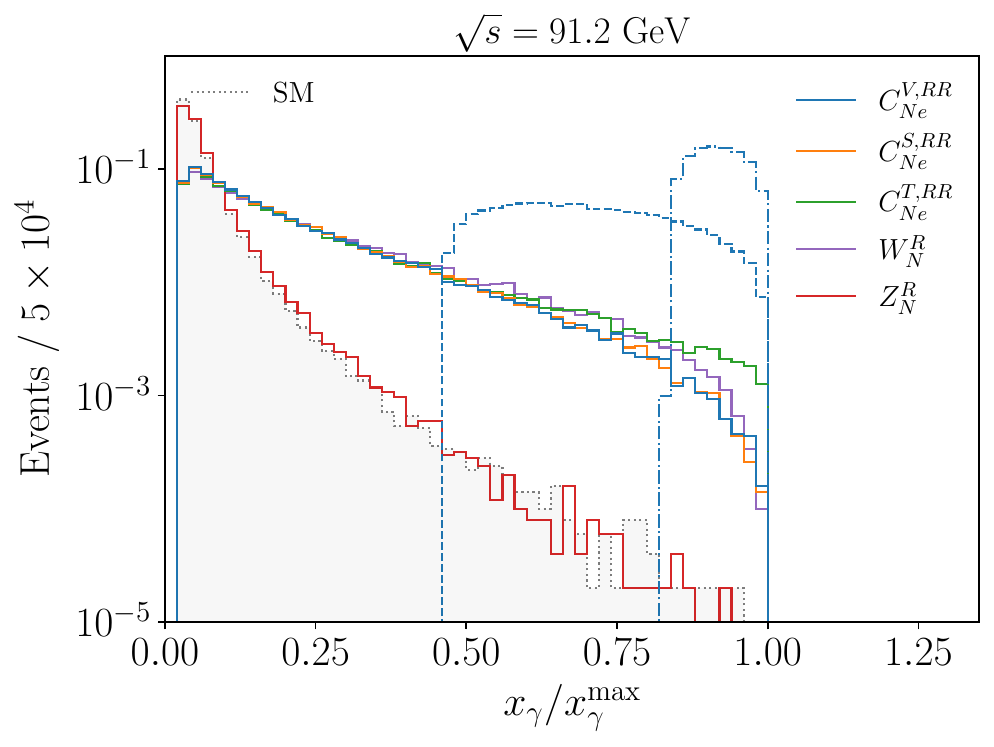}
\includegraphics[width=0.49\textwidth]{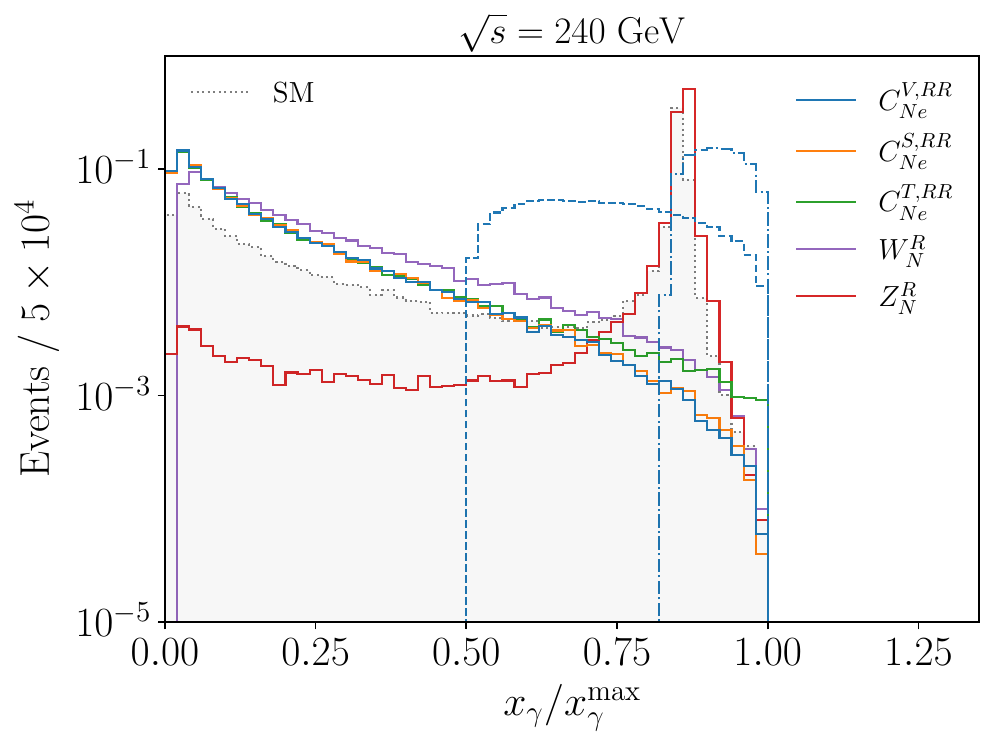}
\caption{Normalised binned distributions in the cosine of the outgoing photon angle $\cos\theta_\gamma$ (above) and energy $x_\gamma = 2E_\gamma/\sqrt{s}$ (below) for mono-$\gamma$ processes induced by the four-fermion, effective $W^\pm$ and $Z$ interactions and SM background in the Dirac HNL scenario. Distributions are shown for $\sqrt{s} = 91.2$~GeV (left) and $\sqrt{s} = 240$~GeV (right). Solid lines indicate the scenario with $m_{N_2} = 10$~GeV. For the vector four-fermion operator, we also show the distributions for $m_{N_2}$ close to the kinematic threshold, see text for details.}
\label{fig:kinematics_ff_monophoton}
\end{figure}

We generate signal and background samples with a generator level cut of $p_T^\gamma > 1 $~GeV. For the signal processes, we repeat the simulation for HNL masses up to the kinematic threshold $m_{N_2} \leq \sqrt{s}/(2 - \delta)$, with $\delta = 0$ for the diagonal WCs. In Tab.~\ref{tab:universal_cuts}, we show the universal kinematic cuts applied to the signals and backgrounds for $\sqrt{s} = 91.2$~GeV and $\sqrt{s} = 240$~GeV. These cuts are informed by the distributions in the outgoing photon energy $E_\gamma$ and angle $\theta_\gamma$. In the third column of this table, we show the improvement in the signal-to-background ratios after implementing the kinematic cuts for
the diagonal four-fermion, effective $W^+$, and diagonal effective $Z$ interactions involving a Dirac HNL with $m_{N_2} = 10$~GeV. Similar to the active-sterile mixing scenario, the implemented cuts eliminate a significant fraction of background events, except for $Z_N^R$ at $\sqrt{s} = 240$~GeV. In Fig.~\ref{fig:kinematics_ff_monophoton}, we illustrate the normalised $\cos\theta_\gamma$ (top panel) and $x_\gamma \equiv 2E_\gamma/\sqrt{s}$ (bottom panel) distributions in the Dirac HNL scenario for $\sqrt{s} = 91.2$~GeV (left) and $\sqrt{s} = 240$~GeV (right). As in Fig.~\ref{fig:kinematics_active_sterile_monophoton}, the distributions in $x_\gamma$ are shown in units of the maximum possible photon energy,
\begin{align}
x_\gamma^{\text{max}} \equiv \frac{2E_\gamma^\text{max}}{\sqrt{s}}=  1 - \frac{m_{N_2}^2(2 -\delta)^2}{s}\,.
\end{align}
In Fig.~\ref{fig:kinematics_ff_monophoton}, the solid lines show the normalised distributions for $m_{N_2} = 10$~GeV and non-zero values of the diagonal WCs in Eq.~\eqref{eq:diagonal}, which induce $e^+e^- \to N_2 \bar{N}_2\gamma$, and a non-zero value of the coefficient $[W_N^R]_{2e}$, leading to $e^+e^- \to \sum_{i}\nu_i \bar{N}_2\gamma + \bar{\nu}_i N_2\gamma$. These are compared to the distributions for the SM background (grey dashed and shaded).

The distributions in $\cos\theta_\gamma$ show the expected behaviour for initial state radiation, with peaks in the forwards and backwards directions. As discussed in App.~\ref{app:2to2_xsec}, the distributions can be approximated by multiplying the $2\to 2$ cross sections by a radiator function, as in Eq.~\eqref{eq:xsec_monophoton}. This predicts identical $\cos\theta_\gamma$ distributions for all of the processes in question, and we see that this is the case for the four-fermion and effective $W^\pm$ interactions. However, we observe that for the SM background and the effective $Z$ interaction, the $\cos\theta_\gamma$ distribution is less prominent for $|\cos\theta_\gamma| \gtrsim 0.8$ and $|\cos\theta_\gamma| \lesssim 0.95$ for $\sqrt{s} = 91.2$~GeV and $\sqrt{s} = 240$~GeV, respectively; as a consequence of the $p_T^\gamma > 1$~GeV cut on these $s$-channel processes. Thus, for the four-fermion and effective $W^\pm$ interactions, $|\cos\theta_\gamma| > 0.8$ and $|\cos\theta_\gamma| < 0.95$ are sensible universal cuts to minimise the SM background. 

However, we note that for HNL masses near the kinematic threshold, the distributions in $\cos\theta_\gamma$ change considerably. In Fig.~\ref{fig:kinematics_ff_monophoton}, we plot the normalised distributions for non-zero values of the vector four-fermion coefficient $C_{Ne}^{V,RR}$ with $m_{N_2} = 44.5$~GeV (dashed) and $45$~GeV (dot-dashed) for $\sqrt{s} = 91.2$~GeV and $m_{N_2} = 119$~GeV (dashed) and $119.4$~GeV (dot-dashed) for $\sqrt{s} = 240$~GeV. The distributions now peak at $\cos\theta_\gamma = 0$ and lose support at $\cos\theta_\gamma \sim \pm 1$. For $\sqrt{s}  = 91.2$~GeV, the cut $|\cos\theta_\gamma| > 0.8$ therefore removes too much of the signal. To avoid this, we also allow events with $|\cos\theta_\gamma| < 0.4$.

The distributions in $x_\gamma$ also agree with the approximate expression in Eq.~\eqref{eq:xsec_monophoton}, with the distributions for the four-fermion and effective $W^\pm$ interactions all decreasing up to the threshold $x_\gamma = x_\gamma^{\text{max}}$. For the SM background and effective $Z$ interactions, the replacement $s\to s(1-x_\gamma)$ in the $2\to 2$ cross section results in a peak at $s(1-x_\gamma) = M_Z^2$, or $x_\gamma = 1 - M_Z^2/s$. As there is no peak for $\sqrt{s} = 91.2$~GeV, we find that a further cut on $E_\gamma$ does not noticeably improve the signal-to-background ratio. For $\sqrt{s} = 240$~GeV case, the peak occurs at $x_\gamma \approx 0.85$; we therefore apply the universal cut $E_\gamma < 40$~GeV. As mentioned previously, this improves the signal-to-background ratio for all scenarios except the effective $Z$ interaction. In Fig.~\ref{fig:kinematics_ff_monophoton}, the $E_\gamma$ distribution for $E_\gamma < 40$~GeV (or $x_\gamma/x_\gamma^{\text{max}}< 0.33$) can be seen to be suppressed with respect to the SM background. While different cuts could be placed on $\cos\theta_\gamma$ and $E_\gamma$ to obtain increased kinematic efficiencies in this case, we do not expect a large improvement in the sensitivity. The universal cuts in Table~\ref{tab:universal_cuts} are applied in both the Majorana and Dirac HNL scenarios, as there are only minor differences in the $\cos\theta_\gamma$ and $E_\gamma$ distributions. They are also applied for all considered HNL masses; for each HNL mass, tailored cuts can be used to further improve the signal-to-background ratios. However, we find that these do not significantly increase the sensitivities with respect to the universal cuts. 

The method used to obtain the geometric acceptance in this analysis is identical to that used in Sec.~\ref{sec:mixing_sim}. For the diagonal WCs in Eq.~\eqref{eq:diagonal}, $N_2$ cannot decay and is therefore stable, giving $\mathcal{P}_{\text{out}} = 1$. For the off-diagonal WCs in Eqs.~\eqref{eq:off-diagonal} and~\eqref{eq:one-light-one-heavy}, however the heavier HNL $N_2$ can decay. For example, the decays $N_2 \to \nu e^- e^+/N_1 e^-e^+$ are possible, as shown in Fig.~\ref{fig:feynman-diagrams-decay}. Additional decay modes are present for the effective $W^\pm$ and $Z$ interactions, such as $N_2 \to \nu \nu \bar{\nu}/N_1 \nu \bar{\nu}$, $N_2 \to \nu q \bar{q}/N_1 q \bar{q}$ and $N_2 \to \ell^- u \bar{d}$. In App.~\ref{app:decays}, we give expressions for the total decay width of $N_2$ in the Majorana and Dirac HNL scenarios. These can be used in Eq.~\eqref{eq:prob_decay} to calculate the probability of the HNL decaying outside the detector on an event-by-event basis, and finally the overall geometric acceptance $\mathcal{P}_{\text{out}}$ for the exclusive mono-$\gamma$ plus $\slashed{E}$ signal. As in Sec.~\ref{sec:active_sterile_mixing}, we also consider an inclusive mono-$\gamma$ search, with $\mathcal{P}_{\text{out}} = 1$.

To estimate the sensitivity of mono-$\gamma$ plus $\slashed{E}$ searches at FCC-ee to the EFT operator WCs in Eqs.~\eqref{eq:diagonal}--\eqref{eq:one-light-one-heavy}, the total number of signal events, $S$, after kinematic and geometric cuts, can be found as in Eq.~\eqref{eq:sig_events} for each simulated HNL mass. Combining with the surviving SM background after kinematic cuts, $B$, the median sensitivity $\mathcal{S}$ is calculated as in Eq.~\eqref{eq:simp_sigsens}. In the next section, we show the bounds at 90\%~CL by excluding the $(m_{N_2}, C_i)$ parameter space with $\mathcal{S} > 1.28$.

In this analysis, we take the active-sterile mixing of $N_1$ and $N_2$ to be negligible. There are three main advantages of this limit:
\begin{itemize}

\item Firstly, to simplify the matching in Sec.~\ref{sec:discussion} of the coefficients $C_i$ in Eqs.~\eqref{eq:diagonal}--\eqref{eq:two-light} to the coefficients of the $\nu$SMEFT operators in Tables~\ref{tab:vSMEFT-operators} and~\ref{tab:vSMEFT-operators-2}. In the limit of vanishingly small active-sterile mixing, $|V_{\alpha N_i}| \ll 1$, the weak and mass eigenstate fields are related by $\nu = P_L \nu'$ and $N = P_R N'$ in the Majorana scenario and $\nu = P_L \nu'$, $N = P_R N'$ and $S = P_L N'$ in the Dirac scenario. Accordingly, there is a one-to-one mapping of the WCs $C_i$ and the $\nu$SMEFT WCs of interest, as described in more detail in Sec.~\ref{sec:discussion}.

\item Secondly, to ensure that the EFT operators dominate the production of HNLs. This is practically assured for the pair production process, $e^+e^-\to N N (\gamma)$, because the contributions from the active-sterile mixing via the SM charged- and neutral-current interactions are suppressed by two powers of $|V_{\alpha N_i}| \ll 1$ in the amplitude. However, for the single production process, $e^+e^-\to \nu N (\gamma)$, the active-sterile mixing can play a more important role; now, the contributions from the SM interactions are proportional to a single power of $|V_{\alpha N_i}|$ and can compete in size with the EFT WCs in Eq.~\eqref{eq:one-light-one-heavy}. For example, taking $C_i = C_{\nu Ne}^{V,RR} \neq 0$ and $V_{eN_2} \neq 0$, we can make use of the expressions for the $2\to 2$ cross sections\footnote{In principle, the $2\to 3$ cross section with a final-state photon should be used, by applying Eq.~\eqref{eq:xsec_monophoton} to Eqs.~\eqref{eq:tot_Dirac_cs} and~\eqref{eq:tot_Maj_cs} and integrating over $\cos\theta_\gamma$ and $E_\gamma$ according to the cuts in Table~\ref{tab:universal_cuts}. However, this gives qualitatively the same result as in Eq.~\eqref{eq:VeN_not_dominate}.} in Eqs.~\eqref{eq:tot_Dirac_cs} and~\eqref{eq:tot_Maj_cs} of App.~\ref{app:2to2_xsec} to compare the sizes of the contributions. At the $Z$ pole, the active-sterile mixing does not dominate the cross section for
\begin{align}
\label{eq:VeN_not_dominate}
|V_{eN_2}| \lesssim \frac{|C_i|}{2\sqrt{2}G_F \sqrt{(g_R^e)^2 + (g_L^e)^2}}\frac{\Gamma_Z}{M_Z} = 2\times 10^{-3}\bigg(\frac{|C_i|}{10^{-6}~\text{GeV}^{-2}}\bigg)\,,
\end{align}
where we have neglected interference terms, which is just below the current upper limits on $|V_{eN_2}|^2$ shown in Fig.~\ref{fig:mixing_exclusions}. Similar conditions can be found for the other EFT WCs in Eq.~\eqref{eq:one-light-one-heavy}. For $\sqrt{s} = 240$~GeV and away from the $Z$ pole in general, the condition in Eq.~\eqref{eq:VeN_not_dominate} is further relaxed.

\item Finally, for non-negligible values of the active-sterile mixing strength, the decay width of $N_2$ is increased. This is inconsistent with our assumption that $N_2$ is stable for the diagonal WCs in Eq.~\eqref{eq:diagonal}. Likewise, the results of the exclusive mono-$\gamma$ plus $\slashed{E}$ analysis for the off-diagonal WCs in Eqs.~\eqref{eq:off-diagonal} and~\eqref{eq:one-light-one-heavy} should be modified. To obtain a qualitative understanding of the size of this effect, in the following section we examine how the electron-flavour mixing strength $V_{eN_2}$ modifies the HNL decay length in the $(m_{N_2}, C_i)$ parameter space.
\end{itemize}
%

%%%%%%%%%%%%%%%%%%%%%%%%%%%%%%%%%%%%%%%%
\subsubsection{Results}
\label{sec:EFT_results}
%%%%%%%%%%%%%%%%%%%%%%%%%%%%%%%%%%%%%%%%

%
\begin{figure}[t!]
\centering
\includegraphics[width=0.49\textwidth]{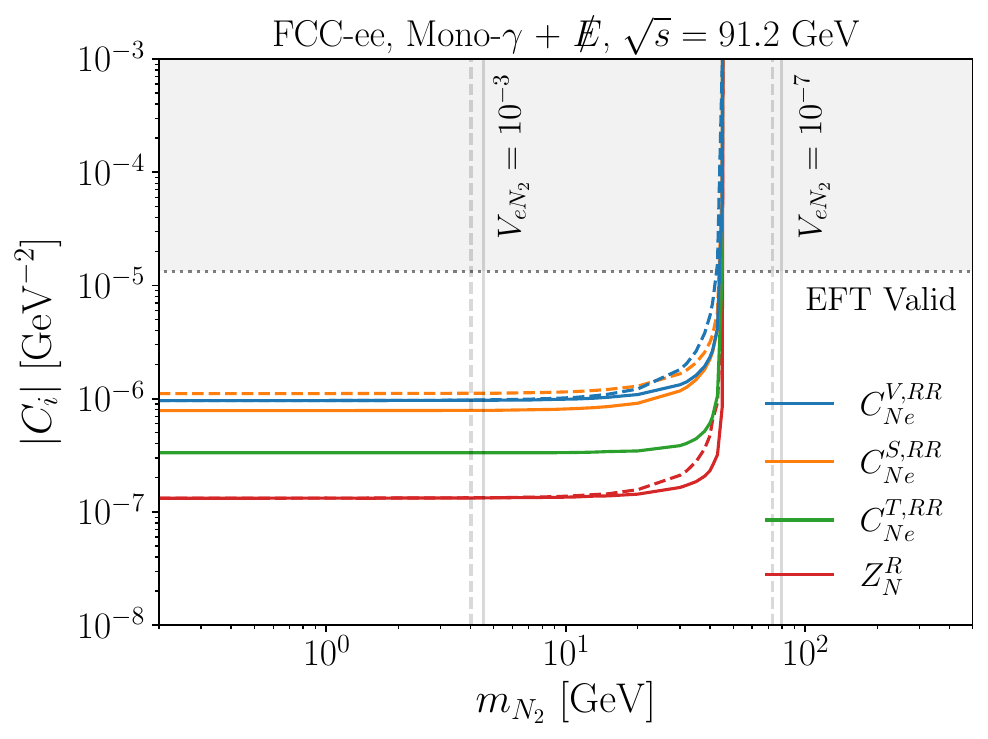}
\includegraphics[width=0.49\textwidth]{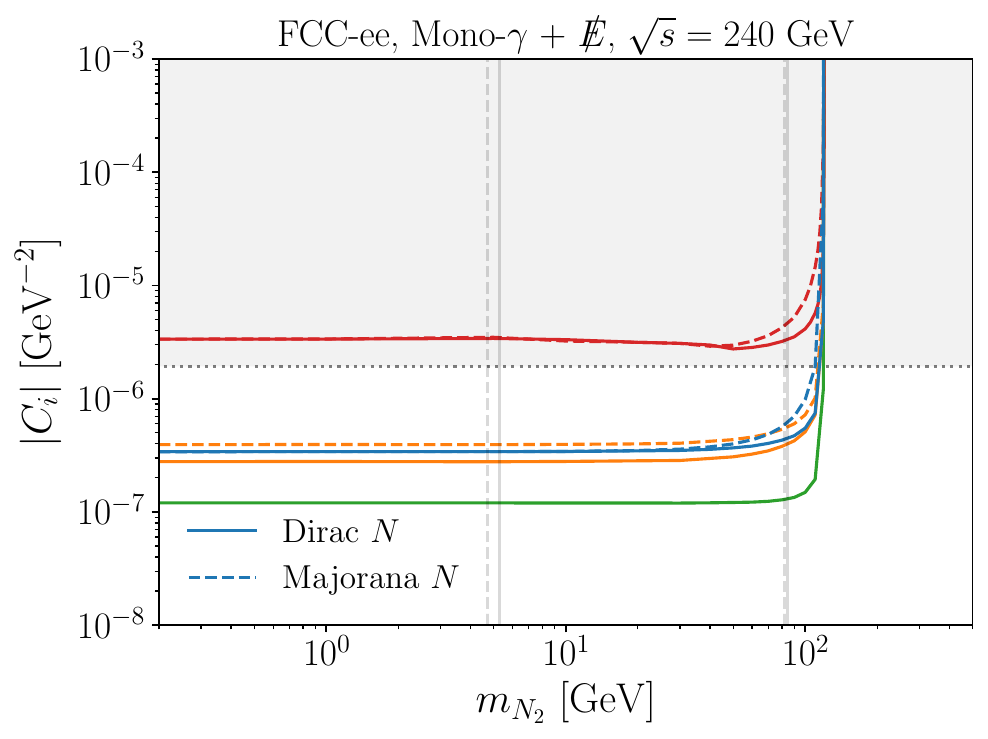}
\caption{Sensitivities of mono-$\gamma$ plus $\slashed{E}$ searches at FCC-ee to the diagonal four-fermion and effective $Z$ interaction WCs as a function of the HNL mass at $90\%$ CL, for $\sqrt{s} = 91.2$~GeV (left) and $\sqrt{s}
= 240$~GeV (right). Limits are shown for $N_2$ being a Majorana (dashed) or Dirac (solid) HNL. The parameter space where the EFT is not valid is indicated by the grey shaded region.}
\label{fig:diag-constraints}
\end{figure}

In Fig.~\ref{fig:diag-constraints}, we present the estimated sensitivities of mono-$\gamma$ plus $\slashed{E}$ searches at FCC-ee to the diagonal four-fermion and effective $Z$ interaction WCs in Eq.~\eqref{eq:diagonal}, for masses of $N_2$ between 200~MeV and 500~GeV. The 90\%~CL sensitivities for the vector (blue), scalar (orange) and tensor (green) four-fermion and effective $Z$ (red) interactions are shown for $\sqrt{s} = 91.2$~GeV (left) and $\sqrt{s} = 240$~GeV (right) in the Majorana (dashed) and Dirac (solid) HNL scenarios. For the effective $Z$ interaction, the sensitivity is shown for $C_i = \frac{2}{v^2}[Z_N^R]_{ii}$ to maintain units of $\text{GeV}^{-2}$. The grey shaded region indicates where the EFT description is no longer valid. As explained in App.~\ref{app:2to2_xsec}, we assume this to be the case for $\Lambda < 3\sqrt{s}$, or $C_i \equiv 1/\Lambda^2 > 1/(9s)$.

Firstly, we see that $\sqrt{s} = 240$~GeV is more sensitive to the four-fermion operators by a factor of $\sim 3$ with respect to $\sqrt{s} = 91.2$~GeV, as the reduced SM background and increased cross section compensate for the lower luminosity. The opposite is true for the effective $Z$ interaction, which benefits from the resonant behaviour of the cross section at $\sqrt{s} = 91.2$~GeV. For the vector four-fermion and effective $Z$ interactions, the sensitivities in the Majorana case fall off faster as a function of $m_{N_2}$ compared to the Dirac case. This is a result of interference terms further suppressing the Majorana cross sections near the kinematic threshold, $m_{N_2} < \sqrt{s}/2$, seen also in Fig.~\ref{fig:xsec_plot_2}. Finally, no constraint is shown for the tensor four-fermion coefficient in the Majorana scenario, which vanishes identically.

The future sensitivities in Fig.~\ref{fig:diag-constraints} assume $V_{\alpha N_i} = 0$ and therefore that $N_2$ is stable. To determine how large the active-sterile mixing can be for these results to remain valid, we consider two values of the electron-flavour mixing strength\footnote{As seen in Fig.~\ref{fig:mixing_exclusions}, $|V_{eN_2}| = 10^{-3}$ is roughly the maximum active-sterile mixing strength still allowed by current experiments in the HNL mass range of interest.}: $|V_{eN_2}| = 10^{-3}$ and~$10^{-7}$. For these two values, we calculate the HNL lifetime $\tau_{N_2} = 1/\Gamma_{N_2}$ and decay length in the lab frame $L = \beta\gamma \tau_{N_2}$, taking the boost factor $\beta\gamma\sim \sqrt{s}/(2m_{N_2})$. In Fig.~\ref{fig:diag-constraints}, the grey dashed (solid) lines indicate where the Majorana (Dirac) HNL decay length satisfies $L = 5$~m, with $L < 5$~m to the right. For $|V_{eN_2}| = 10^{-3}$ and $10^{-7}$, respectively, the mono-$\gamma$ plus $\slashed{E}$ bounds are therefore no longer applicable for $m_{N_2} \gtrsim 5$~GeV and $80$~GeV. Ultimately, for $|V_{eN_2}| \lesssim 10^{-6}$ ($10^{-8}$), we find that the sensitivities for $\sqrt{s} = 91.2$~GeV ($240$~GeV) are unaffected by the mixing-induced decays of $N_2$. For the active-sterile mixing strength expected in the type-I seesaw, $|V_{eN_2}| \sim \sqrt{m_\nu/m_{N_2}}$, which is a well motivated scenario for Majorana HNLs, $L > 5$~m is satisfied for $m_{N_2} \lesssim 50$~GeV.

\begin{figure}[t!]
\centering
\includegraphics[width=0.49\textwidth]{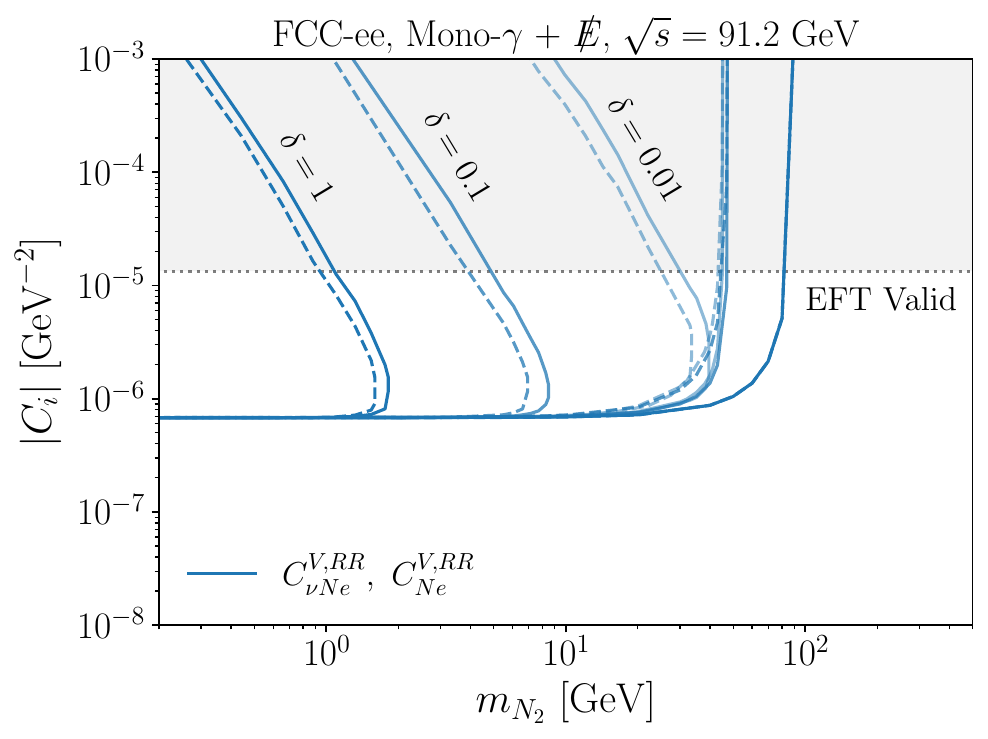}
\includegraphics[width=0.49\textwidth]{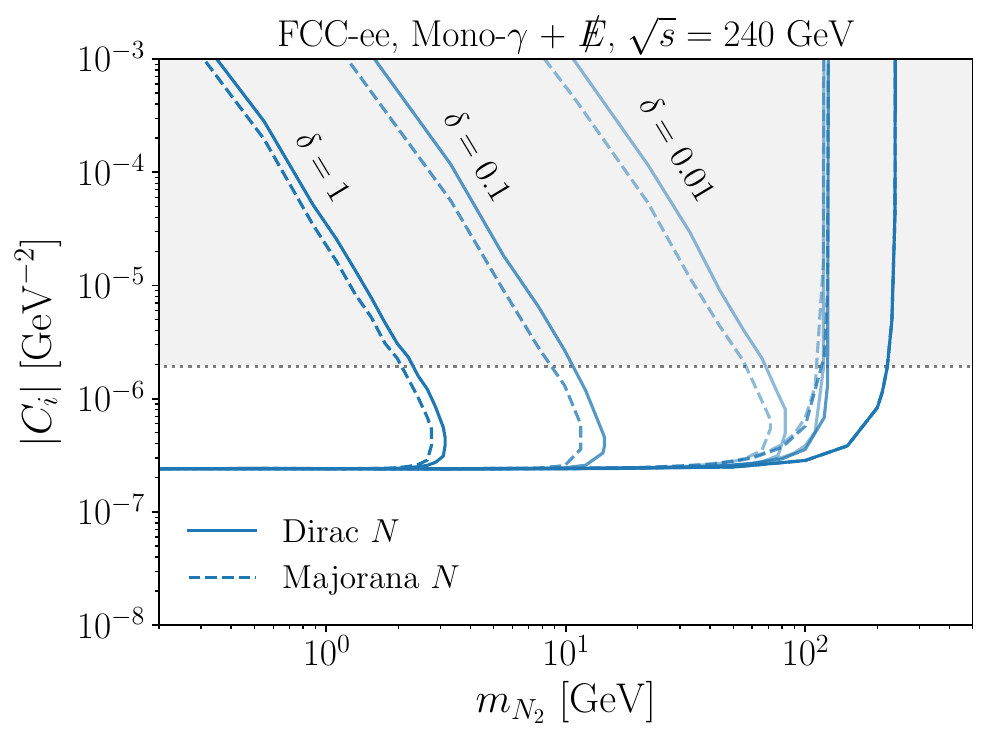}
\includegraphics[width=0.49\textwidth]{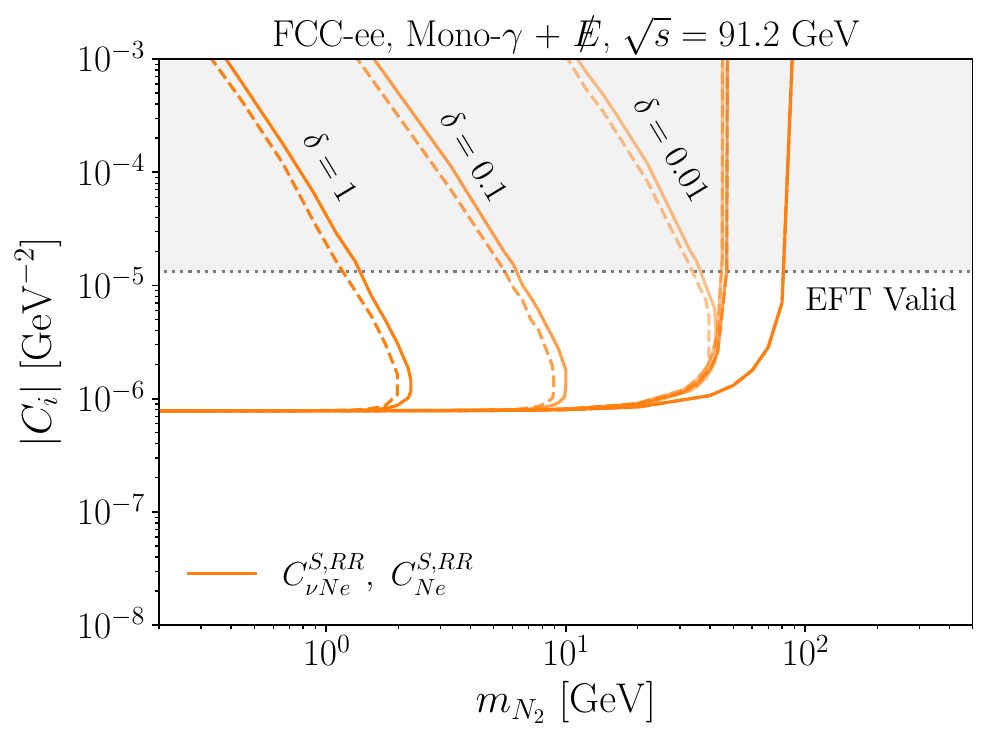}
\includegraphics[width=0.49\textwidth]{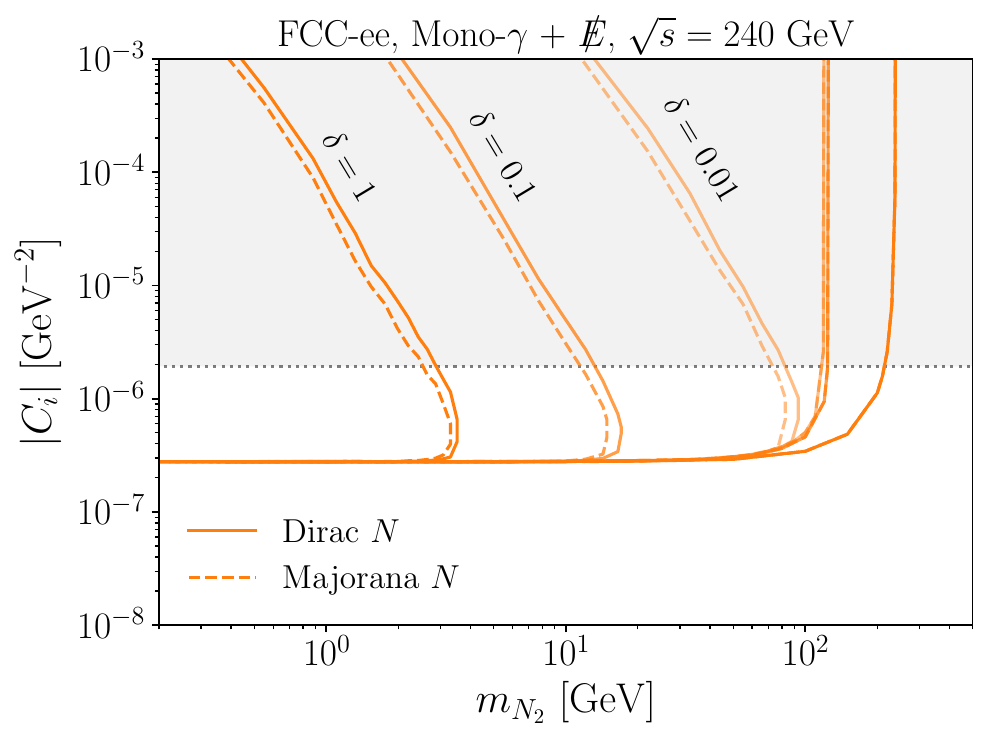}
\includegraphics[width=0.49\textwidth]{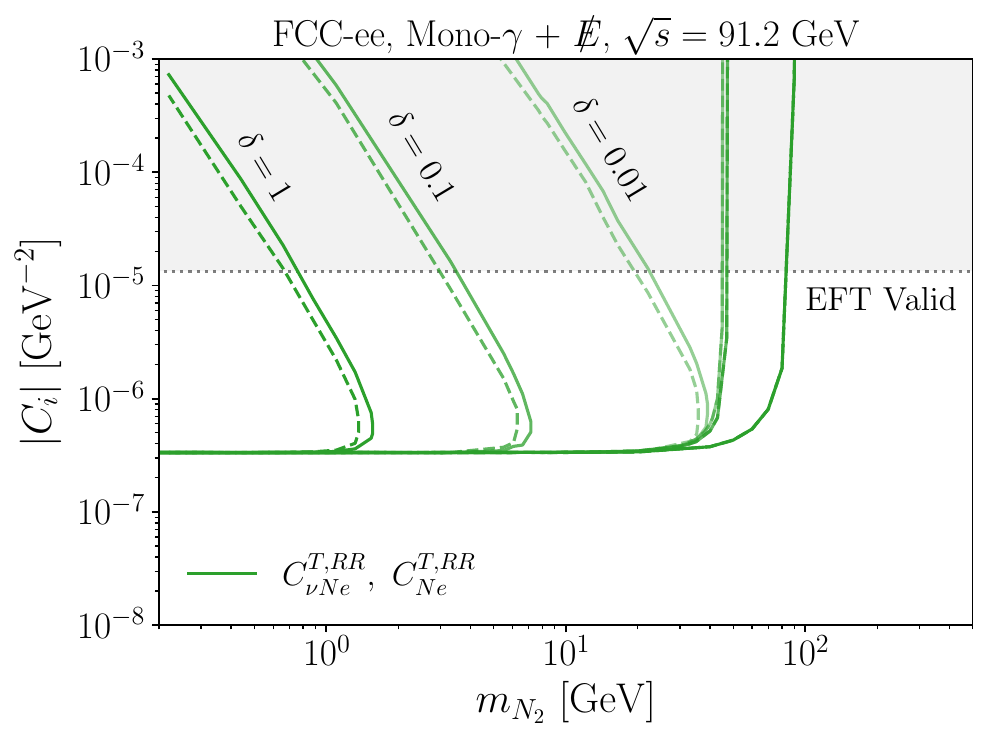}
\includegraphics[width=0.49\textwidth]{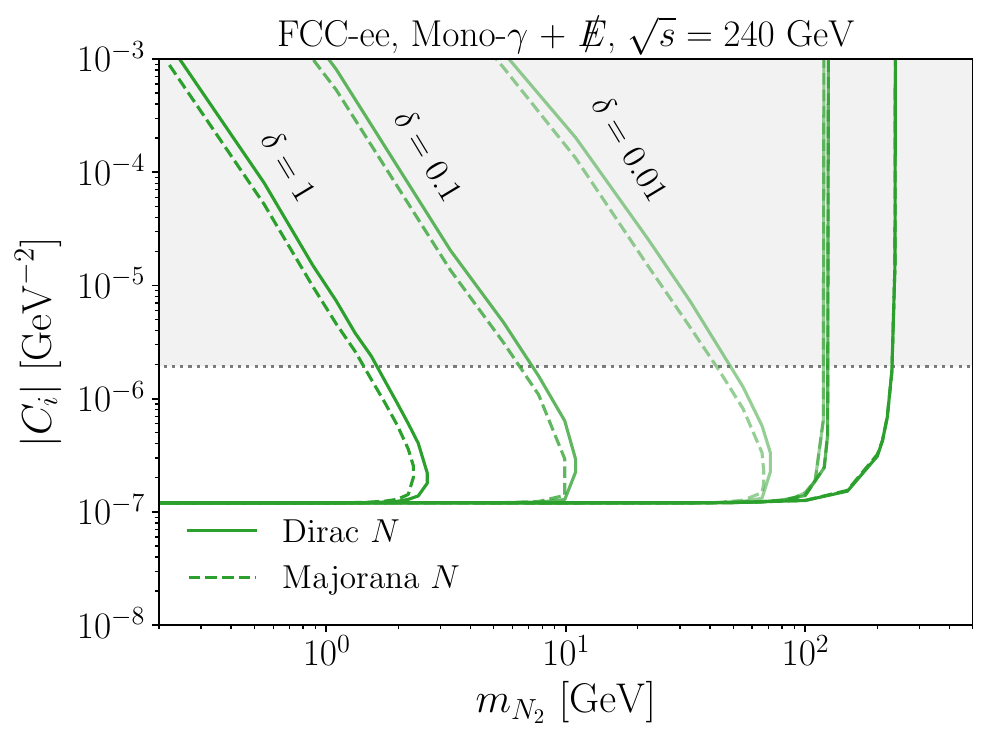}
\caption{Sensitivities of the exclusive and inclusive mono-$\gamma$ plus $\slashed{E}$ searches at FCC-ee to the off-diagonal four-fermion interaction WCs as a function of the HNL mass at $90\%$ CL, for $\sqrt{s} = 91.2$~GeV (left) and $\sqrt{s}
= 240$~GeV (right). Limits are shown for Majorana (dashed) and Dirac (solid) HNLs for three different mass splitting ratios $\delta = (m_{N_2}-m_{N_1})/m_{N_2}$. The sensitivity of the inclusive search is also shown for the Majorana (dotted) and Dirac (dot-dashed) cases. The parameter space where the EFT is not valid is indicated by the grey shaded region.}
\label{fig:off-diag-constraints}
\end{figure}

\begin{figure}
\centering
\includegraphics[width=0.49\textwidth]{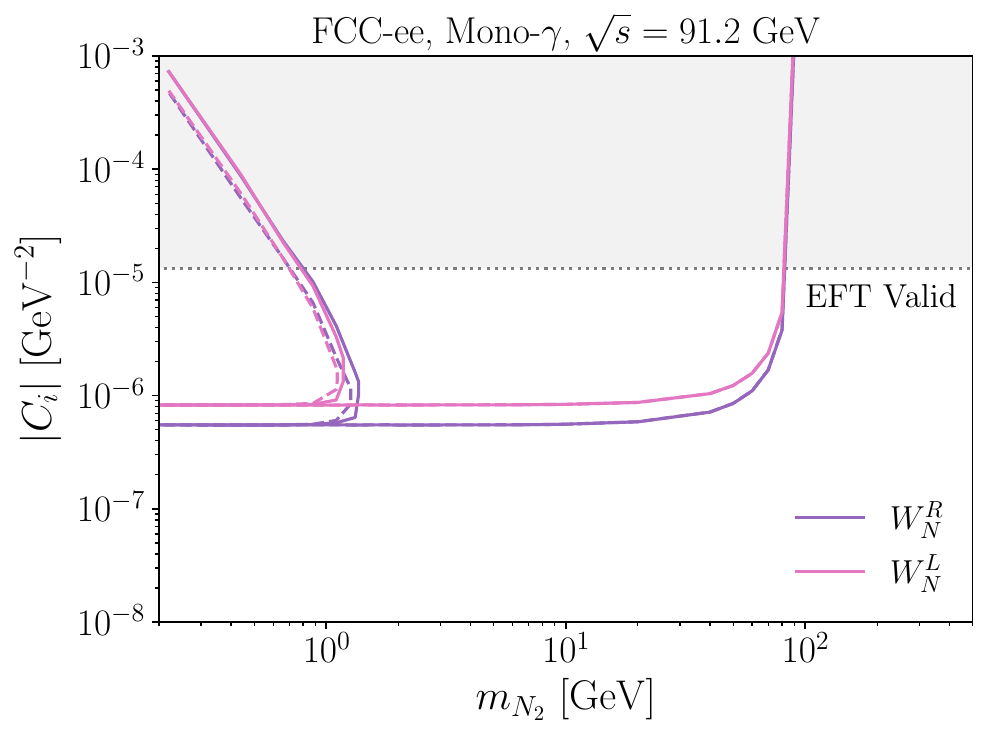}
\includegraphics[width=0.49\textwidth]{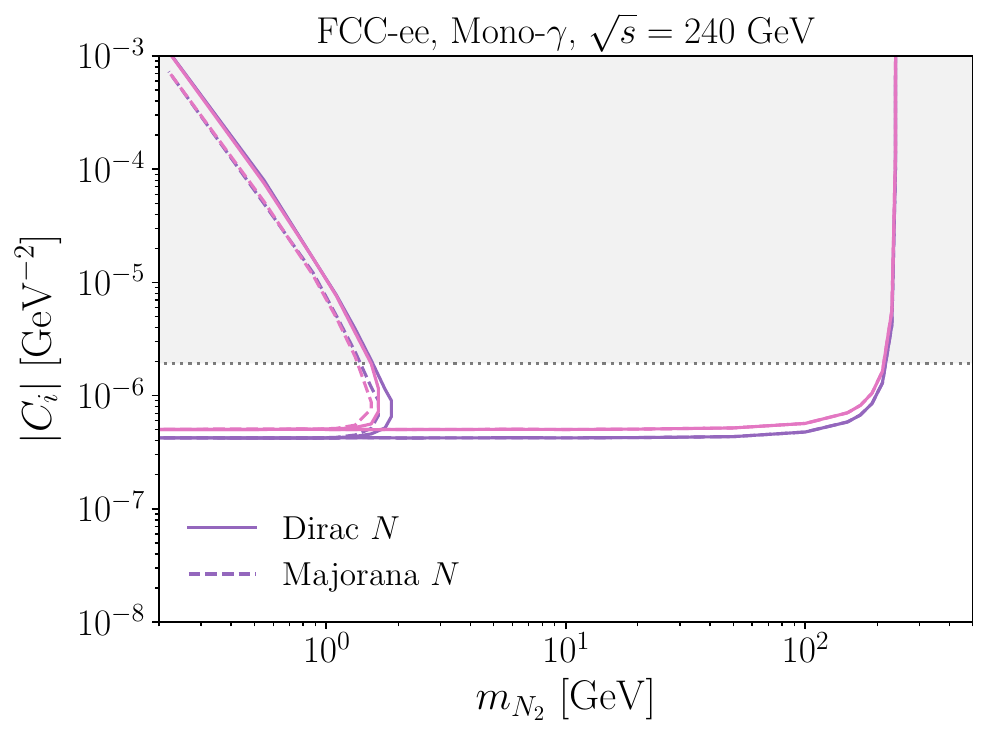}
\includegraphics[width=0.49\textwidth]{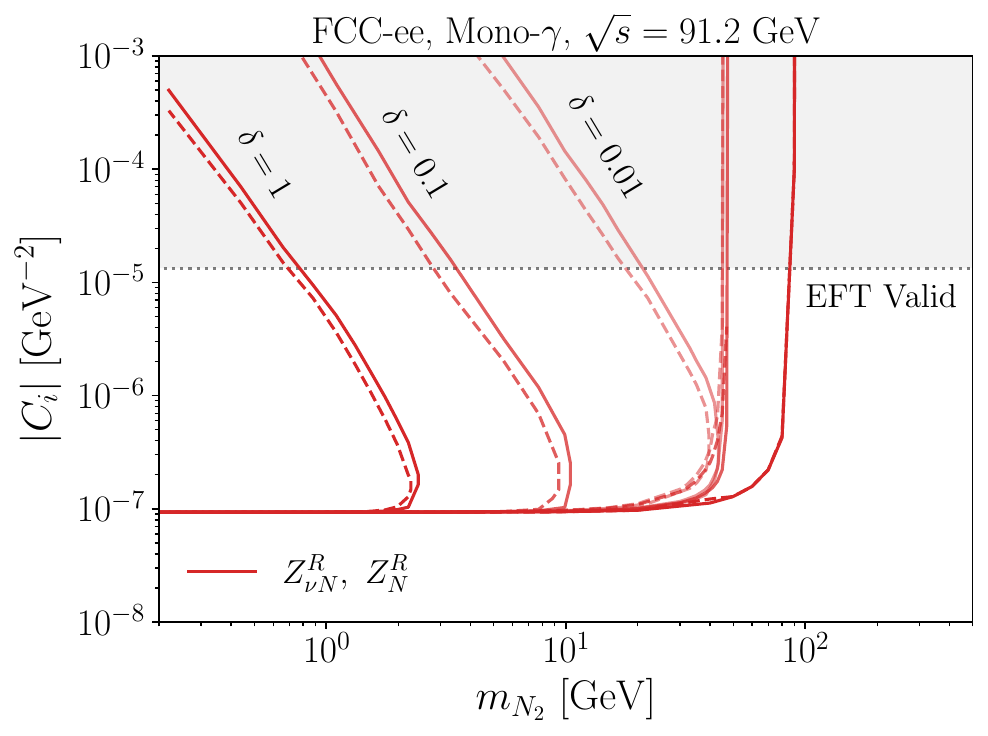}
\includegraphics[width=0.49\textwidth]{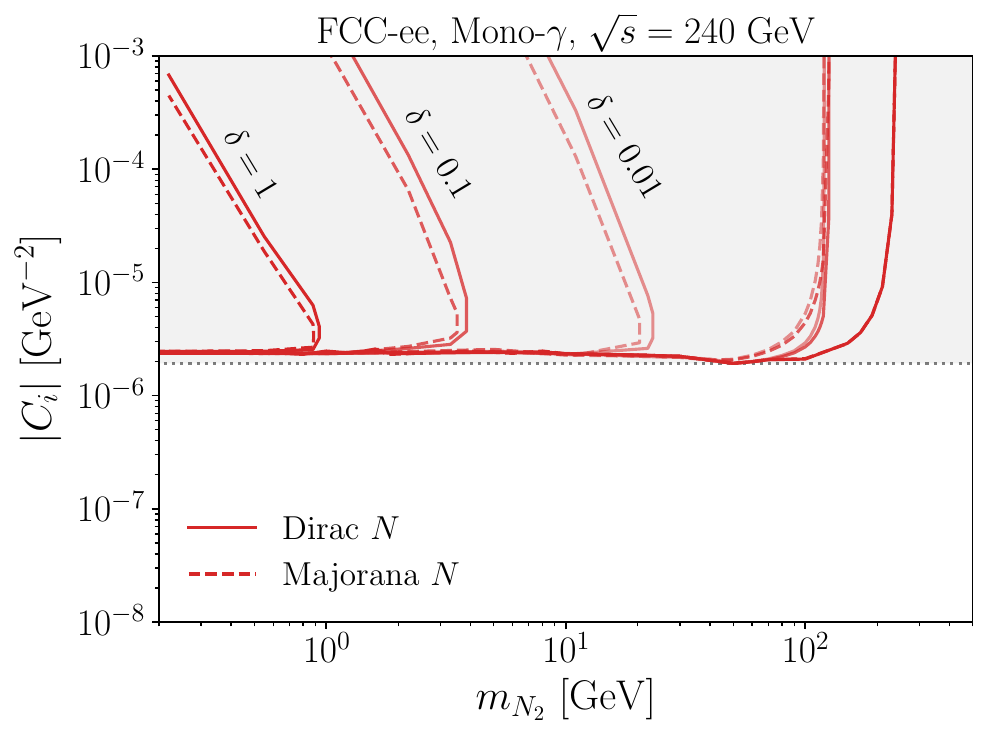}
\caption{Sensitivities of the exclusive and inclusive mono-$\gamma$ plus $\slashed{E}$ searches at FCC-ee to the effective $W^\pm$ and (off-diagonal) $Z$ interactions as a function of the HNL mass, at $90\%$ CL, for $\sqrt{s} = 91.2$~GeV (left) and $\sqrt{s}
= 240$~GeV (right). The benchmark scenarios for the mass splitting ratio $\delta$ are the same as in Fig.~\ref{fig:off-diag-constraints}.}
\label{fig:off-diag-constraints-2}
\end{figure}

For the off-diagonal WCs in Eqs.~\eqref{eq:off-diagonal} and~\eqref{eq:one-light-one-heavy}, we show the 90\%~CL sensitivities of the exclusive and inclusive FCC-ee mono-$\gamma$ plus $\slashed{E}$ searches in Figs.~\ref{fig:off-diag-constraints} and~\ref{fig:off-diag-constraints-2}. We give the results of the $\sqrt{s} = 91.2$~GeV (left) and $\sqrt{s} = 240$~GeV (right) analyses for the vector (blue), scalar (orange) and tensor (green) four-fermion operators in Fig.~\ref{fig:off-diag-constraints} and the effective $W^\pm$ (purple and pink) and $Z$ (red) interactions in Fig.~\ref{fig:off-diag-constraints-2}. For comparison with Fig.~\ref{fig:diag-constraints}, the normalisations $C_i = \frac{2}{v^2}[W_N^R]_{je}$, $\frac{2}{v^2}[W_N^L]_{je}$ and $\frac{2}{v^2}[Z_N^R]_{ij}$ are also used in Fig.~\ref{fig:off-diag-constraints-2}. For the exclusive (inclusive) search, the Majorana and Dirac HNL scenarios are shown as dashed (dotted) and solid (dot-dashed) lines, respectively. The sensitivities for the three values $\delta = 0.01$,  $0.1$ and $1$ are depicted as light, medium, and dark shaded lines, respectively.

The results of the inclusive mono-$\gamma$ plus $\slashed{E}$ search appear similar to the sensitivities for the diagonal WCs in Fig.~\ref{fig:diag-constraints}, with the upper bounds on $|C_i|$ now extending up to the kinematic threshold, $m_{N_2} < \sqrt{s}/(2 - \delta)$. The $\sqrt{s}$ dependencies of the cross sections again ensure that $\sqrt{s} = 91.2$~GeV and $240$~GeV are more constraining for the effective $Z$ and four-fermion interactions, respectively. For the vector four-fermion and effective $Z$ interactions, the sensitivities in the Majorana case again fall off faster as a function of $m_{N_2}$ compared to the Dirac case. Finally, for the effective $W^\pm$ interactions, the sensitivities are marginally stronger for $\sqrt{s} = 240$~GeV, reflecting the logarithmic scaling of the cross sections with $\sqrt{s}$. The bounds on the WC $[W_N^R]_{je}$ are slightly more stringent than for $[W_N^L]_{je}$, as the cross section for the former is larger if the WCs are of equal size.

For the exclusive mono-$\gamma$ plus $\slashed{E}$ search, we see that the constraints on the off-diagonal WCs are impacted substantially by the decays of $N_2$, similar to the active-sterile mixing sensitivity in Fig.~\ref{fig:mixing_exclusions}. This is most pronounced in the $\delta = 1$ case, which corresponds to the WCs in Eq.~\eqref{eq:off-diagonal} for $m_{N_1} = 0$ and the WCs in Eq.~\eqref{eq:one-light-one-heavy}. For HNL masses up to certain size, a range of $|C_i|$ values are excluded, because the HNL is very unlikely to decay outside the detector if $|C_i|$ is too large. However, for $\delta \ll 1$, the sensitivities can extend to much larger values of $m_{N_2}$, because the HNL decay rate via $N_2\to N_1 e^-e^+$ (and additional decays such as $N_2\to N_1 \nu\bar{\nu}$, $N_2\to N_1 \mu^-\mu^+$, $N_2\to N_1 \tau^-\tau^+$ and $N_2\to N_1 q\bar{q}$ for the effective $Z$ interaction) is suppressed, increasing the HNL lifetime sufficiently for heavier HNLs to be long-lived.

As for the diagonal WCs, we have assumed $V_{\alpha N_i} = 0$ to determine the exclusive mono-$\gamma$ plus $\slashed{E}$ sensitivities in Figs.~\ref{fig:off-diag-constraints} and~\ref{fig:off-diag-constraints-2}. To explore the possible impact of the active-sterile mixing, we again compute the HNL lifetime $\tau_{N_2}$ and decay length $L = \beta\gamma\tau_{N_2}$, with $\beta\gamma\sim \sqrt{s}/(2m_{N_2})$, as a function of $m_{N_{2}}$ and $C_i$, for $V_{\alpha N_i} \neq 0$. The exclusive mono-$\gamma$ plus $\slashed{E}$ sensitivities for the $\delta = 0.1$~and~$1$ scenarios, which do not reach up to $m_{N_2} \sim 20$~GeV, remain unaffected for $|V_{eN_2}| \lesssim 10^{-4}$ and $10^{-2}$, respectively. The effective $W^\pm$ interactions are likewise unaffected for $|V_{eN_2}| \lesssim 10^{-2}$. The $\delta = 0.01$ sensitivities meanwhile remain valid for $|V_{eN_2}| \lesssim 10^{-6}$ ($10^{-7}$) for $\sqrt{s} = 91.2$~GeV (240~GeV). All exclusive sensitivities remain unaffected by the seesaw prediction for the active-sterile mixing, $|V_{eN_2}| \sim \sqrt{m_\nu/m_{N_2}}$.

%%%%%%%%%%%%%%%%%%%%%%%%%%%%%%%%%%%%%%%%
\section{Displaced Vertex Constraints at FCC-ee}
\label{sec:displacedvertex}
%%%%%%%%%%%%%%%%%%%%%%%%%%%%%%%%%%%%%%%%

In this section, we examine the sensitivity of DV searches at FCC-ee to the EFT operators in Sec.~\ref{sec:model}. Again, we consider the presence of one HNL interaction at a time. The scenario where all EFT WCs are zero, $C_i = 0$, and the active-sterile mixing is non-negligible, $V_{\alpha N_i} \neq 0$, has already been considered in~\cite{Blondel:2014bra,Antusch:2016vyf,Blondel:2022qqo,Ajmal:2024kwi}. Here, we consider $V_{\alpha N_i} = 0$ and $C_i \neq 0$ for Majorana and Dirac HNLs. As for the mono-$\gamma$ plus $\slashed{E}$ analysis in Sec.~\ref{sec:monophoton}, we consider the $\sqrt{s} = 91.2$~GeV and $240$~GeV runs at FCC-ee with the integrated luminosities $\mathcal{L} = 100~\rm{ab}^{-1}$ and $5~\rm{ab}^{-1}$, respectively.

%%%%%%%%%%%%%%%%%%%%%%%%%%%%%%%%%%%%%%%%
\subsection{EFT Operators}
\label{subsec:ff_DV}
%%%%%%%%%%%%%%%%%%%%%%%%%%%%%%%%%%%%%%%%

The following benchmark scenarios are taken in this analysis. As discussed in Sec.~\ref{sec:ff_monophoton}, the diagonal four-fermion and effective $Z$ interaction WCs in Eq.~\eqref{eq:diagonal} cannot induce decays of $N_2$ in the $V_{\alpha N_i} = 0$ limit, and therefore are not considered further here. Likewise, the WCs of the operators in Eq.~\eqref{eq:two-light} containing two active neutrino fields cannot be probed by DV searches. Therefore, we only consider the off-diagonal four-fermion and effective $Z$ interaction WCs in Eqs.~\eqref{eq:off-diagonal} and~\eqref{eq:one-light-one-heavy} and the effective $W^\pm$ interaction WCs in Eq.~\eqref{eq:one-light-one-heavy}. For the off-diagonal WCs involving $N_1$ and $N_2$ in Eq.~\eqref{eq:off-diagonal}, we consider three mass splitting ratios; $\delta = 0.1$, $0.5$, and $1$. The results for $\delta = 1$ are equally applicable for the WCs of the off-diagonal operators containing an active neutrino and $N_2$ in Eq.~\eqref{eq:one-light-one-heavy}.

%%%%%%%%%%%%%%%%%%%%%%%%%%%%%%%%%%%%%%%%
\subsubsection{Sensitivity Estimate}
\label{sec:DV_sim}
%%%%%%%%%%%%%%%%%%%%%%%%%%%%%%%%%%%%%%%%

This sensitivity analysis proceeds as follows. Using \texttt{MadGraph5\_aMC@NLO}, we simulate for the four-fermion operators the $2\to 2$ signal process $e^+ e^- \to N_1 \bar{N}_2 + \bar{N}_1 N_2$, in both the Majorana and Dirac HNL scenarios. As in Sec.~\ref{sec:EFT_sim}, the definition of the WCs in the Majorana \texttt{FeynRules} model file reproduces in the expected behaviour of the cross section in \texttt{MadGraph5\_aMC@NLO} for the Majorana four-fermion operators. The HNLs can instead be treated as Majorana fermions for the effective $W^\pm$ and $Z$ interactions, for which we simulate $e^+ e^- \to \nu_e N_2 + \bar{\nu}_e N_2$ and $e^+ e^- \to N_1 N_2$ in the Majorana scenario, respectively, and $e^+ e^- \to \nu_e \bar{N}_2 + \bar{\nu}_e N_2$ and $e^+ e^- \to N_1 \bar{N}_2 + \bar{N}_1 N_2$ in the Dirac scenario. All signal processes are simulated with $N_{\text{tot}} = 5\times 10^4$ events.

In the simulation, we further require $N_2$ (and $\bar{N}_2$ in the Dirac HNL scenario) to decay via the operators of interest to a di-electron final state; specifically, $N_2 \to N_1 e^- e^+$ for the four-fermion and effective $Z$ interactions and $N_2 \to \nu e^- e^+$ for the effective $W^\pm$ interactions. The total width of $N_2$ and the branching ratios of these channels are calculated using the expressions of App.~\ref{app:decays} and inputted by hand.

We estimate the DV final state reach in a background-free approach. The SM backgrounds are predominantly prompt and, if necessary, can be reduced by using a cut on the electron-track transverse impact parameter, e.g. $|d_0| > 0.6$~mm~\cite{Blondel:2022qqo, Ajmal:2024kwi}. In addition, we require $p_T^e > 0.7$~GeV as the minimum momentum necessary to identify an electron at the FCC-ee. The kinematic efficiency $\epsilon_{k}$ is found by dividing the number of events surviving the cut by $N_\text{tot}$. For the geometric acceptance, we use Eq.~\eqref{eq:prob_decay} with $f(\sqrt{s}, m_{N_1}, m_{N_2}, b) = \delta(b - b')$ and the boost factor of $N_2$ fixed to $b' = \lambda(s, m_{N_1}^2, m_{N_2}^2)/(2m_{N_2}\sqrt{s})$, which gives the probability of $N_2$ decaying inside the detector, $\mathcal{P}_{\text{in}}$. The FCC-ee detector is taken as spherical, with the minimum and maximum radii $L_1 = 0.1$~mm and $L_2 = 5$~m, respectively.

With the kinematic cuts and geometric acceptance described above, we estimate the total number of DV signal events as,
\begin{align}
\label{eq:sig_events_DV}
S = \mathcal{L} \times \sigma \times \text{BR}(N_2\to \nu e^-e^+/N_1e^-e^+)\times  \mathcal{P}_{\text{in}}\times \epsilon_k \,.
\end{align}
Given zero background, we determine the excluded regions at 90\%~CL in the $(m_{N_2}, C_i)$ parameter space by identifying where the condition $S > 2.3$ is met for the signal events. 

%%%%%%%%%%%%%%%%%%%%%%%%%%%%%%%%%%%%%%%%
\subsubsection{Results}
\label{sec:DV_results}
%%%%%%%%%%%%%%%%%%%%%%%%%%%%%%%%%%%%%%%%

%
\begin{figure}[t!]
\centering
\includegraphics[width=0.49\textwidth]{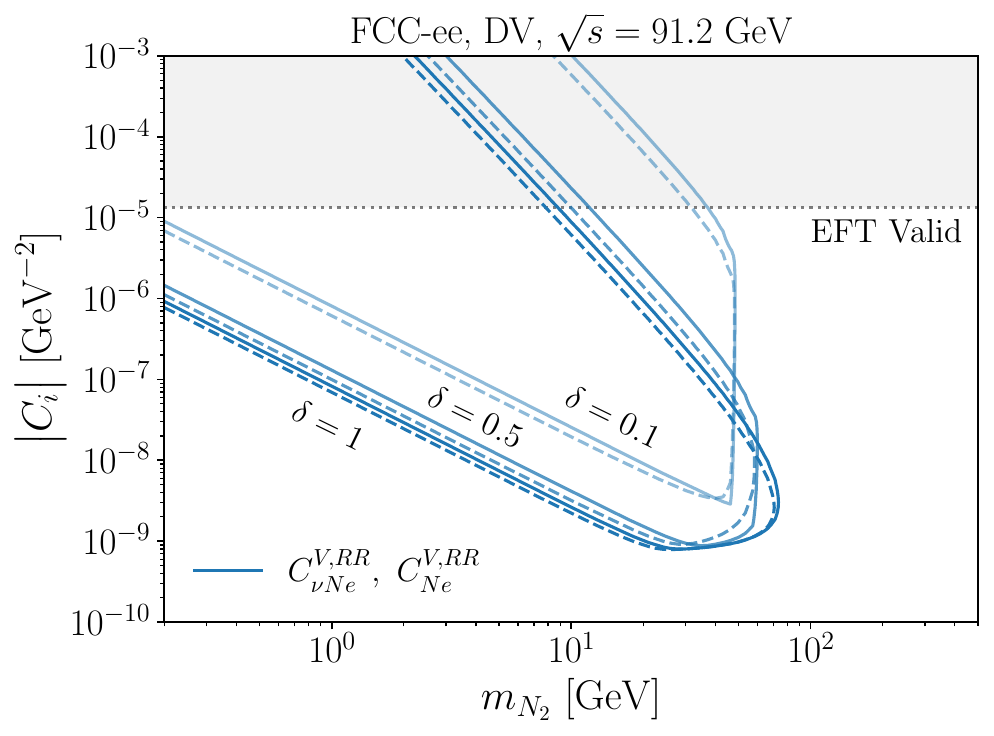}
\includegraphics[width=0.49\textwidth]{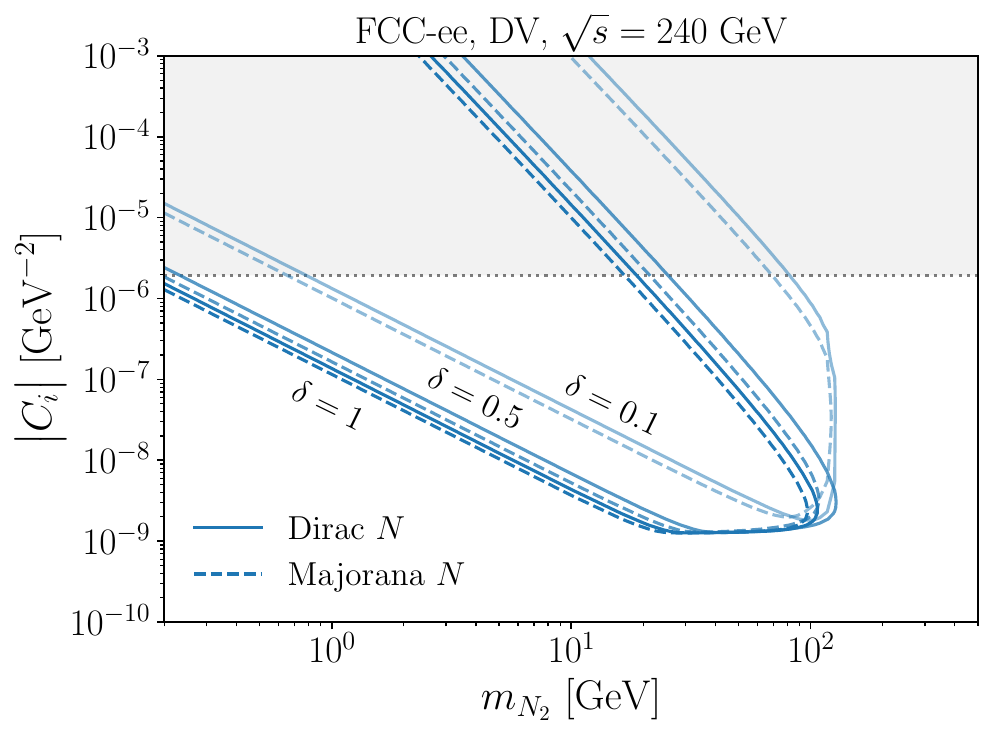}
\includegraphics[width=0.49\textwidth]{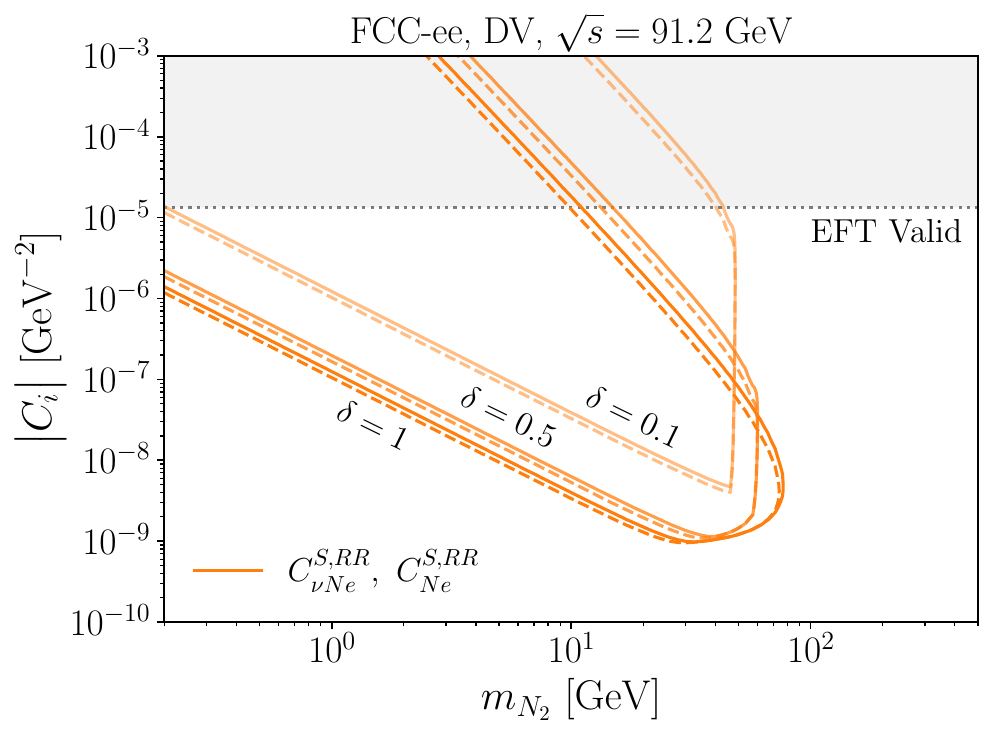}
\includegraphics[width=0.49\textwidth]{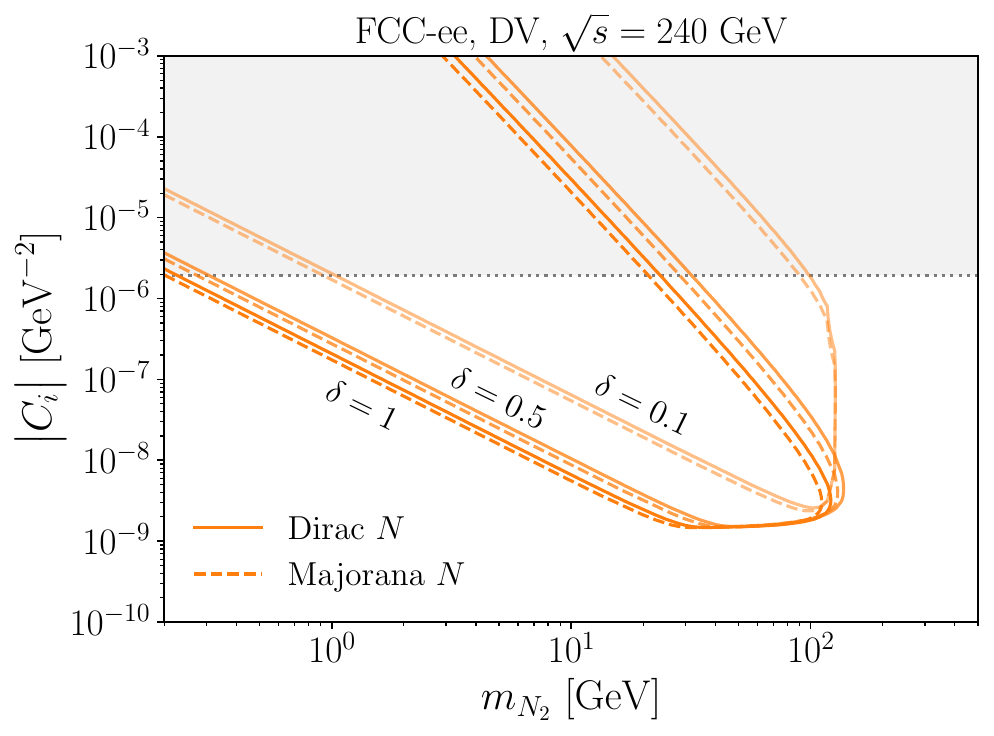}
\includegraphics[width=0.49\textwidth]{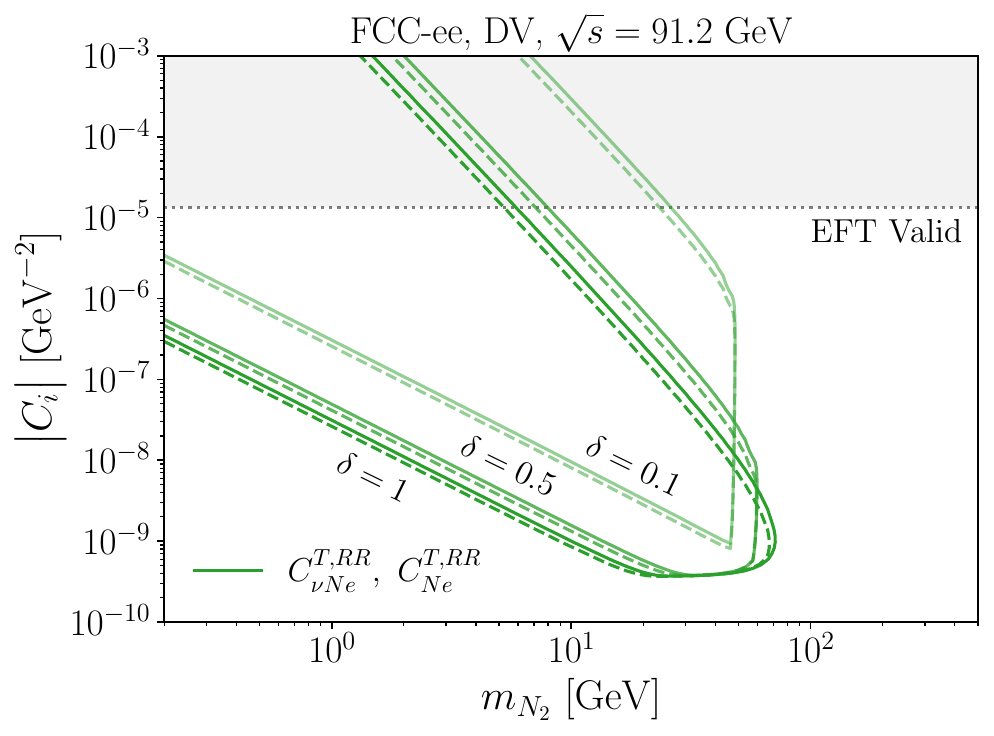}
\includegraphics[width=0.49\textwidth]{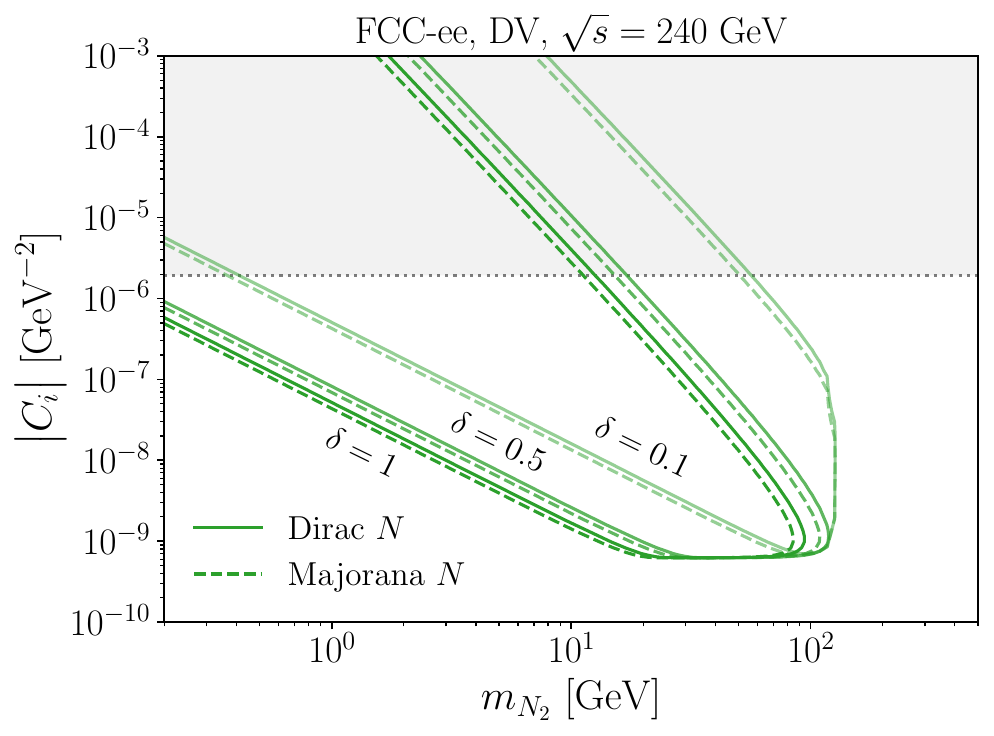}
\caption{Sensitivities of the of DV search at FCC-ee to the off-diagonal four-fermion interaction WCs as a function of the HNL mass at $90\%$ CL, for $\sqrt{s} = 91.2$~GeV (left) and $\sqrt{s}
= 240$~GeV (right). Limits are shown for Majorana (dashed) and Dirac (solid) HNLs for three different mass splitting ratios $\delta = (m_{N_2}-m_{N_1})/m_{N_2}$. The parameter space where the EFT is not valid is indicated by the grey shaded region.}
\label{fig:ff_DV_constraints}
\end{figure}
\begin{figure}[t!]
\centering
\includegraphics[width=0.49\textwidth]{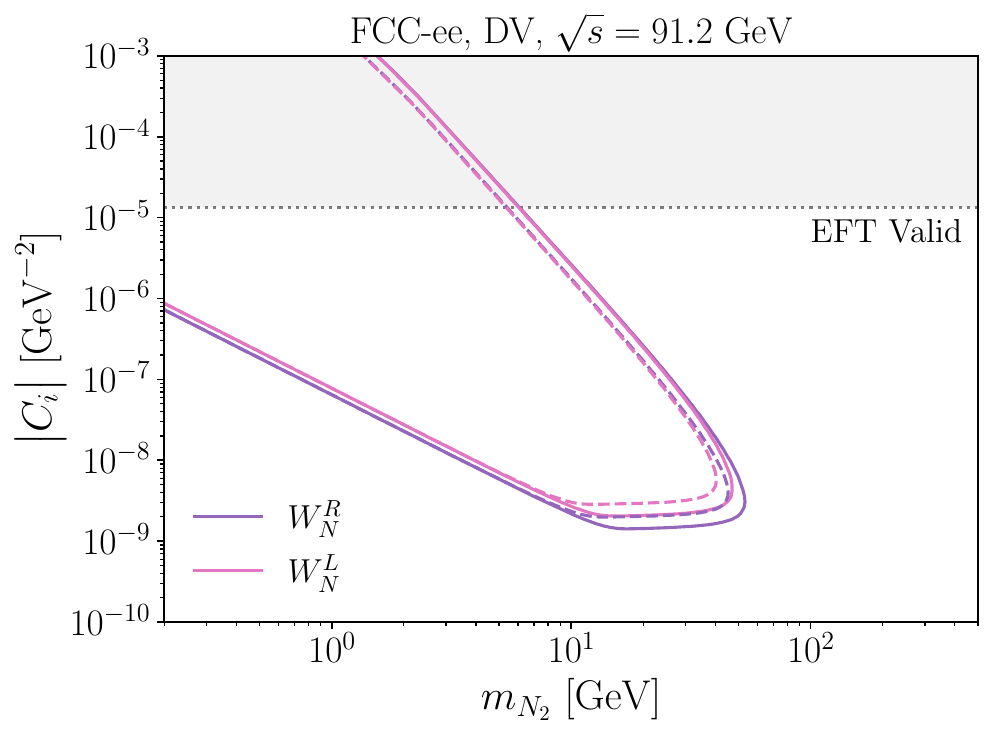}
\includegraphics[width=0.49\textwidth]{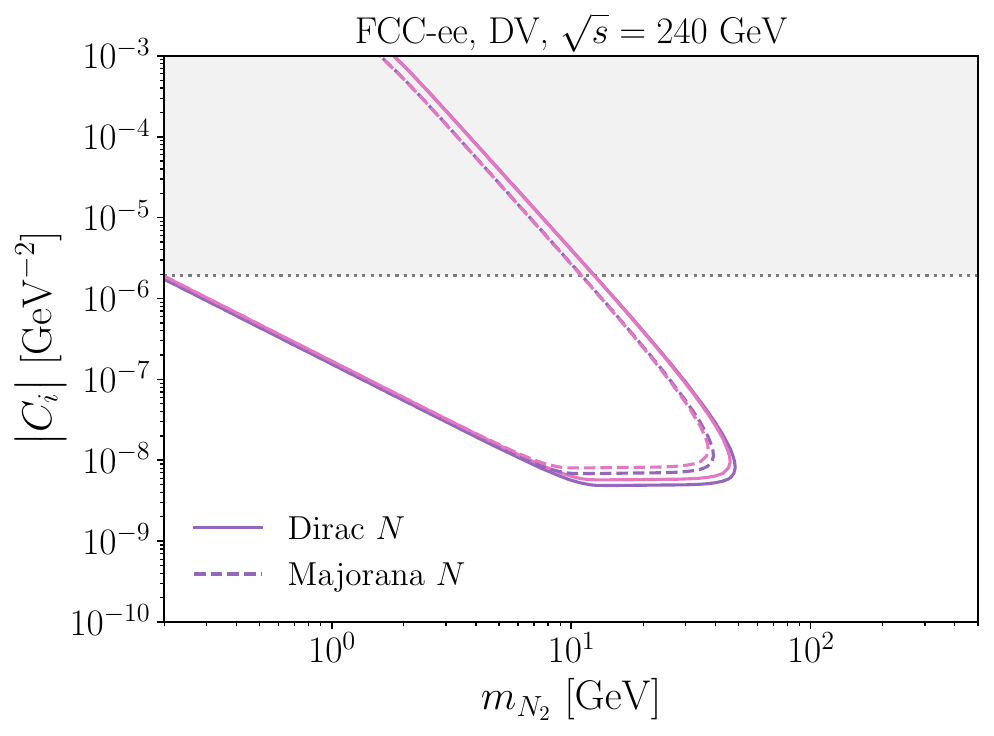}
\includegraphics[width=0.49\textwidth]{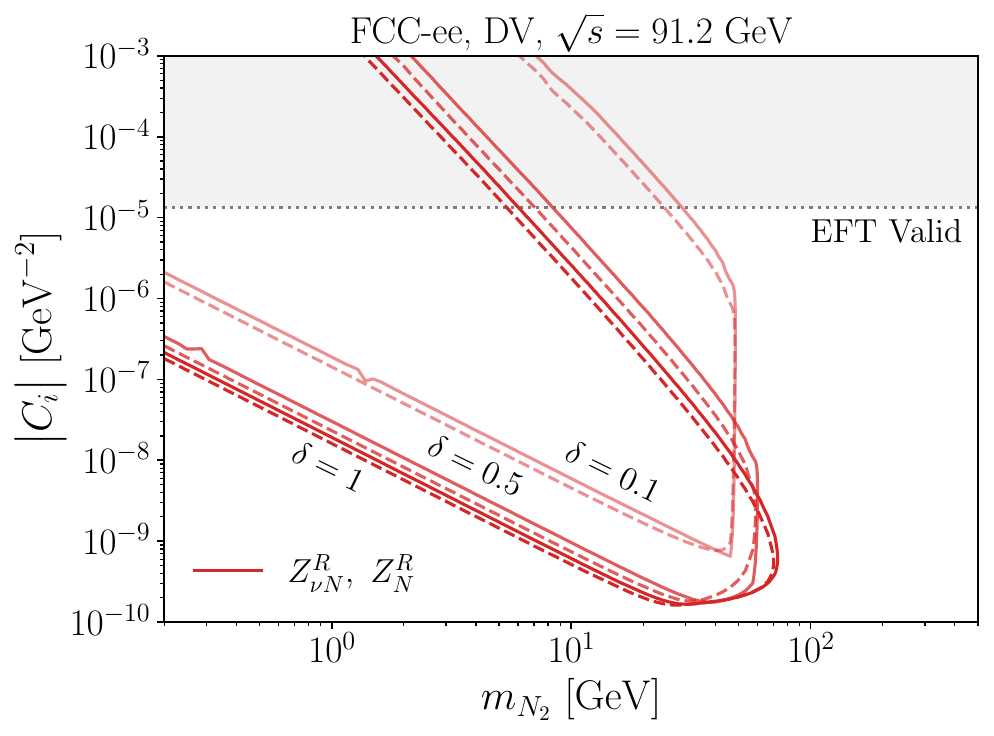}
\includegraphics[width=0.49\textwidth]{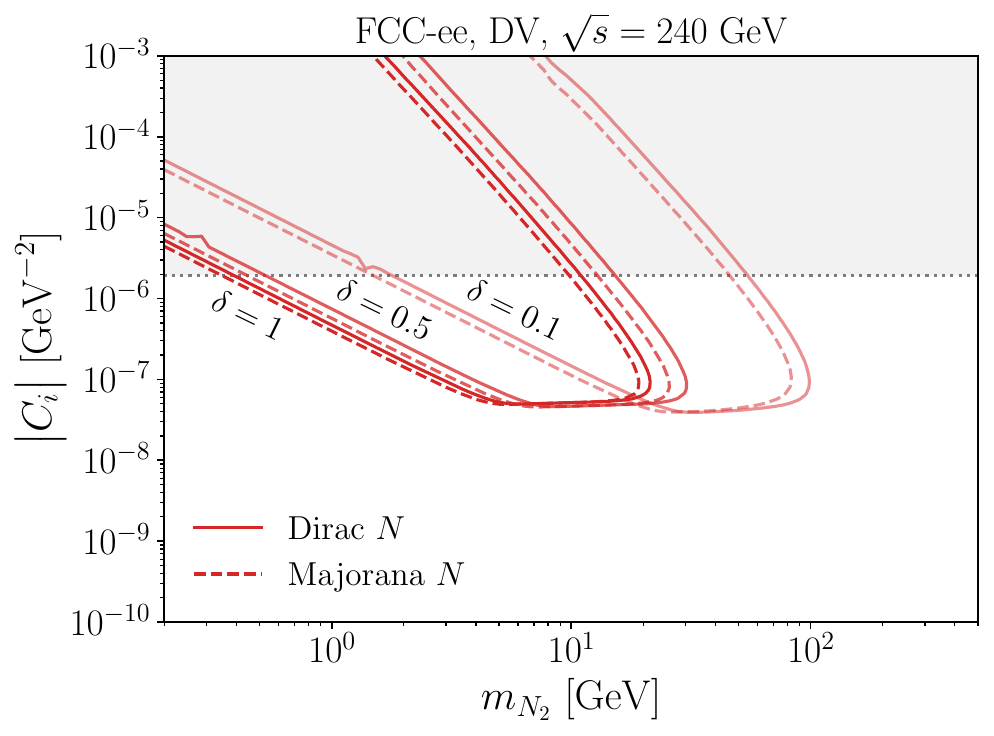}
\caption{Sensitivities of the DV search at FCC-ee to the effective $W^\pm$ and (off-diagonal) $Z$ interactions as a function of the HNL mass at $90\%$ CL, for $\sqrt{s} = 91.2$~GeV (left) and $\sqrt{s}
= 240$~GeV (right). The benchmark scenarios for the mass splitting ratio $\delta$ are the same as in Fig.~\ref{fig:ff_DV_constraints}.}
\label{fig:gauge_DV_constraints}
\end{figure}

In Figs.~\ref{fig:ff_DV_constraints} and~\ref{fig:gauge_DV_constraints}, we show the resulting 90\%~CL sensitivities of the DV search at FCC-ee to the WCs in Eqs.~\eqref{eq:off-diagonal} and~\eqref{eq:one-light-one-heavy}, for masses of $N_2$ between 200~MeV and $\sqrt{s}$. For $\sqrt{s} = 91.2$~GeV (left) and $\sqrt{s} = 240$~GeV (right), we show the sensitivities for the vector (blue), scalar (orange) and tensor (green) four-fermion operators in Fig.~\ref{fig:ff_DV_constraints} and the effective $W^\pm$ (purple and pink) and $Z$ (red) interactions (with the same normalisations as in Sec.~\ref{sec:EFT_results}, $C_i = \frac{2}{v^2}[W_N^R]_{je}$, $\frac{2}{v^2}[W_N^L]_{je}$ and $\frac{2}{v^2}[Z_N^R]_{ij}$) in Fig.~\ref{fig:gauge_DV_constraints}. The sensitivities for the three values $\delta = 0.1$,  $0.5$ and $1$ are depicted as light, medium, and dark shaded lines, respectively. The same EFT validity regions are indicated as in Figs.~\ref{fig:off-diag-constraints} and~\ref{fig:off-diag-constraints-2}.

We first note that, like the active-sterile mixing, the DV search can probe much smaller values of the WCs $|C_i|$ with respect to the mono-$\gamma$ plus $\slashed{E}$ search. Firstly, for the four-fermion operators with $\delta = 1$, both the $\sqrt{s} = 91.2$~GeV and $240$~GeV runs can reach $|C_i|\sim 10^{-9}~\text{GeV}^{-2}$ for $m_{N_2} \sim 20$--$70$~GeV and $m_{N_2} \sim 20$--$90$~GeV, respectively. Instead, for $\delta = 0.1$ and $0.5$, the sensitivities become bounded for $\sqrt{s} = 91.2$~GeV by the kinematic threshold, $m_{N_2} < \sqrt{s}/(2-\delta)$, which can be seen as vertical slices on the right-hand side of the excluded regions. For $\sqrt{s} = 240$~GeV, the opposite behaviour can be seen, with the sensitivities reaching larger values of $m_{N_2}$ for successively smaller values of $\delta$, until the bounds again hit the kinematic threshold. Unlike the mono-$\gamma$ plus $\slashed{E}$ sensitivities, the bounds weaken as $m_{N_2} \to 0$, because in this limit $N_2$ becomes too long-lived, with the decay length in the lab frame $L = \beta\gamma\tau_{N_2} > 5$~m. On the upper side of the excluded regions, $N_2$ instead decays too promptly, with $L < 0.1$~mm.

The sensitivities for the effective $W^\pm$ interactions, like the four-fermion operators, are slightly more stringent for $\sqrt{s} = 91.2$~GeV compared to $240$~GeV. Even though the cross sections are larger for these operators at higher $\sqrt{s}$, the reduced luminosity at $\sqrt{s} = 240$~GeV does not compensate enough. Furthermore, the sensitivities for $\sqrt{s} = 240$~GeV do not benefit from the reduced SM background with respect to $\sqrt{s} = 91.2$~GeV, like the mono-$\gamma$ plus $\slashed{E}$ search. Smaller values of the WC $[W_N^R]_{je}$ are probed compared to $[W_N^L]_{je}$, as the former induces a larger cross section for equal values of the WCs. The sensitivities overlap at larger values of $|C_i|$ because the rates for $N_2 \to \nu e^-e^+$ are equal. For the effective $Z$ interaction, the resonant cross section at $\sqrt{s} = 91.2$~GeV clearly provides a more stringent limit compared to $\sqrt{s} = 240$~GeV, reaching almost to $|C_i|\sim 10^{-10}~\text{GeV}^{-2}$ for $m_{N_2}\sim 30$~GeV. The cross section is insufficient for the $\delta = 0.1$ scenario to reach the kinematic threshold, as was the case for the four-fermion operators.

Finally, as in Sec.~\ref{sec:monophoton}, we have considered the impact of non-zero active-sterile mixing on the results of this analysis. For the $\delta = 1$ sensitivities, $|V_{eN_2}| \lesssim 10^{-5}$ is typically required for the active-sterile mixing to not significantly alter the results. Generally, we find that $|V_{eN_2}| \lesssim 10^{-7}$ and $10^{-6}$ must be satisfied for the $\delta = 0.1$ and $0.5$ sensitivities in Figs.~\ref{fig:ff_DV_constraints} and~\ref{fig:gauge_DV_constraints} to remain fully applicable.

%%%%%%%%%%%%%%%%%%%%%%%%%%%%%%%%%%%%%%%%
\section{Discussion}
\label{sec:discussion}
%%%%%%%%%%%%%%%%%%%%%%%%%%%%%%%%%%%%%%%%

Having presented the sensitivities of FCC-ee to the EFT WCs $C_i$ from mono-$\gamma$ plus $\slashed{E}$ searches in Sec.~\ref{sec:ff_monophoton} and DV searches in Sec.~\ref{subsec:ff_DV}, we now assess the corresponding constraints on the WCs of the $\nu$SMEFT operators in Tables~\ref{tab:vSMEFT-operators} and~\ref{tab:vSMEFT-operators-2}, which can further be translated to lower bounds on the scale of new physics $\Lambda$, shown in Figs.~\ref{fig:two-heavy-barchart},~\ref{fig:one-light-one-heavy-barchart} and~\ref{fig:two-light-barchart}. We also compare the sensitivities of FCC-ee to existing constraints on the $\nu$SMEFT operators in Fig.~\ref{fig:nuSMEFT_constraints_comparison}.

As discussed previously, the analyses of Secs.~\ref{sec:ff_monophoton} and \ref{subsec:ff_DV} assume the active-sterile mixing to be negligible, such that the EFT operators dominate the production and decay of the Majorana or Dirac HNLs $N_1$ and $N_2$. This assumption furthermore streamlines the matching of the WCs $C_i$ in the HNL mass basis and the $\nu$SMEFT WCs in the weak basis. For non-negligible $V_{\alpha N_i}$, multiple $\nu$SMEFT operators can contribute to an operator $C_i$ after EW symmetry breaking, albeit with the contributions arising from $V_{\alpha N_i} \neq 0$ being suppressed both by $|V_{\alpha N_i}| \ll 1$ and the scale of new physics $\Lambda \gg v$. Therefore, in the following, we simply take the leading contributions to $C_i$ with $|V_{\alpha N_i}| = 0$.

%%%%%%%%%%%%%%%%%%%%%%%%%%%%%%%%%%%%%%%%
\subsection{$d = 6$ $\nu$SMEFT Operators}
%%%%%%%%%%%%%%%%%%%%%%%%%%%%%%%%%%%%%%%%

Using the matching relations in App.~\ref{app:matching}, the $d = 6$ $\nu$SMEFT operators in Table~\ref{tab:vSMEFT-operators} contribute to the WCs $C_i$ in Eqs.~\eqref{eq:diagonal}--\eqref{eq:two-light}. In the $|V_{\alpha N_i}| = 0$ limit, we have the trivial relations between the weak and mass eigenstates $N = P_R N'$ in the Majorana scenario and $N = P_R N'$ and $S = P_L N'$ in the Dirac scenario. The WCs $C_i$ in Eqs.~\eqref{eq:one-light-one-heavy}--\eqref{eq:two-light} are also defined for the flavour eigenstate active neutrinos, which by definition are the weak eigenstates. Thus, the indices of the $\nu$SMEFT operators can be interchanged with the flavour indices $\rho, \sigma = e,\mu,\tau$ for the active neutrinos and the mass indices $i,j = 1,2$ for the HNLs. 

For the WCs $C_i$ in Eqs.~\eqref{eq:diagonal} and~\eqref{eq:off-diagonal}, the matching with the $d = 6$ $\nu$SMEFT operators is therefore given by,
\begin{align}
\label{eq:matching_example_1}
C_{\underset{ijee}{Ne}}^{V,RR} = C_{\underset{eeij}{eN}}\,, \quad \frac{2}{v^2}[Z_N^R]_{ij} = C_{\underset{ij}{HN}}\,,
\end{align}
and for the WCs $C_i$ in Eq.~\eqref{eq:one-light-one-heavy} by,
\begin{align}
\label{eq:matching_example_2}
C_{\underset{\alpha jee}{\nu Ne}}^{S,RR} = C_{\underset{\alpha j ee}{lNle}} + \frac{1}{2}C_{\underset{ej\rho e}{lNle}}\,,\quad C_{\underset{\alpha jee}{\nu Ne}}^{T,RR} = \frac{1}{8}C_{\underset{ej\rho e}{lNle}}\,, \quad \frac{2}{v^2}[W_N^R]_{je} = C_{\underset{je}{HNe}}\,.
\end{align}
The constraints on the WCs $C_i$ in Figs.~\ref{fig:diag-constraints}--\ref{fig:gauge_DV_constraints} can now be translated to bounds on the $d = 6$ $\nu$SMEFT operators. 

\begin{figure}[t!]
\centering
\includegraphics[width=0.49\textwidth]{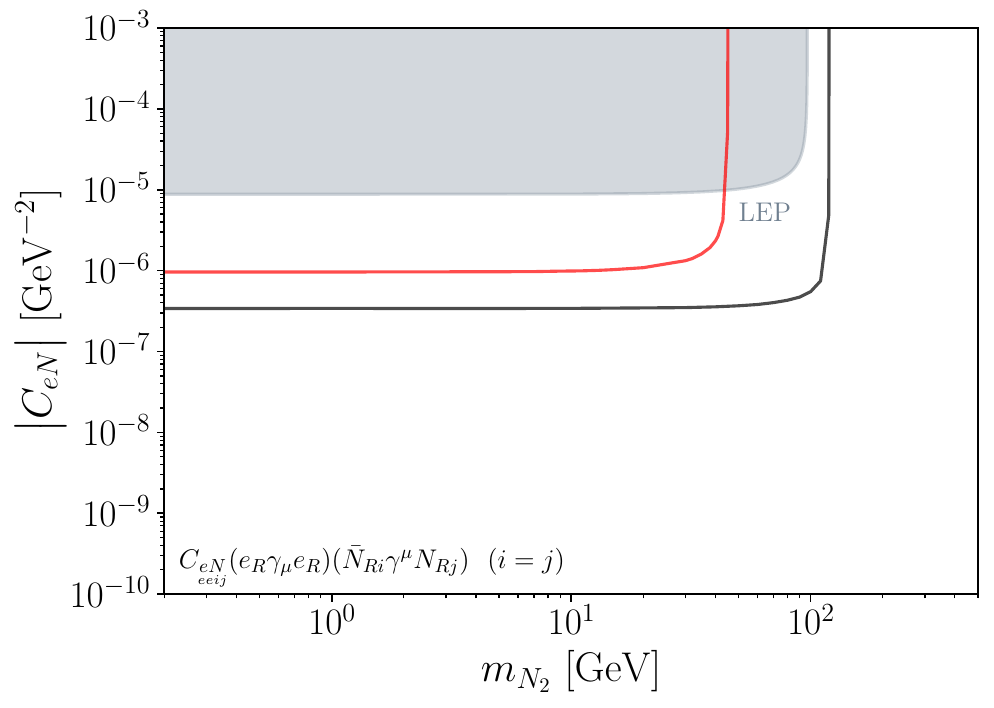}
\includegraphics[width=0.49\textwidth]{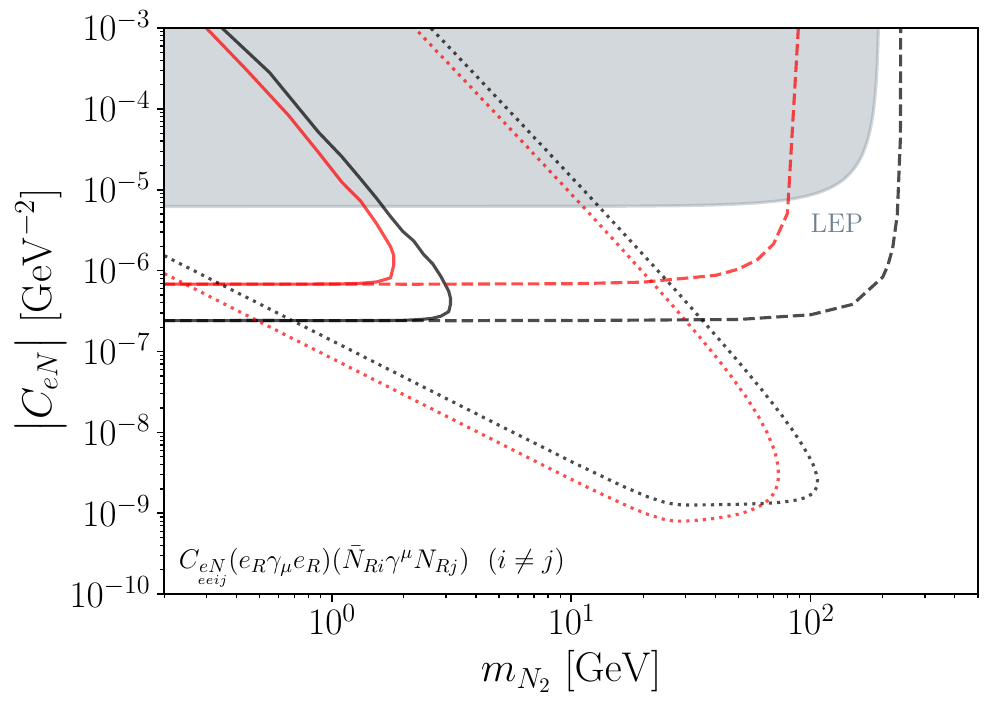}
\includegraphics[width=0.49\textwidth]{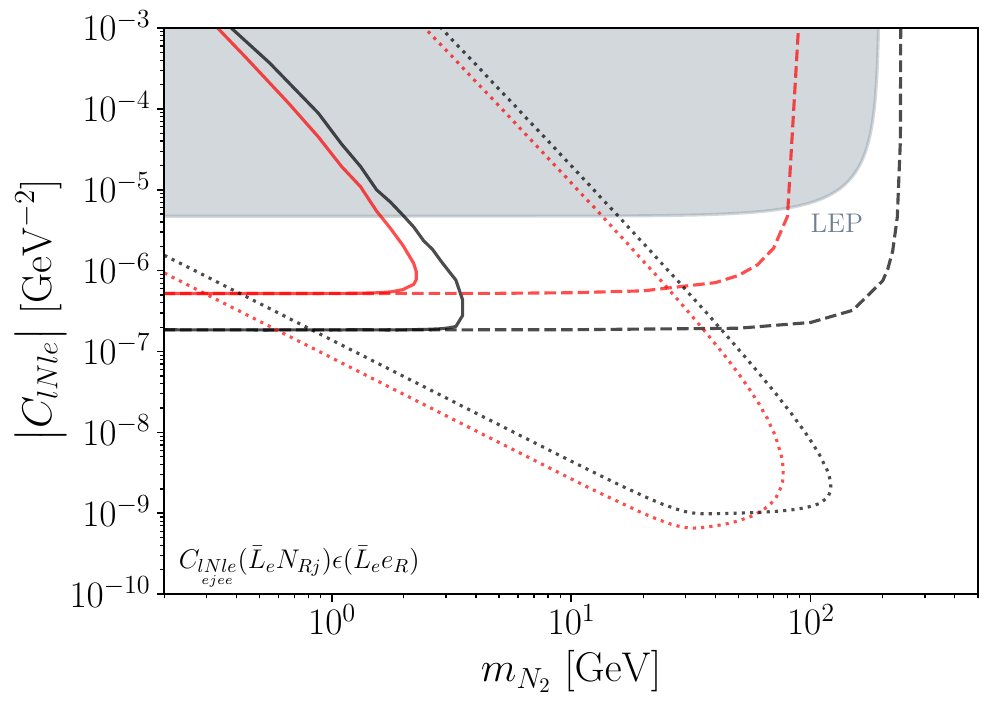}
\includegraphics[width=0.49\textwidth]{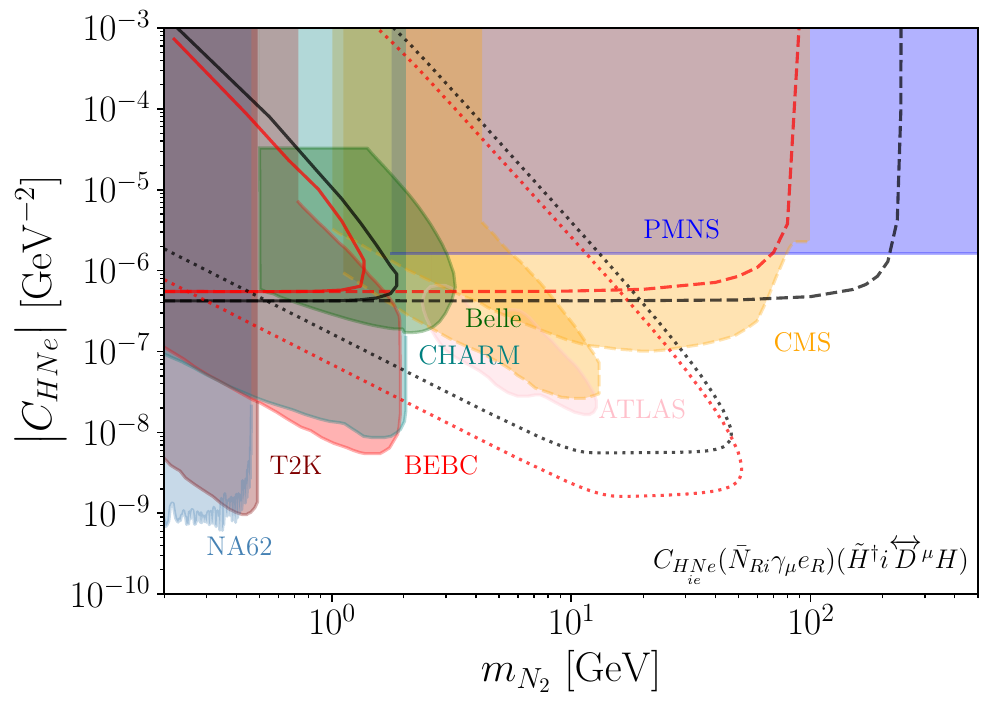}
\includegraphics[width=0.49\textwidth]{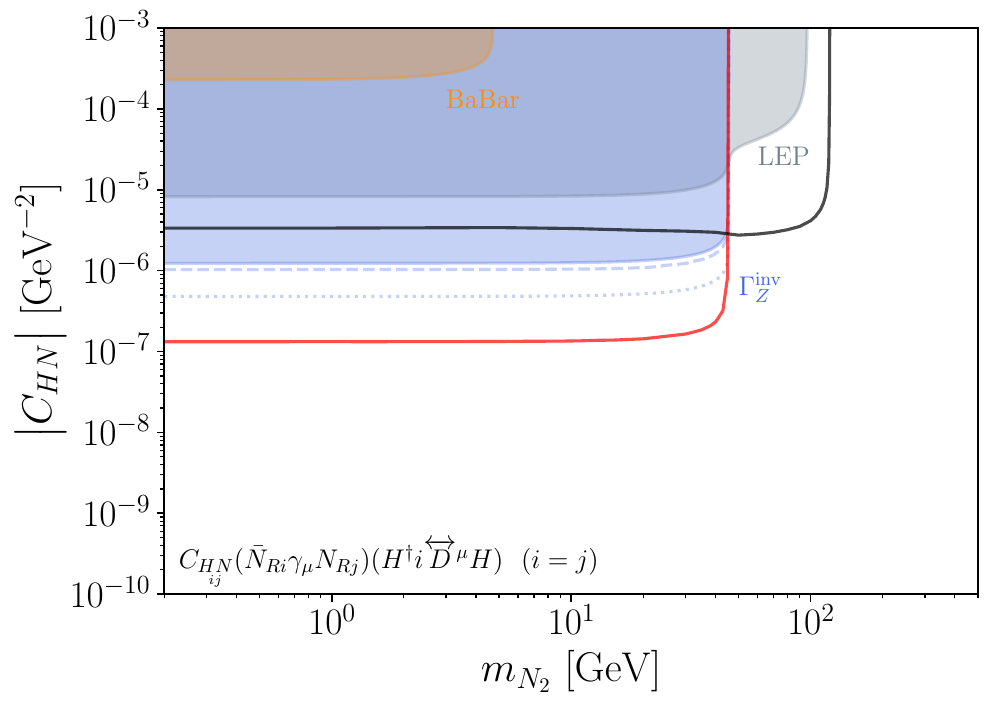}
\includegraphics[width=0.49\textwidth]{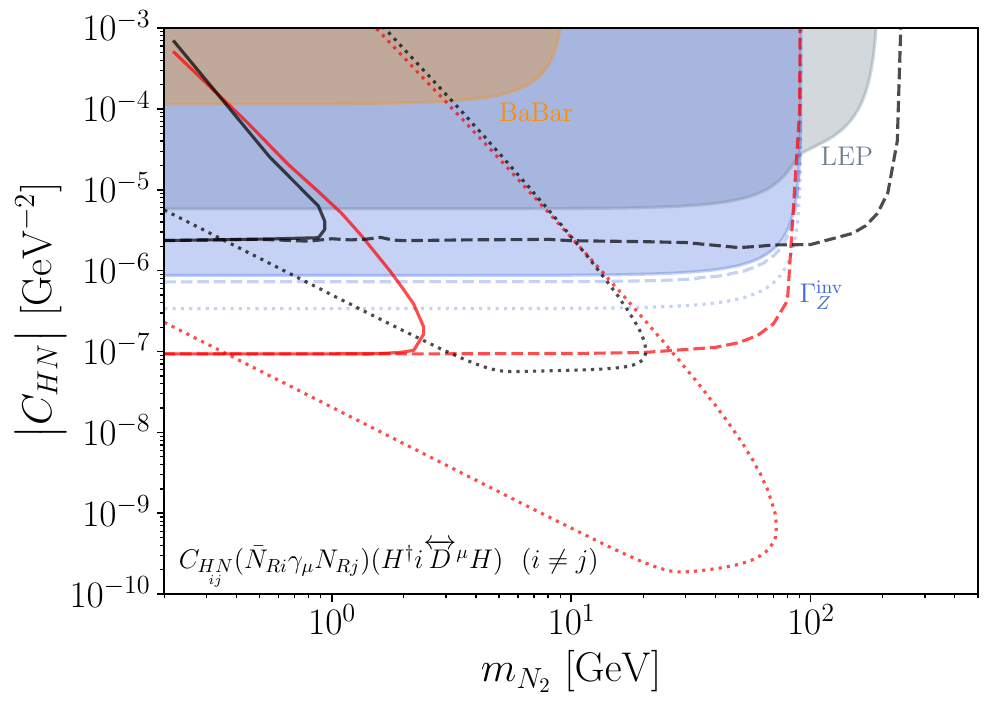}
\caption{FCC-ee sensitivities to the $d = 6$ $\nu$SMEFT operator WCs $C_{eN}$ (top), $C_{lNle}$ and $C_{HNe}$ (centre) and $C_{HN}$ (bottom), in the Dirac HNL scenario, compared to existing constraints. The red (black) curves correspond to $\sqrt{s} = 91.2$~GeV (240~GeV), while the solid (dashed) and dot-dashed lines show the exclusive (inclusive) mono-$\gamma$ plus $\slashed{E}$ and DV analyses, respectively. Existing bounds (shaded) are discussed in the main text.}
\label{fig:nuSMEFT_constraints_comparison}
\end{figure}

In Fig.~\ref{fig:nuSMEFT_constraints_comparison}, we show in the upper two panels the bounds on the coefficient $C_{eN}$ of the $d = 6$ $\nu$SMEFT operator $Q_{eN} = (\bar{e} \gamma_\mu e)(\bar{N}_{Ri} \gamma^\mu N_{Rj})$ as a function of $m_{N_2}$, for $i = j = 2$ (left) and $i = 1$, $j =2$ (right), in the Dirac HNL scenario, with $N_1$ assumed to be massless (i.e., $\delta = 1$). In all panels of Fig.~\ref{fig:nuSMEFT_constraints_comparison}, the same colour scheme is used as in Fig.~\ref{fig:mixing_exclusions}, with the $\sqrt{s} = 91.2$~GeV (red) and $\sqrt{s} = 240$~GeV (black) FCC-ee runs shown for the exclusive (solid) and inclusive (dashed) mono-$\gamma$ plus $\slashed{E}$ searches and the DV search (dotted), which are taken directly from the constraints in Figs.~\ref{fig:diag-constraints},~\ref{fig:off-diag-constraints}~and~\ref{fig:ff_DV_constraints}, according to the matching relation in $C_{Ne}^{V,RR} = C_{eN}$.

An existing bound on $C_{eN}$ originates from mono-$\gamma$ plus $\slashed{E}$ searches at LEP. As in~\cite{Fernandez-Martinez:2023phj}, we obtain these bounds by recasting the constraint in~\cite{Fox:2011fx} on a purely vector four-fermion operator coupling a Dirac fermion dark matter candidate to electrons, $(\bar{N} \gamma_\mu N)(\bar{e} \gamma^\mu e)$, using DELPHI data for $\sqrt{s}$ values between 180~GeV and 209~GeV~\cite{DELPHI:2003dlq,DELPHI:2008uka}. We follow the same procedure as~\cite{Fernandez-Martinez:2023phj}, rescaling the bound on the purely vector operator by equating the cross sections for $e^+e^- \to N N\gamma$, i.e.
\begin{align}
\label{eq:xsec_rescale}
|C_i|^2\hat{\sigma}_i = |C_j|^2\hat{\sigma}_j\,,
\end{align}
where the hat denotes that the dependence on the coefficient $C_i$ is removed from the cross section $\sigma_i$. The cross section for the purely vector operator can be found by setting the WCs to $C_{Ne}^{V,RR} = C_{Ne}^{V,RL} = C_{Ne}^{V,LR} = C_{Ne}^{V,LL} \equiv C_V$ in Eq.~\eqref{eq:tot_Dirac_cs} and inserting the result into Eq.~\eqref{eq:xsec_monophoton}, integrating over the photon signal region $0.06 < x_\gamma < 1 - m_{N_2}^2(2-\delta)^2/s$ and $|c_\gamma| < 1/\sqrt{2}$, for $\sqrt{s} = 200$~GeV. The cross section for $C_{eN} \neq 0$ can be found by repeating the same procedure, with $C_{Ne}^{V,RR} = C_{eN}$. In Fig.~\ref{fig:nuSMEFT_constraints_comparison}, the resulting constraints on $|C_{eN}|$ are shown as grey shaded regions. We note that for $i \neq j$, we do not take into account the decay $N_2 \to N_1 e^-e^+$ via $C_{eN}$, which would modify the LEP bound similarly to the exclusive mono-$\gamma$ plus $\slashed{E}$ bounds at FCC-ee. For both $i = j$ and $i \neq j$, the FCC-ee mono-$\gamma$ plus $\slashed{E}$ search at $\sqrt{s} = 240$~GeV improves on the LEP bound by over an order of magnitude, and also extends the sensitivity to larger values of $m_{N_2}$. For $i \neq j$, the DV searches at FCC-ee increase the sensitivity much further for $400~\text{MeV} \lesssim m_{N_2} \lesssim 110~\text{GeV}$.

In the centre-left panel of Fig.~\ref{fig:nuSMEFT_constraints_comparison}, we instead show the bounds on the coefficient $C_{lNle}$ for the $d = 6$ $\nu$SMEFT operator $Q_{lNle} = (\bar{L}_\rho N_{Rj})\epsilon(\bar{L}_\sigma e)$, with $\rho = \sigma = e$ and $j = 2$. From Eq.~\eqref{eq:matching_example_2}, $C_{lNle}$ contributes to both the scalar and tensor four-fermion interaction WCs as $C_{\nu Ne}^{S,RR} = 12C_{\nu Ne}^{T,RR} = 3C_{lNle}/2$. However, in the analyses of Secs.~\ref{sec:ff_monophoton} and~\ref{subsec:ff_DV}, we considered only one non-zero coefficient $C_i \in \{C_{\nu Ne}^{S,RR}, C_{\nu Ne}^{T,RR}\}$ at a time. In general, it is not possible to obtain the exact limits on $C_{lNle}$ from the individual constraints on $C_{\nu Ne}^{S,RR}$ and $C_{\nu Ne}^{T,RR}$ in Figs.~\ref{fig:off-diag-constraints} and~\ref{fig:ff_DV_constraints}, because 
this cannot account for one coefficient dominating the production and another the decay, or both being equally important. 
Fortunately, with the particular relation $C_{\nu Ne}^{S,RR} = 12C_{\nu Ne}^{T,RR}$, the scalar coefficient dominates both the production and decay of $N_2$. In Fig.~\ref{fig:nuSMEFT_constraints_comparison}, we thus show the limits found by rescaling the mono-$\gamma$ plus $\slashed{E}$ and DV bounds on the scalar coefficient only, according to $C_{\nu N e}^{S,RR} = 3C_{lNle}/2$. For the LEP bound, we again rescale the purely vector operator, with the cross section for $C_{lNle}\neq 0$ found by setting $C_{\nu Ne}^{S,RR} = 12C_{\nu Ne}^{T,RR} = 3/(2\Lambda^2)$ in Eq.~\eqref{eq:tot_Dirac_cs}. The improvement of the FCC-ee bounds over LEP can be seen to be similar to that for the coefficient $C_{eN}$ for $i\neq j$.

Before moving on, we note that Ref.~\cite{Biekotter:2020tbd} also derives limits on $C_{lN}$, $C_{eN}$ and $C_{lNle}$ of $\Lambda \gtrsim 1$~TeV from the signal $e^+e^- \to \gamma(\gamma)+E_{\text{miss}}$ at L3~\cite{L3:2003yon}, but assumes that the HNL decay is prompt and proceeds dominantly via $N \to \nu \gamma$, induced by the $d = 6$ dipole operators. The same assumption is used in Ref.~\cite{Barducci:2022hll} to assess the sensitivity of FCC-ee, finding $\Lambda\gtrsim 10~$TeV (7~TeV) for $\mathcal{O}_{lN}$ and $\mathcal{O}_{eN}$ ($\mathcal{O}_{lNle}$) at $\sqrt{s} = 91.2$~GeV and $\Lambda\gtrsim 5$~TeV (4~TeV) at $\sqrt{s} = 240$~GeV. These constraints are not directly comparable to our analysis, which assumes one operator at a time and therefore that the photon originates from an incoming $e^+/e^-$. In Ref.~\cite{Barducci:2022hll}, the authors also consider DV signatures caused by HNL decay via $N \to 3f$, which is analogous to our DV sensitivities with $\delta = 1$. Ref.~\cite{Barducci:2022hll} finds a maximum FCC-ee reach of $\Lambda \sim 30$~TeV, which is qualitatively similar to the DV sensitivities for $C_{eN}$ ($i \neq j$) and $C_{lNle}$ in Fig.~\ref{fig:nuSMEFT_constraints_comparison}.

Next, in the centre-right panel of Fig.~\ref{fig:nuSMEFT_constraints_comparison}, we show the bounds on the coefficient $C_{HNe}$ of the $d = 6$ $\nu$SMEFT operator $Q_{HNe} = (\bar{N}_{Rj}\gamma_\mu e)(\tilde{H}^{\dagger} i D^\mu H)$, for $j = 2$. From the matching condition in Eq.~\eqref{eq:matching_example_2}, the sensitivities of the mono-$\gamma$ plus $\slashed{E}$ and DV searches at FCC-ee can be taken directly from Figs.~\ref{fig:off-diag-constraints-2} and~\ref{fig:gauge_DV_constraints}. We note that any charged-current process that leads to the production and decay of HNLs via the active-sterile mixing can also do so via the effective $W^\pm$ interaction, with the rates for these processes found by replacing $V_{eN}$ by $W_N^R = \frac{v^2}{2}C_{HNe}$. The constraints from signal processes involving only charged-current interactions can therefore be rescaled trivially. Many constraints on the active-sterile mixing also arise from signal processes involving neutral-current HNL decays. These bounds must be rescaled by the ratio of decay rates for the active-sterile mixing and effective $W^\pm$ interaction scenarios, as performed in~\cite{Fernandez-Martinez:2023phj}. From the rescaling of bounds on the electron-flavour mixing $V_{eN_2}$, we show in Fig.~\ref{fig:nuSMEFT_constraints_comparison} the bounds from NA62~\cite{NA62:2020mcv}, T2K~\cite{T2K:2019jwa}, BEBC~\cite{Barouki:2022bkt}, CHARM~\cite{CHARM:1985nku,CHARMII:1994jjr}, Belle~\cite{Belle:2013ytx}, ATLAS~\cite{ATLAS:2022atq}, CMS~\cite{CMS:2024bni} and PMNS unitarity~\cite{Fernandez-Martinez:2016lgt}. As we consider Dirac HNLs in Fig.~\ref{fig:nuSMEFT_constraints_comparison}, we do not show constraints taken from searches for lepton number violating signals, such as $0\nu\beta\beta$ decay and same-sign lepton signatures at colliders. The constraints on $C_{HNe}$ from $0\nu\beta\beta$ decay experiments have been explored in detail in~\cite{Dekens:2020ttz,Dekens:2021qch}.

Finally, the lower two panels of Fig.~\ref{fig:nuSMEFT_constraints_comparison} depict the bounds on the coefficient $C_{HN}$ of the $d = 6$ $\nu$SMEFT operator $Q_{HN} = (\bar{N}_{Ri}\gamma_\mu N_{Rj})(H^{\dagger} i \overleftrightarrow{D}^\mu H)$, for $i = j = 2$ (left) and $i = 1$, $j =2$ (right). Again, from the matching condition in Eq.~\eqref{eq:matching_example_1}, the FCC-ee sensitivities can be directly transferred from Figs.~\ref{fig:off-diag-constraints-2} and~\ref{fig:gauge_DV_constraints}. We compare these to the current limit from LEP, found by rescaling the upper bound on the purely vector four-fermion coefficient by the ratio of cross sections for $e^+ e^- \to NN\gamma$; our result for $i = j$ is in good agreement with the calculation of~\cite{Fernandez-Martinez:2023phj}. Also shown is the constraint from decays $\Upsilon(1S)\to NN$ at BaBar, which enforced the upper limit $\text{BR}(\Upsilon(1S)\to\text{inv}) < 3\times 10^{-4}$ at 90\%~CL~\cite{BaBar:2009gco}. Bounds from the invisible decays of other neutral mesons ($\pi^0$, $\eta$, $\eta'$, $\omega$, $\phi$, $J/\psi$) are less stringent, and therefore do not appear in the parameter space of Fig.~\ref{fig:nuSMEFT_constraints_comparison}. Indirect bounds from supernova cooling~\cite{DeRocco:2019jti,Fernandez-Martinez:2023phj} are also present at smaller HNL masses than those shown in Fig.~\ref{fig:nuSMEFT_constraints_comparison}. 

The coefficient $C_{HN}$ can contribute to the invisible decays of the $Z$ boson at LEP and FCC-ee. At LEP, the invisible $Z$ width was inferred from the peak hadronic cross section, $\sigma_{\text{had}}^{\text{peak},0}$, and ratio of hadronic and leptonic partial widths, $R_\ell^0 = \Gamma_\text{had}/\Gamma_{\ell\ell}$, assuming lepton universality, to be $\Gamma_{\text{inv}}|_{\text{exp}}  = 499.0\pm 1.5$~MeV~\cite{ALEPH:2005ab}. Compared to the SM prediction, $\Gamma_{\text{inv}}|_{\text{SM}}  = 501.48 \pm 0.04$~MeV~\cite{Freitas:2014hra}, this constrains the contribution of additional invisible final states. Using the decay rate for $Z\to NN$ in App.~\ref{app:decays}, we perform a simple $\chi^2$ fit to the LEP invisible width, excluding the blue shaded region at 90\%~CL in Fig.~\ref{fig:nuSMEFT_constraints_comparison}. A similar measurement can be performed at FCC-ee, with a relative precision of $10^{-4}$ expected for the invisible $Z$ width, being limited by the uncertainty in the luminosity~\cite{AlcarazMaestre:2021ssl}. Assuming the measured invisible width to be agreement with the SM, we find the dashed blue lines in Fig.~\ref{fig:nuSMEFT_constraints_comparison} to be the sensitivity of FCC-ee. Taking instead the same experimental value as LEP, we obtain the dotted blue lines. Note that for $i \neq j$, we again do not take into account the impact of $N_2$ decays on the LEP and BaBar constraints, which would render the bounds invalid for large $m_{N_2}$ values. Comparing all of the constraints and sensitivities in Fig.~\ref{fig:nuSMEFT_constraints_comparison}, we can see that the FCC-ee mono-$\gamma$ plus $\slashed{E}$ searches at $\sqrt{s} = 91.2$~GeV and $\sqrt{s} = 240$~GeV provide the most stringent limits below and above $m_{N_2} \lesssim 91.2/(2 - \delta)$, respectively. For $i \neq j$, the DV search at $\sqrt{s} = 91.2$~GeV can extend the limits to much smaller values of $C_{HN}$ for $100~\text{MeV} \lesssim m_{N_2} \lesssim 80~\text{GeV}$.

\begin{figure}[t!]
\centering
\includegraphics[width=0.49\textwidth]{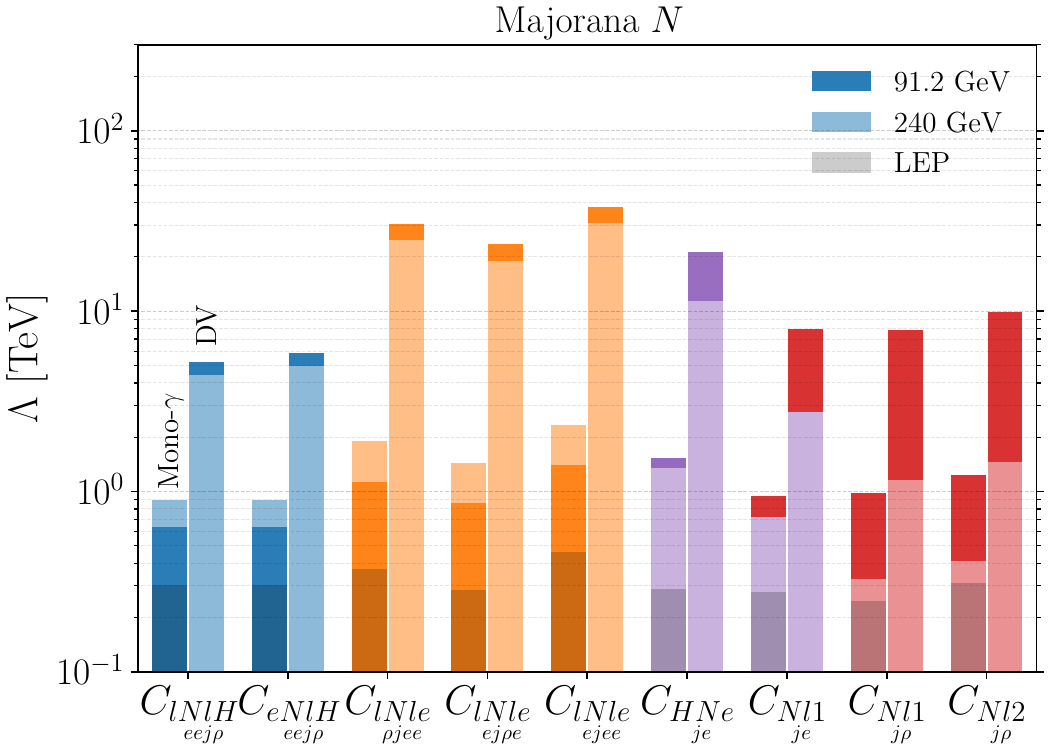}
\includegraphics[width=0.49\textwidth]{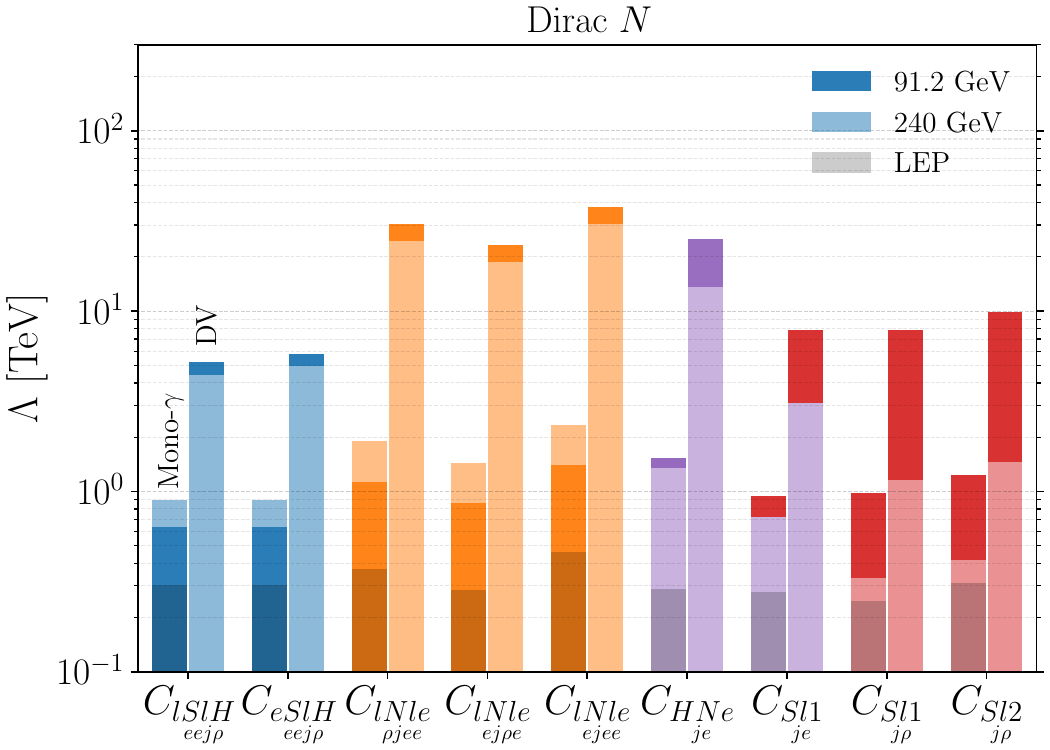}
\caption{Maximum reach to the scale of new physics $\Lambda$ for the $d = 6$ and $d = 7$ $\nu$SMEFT operators involving a single HNL, inducing $e^+ e^- \to \nu N (\gamma)$, in the Majorana (left) and Dirac (right) scenarios. For each operator, the FCC-ee sensitivities from the mono-$\gamma$ plus $\slashed{E}$ (left) and the DV (right) searches are shown. The LEP mono-$\gamma$ plus $\slashed{E}$ constraints are shown as black striped bars.}
\label{fig:one-light-one-heavy-barchart}
\end{figure}

This concludes the review of the constraints on the $d = 6$ operators in Table~\ref{tab:vSMEFT-operators}. However, for completion, we summarise in Figs.~\ref{fig:two-heavy-barchart}--\ref{fig:two-light-barchart} the sensitivities of FCC-ee to all $d \leq 7$ operators considered in this work. In the Majorana HNL scenario, these are the $\Delta L = 0$ and $\Delta L = \pm 2$ operators in Table~\ref{tab:vSMEFT-operators}. Meanwhile, in the Dirac HNL scenario, the $\Delta L = 0$ operators in Tables~\ref{tab:vSMEFT-operators} and~\ref{tab:vSMEFT-operators-2} are relevant. For the FCC-ee sensitivities in Figs.~\ref{fig:diag-constraints}--\ref{fig:gauge_DV_constraints}, we take the smallest value of $|C_i|$ that can be probed, which occurs in the $m_{N_2} \to 0$ limit for the mono-$\gamma$ plus $\slashed{E}$ sensitivities and at the tip of the DV sensitivities. These values are then related to the $d = 6$ and $d = 7$ $\nu$SMEFT operator WCs using the relations in App.~\ref{app:matching}, assuming that the active-sterile mixing is negligible. Then, the $\nu$SMEFT WCs are related to the scale of new physics as $C_i^{(6)} = 1/\Lambda^2$ and $C_i^{(7)} = 1/\Lambda^3$ for the $d = 6$ and $d = 7$ operators, respectively. In Figs.~\ref{fig:two-heavy-barchart}--\ref{fig:two-light-barchart}, we therefore show the maximum reach of FCC-ee for each operator, which may occur at different values of $m_{N_2}$. For comparison, we also show in Figs.~\ref{fig:two-heavy-barchart},~\ref{fig:one-light-one-heavy-barchart} and~\ref{fig:two-light-barchart} the current constraints from mono-$\gamma$ plus $\slashed{E}$ searches at LEP.

%%%%%%%%%%%%%%%%%%%%%%%%%%%%%%%%%%%%%%%%
\subsection{Two HNL $\nu$SMEFT Operators at $d\leq 7$}
%%%%%%%%%%%%%%%%%%%%%%%%%%%%%%%%%%%%%%%%

%
\begin{figure}[t!]
\centering
\includegraphics[width=0.49\textwidth]{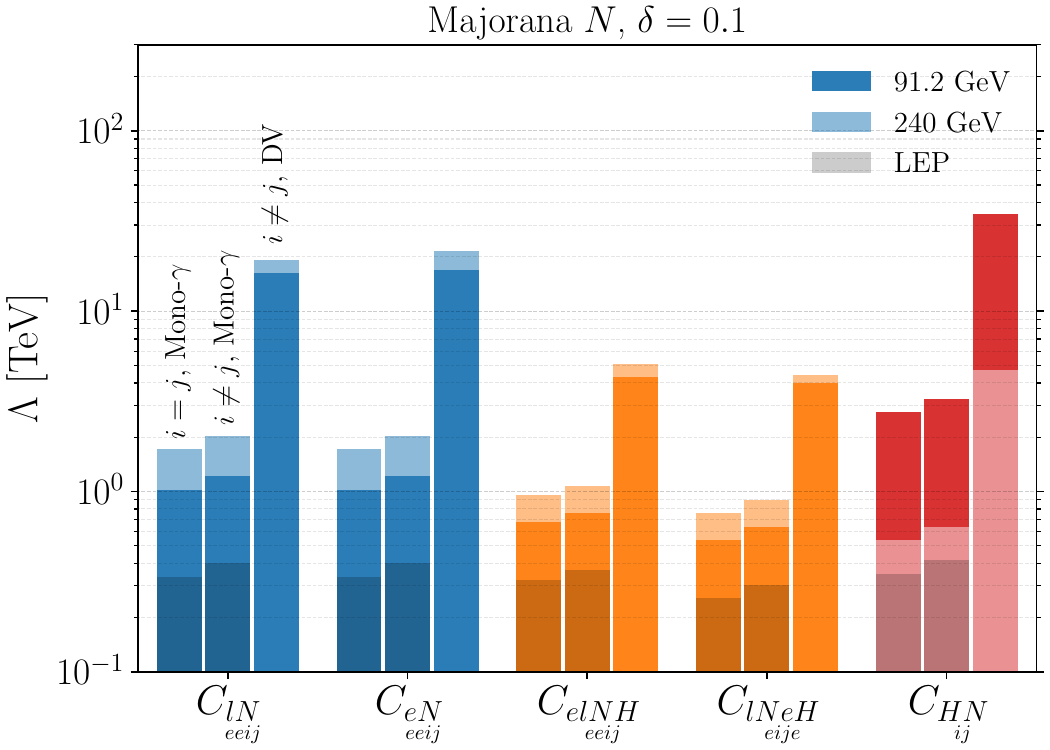}
\includegraphics[width=0.49\textwidth]{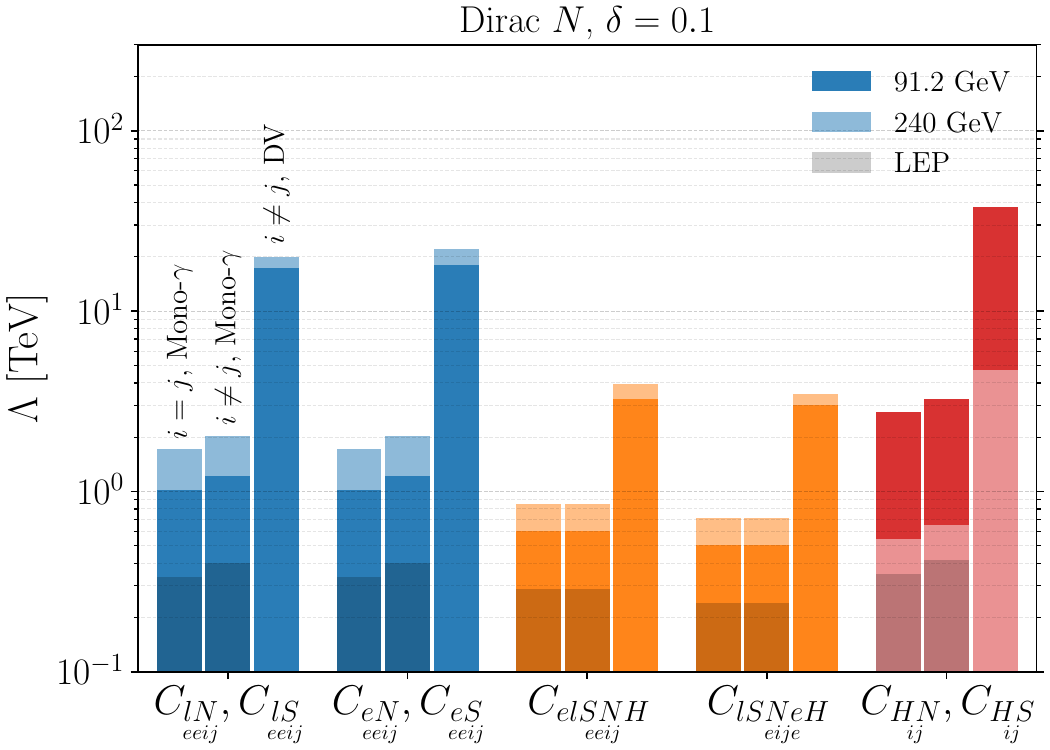} 
\caption{Maximum reach to the scale of new physics $\Lambda$ for the $d = 6$ and $d = 7$ $\nu$SMEFT operators involving two HNLs, inducing $e^+ e^- \to N N (\gamma)$, in the Majorana (left) and Dirac (right) scenarios. For each operator, the FCC-ee sensitivities from the mono-$\gamma$ plus $\slashed{E}$ search are shown for $i = j$ (left) and $i \neq j$ (centre) and the DV search for $i \neq j$ (right). The LEP mono-$\gamma$ plus $\slashed{E}$ constraints are shown as black striped bars.}
\label{fig:two-heavy-barchart}
\end{figure}

Firstly, in Fig.~\ref{fig:two-heavy-barchart}, we show the maximum reach of FCC-ee to the $d = 6$ and $d = 7$ $\nu$SMEFT operators inducing effective interactions at the EW scale involving two HNLs $N_1$ and $N_2$ in the Majorana (left) and Dirac (right) scenarios, assuming the benchmark mass splitting ratio $\delta = (m_{N_2} - m_{N_1})/m_{N_2} = 0.1$. For each operator, the left bar shows the maximum reach of FCC-ee to the $i = j$ coefficient from the mono-$\gamma$ plus $\slashed{E}$ search. The centre and right bars instead indicate the reach to the $i \neq j$ coefficient from the mono-$\gamma$ plus $\slashed{E}$ and DV searches, respectively. The light and dark shaded bars, respectively, illustrate the reaches for $\sqrt{s} = 91.2$~GeV and~$240$~GeV.

Most $\nu$SMEFT operators are matched to a single effective interaction at the EW scale. However, the $d = 6$ operator $Q_{lN}$ generates both of the operators $(\bar{N}\gamma_\mu P_R N)(\bar{e}\gamma^\mu P_L e)$ and $(\bar{N}\gamma_\mu P_R N)(\bar{\nu}_e\gamma^\mu P_L \nu_e)$ at the EW scale, giving the decay mode $N_2 \to N_1 \nu \bar{\nu}$ which was not considered for the bound on $|C_i|$ in Fig.~\ref{fig:ff_DV_constraints}. For the DV bound in Fig~\ref{fig:two-heavy-barchart}, we take this into account, resulting in $\Lambda$ being at a slightly lower value with respect to $C_{eN}$. Additionally, the $d = 7$ operator $Q_{lSNeH}$ contributes to both the scalar and tensor four-fermion operators in the Dirac HNL scenario, with $C_{Ne}^{S,RR} = 4 C_{Ne}^{T,RR} = - v C_{lSNeH}/(2\sqrt{2})$. If we assume that the FCC-ee sensitivities are limited by the production cross section\footnote{This is always true for the maximum mono-$\gamma$ plus $\slashed{E}$ sensitivity in the limit $m_{N_2} \to 0$, where the HNL decay length satisfies $L \gg 5$~m and the probability to decay outside the detector is $\mathcal{P}_{\text{out}} \approx 1$, but not for the maximum DV sensitivity for $\delta = 0.1$. In Fig.~\ref{fig:ff_DV_constraints}, the $\delta = 0.1$ sensitivities are cut off at the kinematic threshold $m_{N_2} = \sqrt{s}/(2 - \delta)$, unlike the $\delta = 0.5$ and $1$ sensitivities which reach the cross section-limited regime at the tips of the excluded regions, where the probability to decay inside the detector satisfies $\mathcal{P}_{\text{in}} \approx 1$. Nevertheless, we have verified that $\Lambda^6 \approx \Lambda_S^6 + \Lambda_T^6$ still approximately holds for $\delta = 0.1$.}, we can convert the scalar and tensor coefficient limits separately as $|C_{Ne}^{S,RR}| = v/(2\sqrt{2}\Lambda_S^3)$ and $|C_{Ne}^{T,RR}| = v/(8\sqrt{2}\Lambda_T^3)$ and determine the scale of new physics for the $\nu$SMEFT coefficient as $C_{lSNeH} = 1/\Lambda^3$, with $\Lambda^6 = \Lambda_S^6 + \Lambda_T^6$. This simple relation arises because the scalar and tensor operators do not interfere at the level of the total cross section $\sigma$. As can be seen in Eq.~\eqref{eq:tot_Dirac_cs_theta}, there is a scalar-tensor interference term in $d\sigma/d\cos\theta$, but it is proportional to $\cos\theta$ and therefore vanishes in $\sigma$. The mono-$\gamma$ plus $\slashed{E}$ (DV) sensitivities can be seen to reach $\Lambda\sim 1$--$2$~TeV ($\Lambda\sim 20$--$30$~TeV) for the $d = 6$ operators and $\Lambda\sim 600$--$900$~GeV ($\Lambda\sim 3$--$5$~TeV) for the $d = 7$ operators.\footnote{We emphasise that the DV reaches are taken at the mass of $N_2$ which maximises the sensitivity. Not all values of $\Lambda$ smaller than the reach shown in Fig.~\ref{fig:two-heavy-barchart} would be excluded for this particular value of $m_{N_2}$, but would instead be ruled out for other values.} The FCC-ee sensitivities are seen to be a large improvement over the mono-$\gamma$ plus $\slashed{E}$ constraints at LEP (black striped bars), which exclude up to $\Lambda \sim 300$--$500$~GeV.

%

%%%%%%%%%%%%%%%%%%%%%%%%%%%%%%%%%%%%%%%%
\subsection{Single HNL $\nu$SMEFT Operators at $d\leq 7$}
%%%%%%%%%%%%%%%%%%%%%%%%%%%%%%%%%%%%%%%%

Next, in Fig.~\ref{fig:one-light-one-heavy-barchart}, we show the maximum reach of FCC-ee to the $d = 6$ and $d = 7$ $\nu$SMEFT operators leading to effective interactions with a single HNL at the EW scale, shown for the Majorana (left) and Dirac (right) scenarios. For each operator, the left and right bars show the maximum sensitivity of the mono-$\gamma$ plus $\slashed{E}$ and DV searches at FCC-ee, respectively. For example, $Q_{lNle} = (\bar{L}_{\rho}N_{Rj})\epsilon(\bar{L}_\sigma e) $ can contribute to both the scalar and tensor four-fermion interaction WCs, as already discussed in the context of Fig.~\ref{fig:nuSMEFT_constraints_comparison}. For $\rho = \mu,\tau$ and $\sigma = e$, only the scalar coefficient is generated, with $C_{\nu N e}^{S,RR} = C_{lNle}$. However, for $\rho = e$ and $\sigma = \mu,\tau$, the matching condition is now given by $C_{\nu Ne}^{S,RR} = 4C_{\nu Ne}^{T,RR} = C_{lNle}/2$. Assuming that the maximum FCC-ee sensitivities are cross section-limited, the scalar and tensor four-fermion interaction WCs are again converted separately as $|C_{\nu Ne}^{S,RR}| = 1/(2\Lambda_S^2)$ and $|C_{\nu Ne}^{T,RR}| = 1/(8\Lambda_T^2)$, and the scale for the $d = 6$ $\nu$SMEFT coefficient found as $C_{lNle} = 1/\Lambda^2$, with $\Lambda^4 = \Lambda_S^4 + \Lambda_T^4$. We note that, again, there is no interference between the scalar and tensor operators. For $\rho = \sigma = e$, we instead convert the scalar coefficient as $C_{\nu Ne}^{S,RR} = 3/(2\Lambda_S^2)$. 

The $d = 7$ $\nu$SMEFT operator $Q_{Nl1}$ also contributes to both the effective $W^\pm$ and $Z$ interactions in the Majorana HNL scenario, with the matching condition $W_N^L = -2Z_{\nu N}^R = -v^3 C_{Nl1}/(2\sqrt{2})$. In this case, we must also take into account that the contributions of $W_N^L$ and $Z_{\nu N}^R$ to the process $e^+e^-\to \nu N (\gamma)$ interfere. Taking the sensitivities to again be cross section limited, the effective $W^\pm$ and $Z$ interaction WCs can be converted to separate scales of new physics as $|W_N^L|= v^3/(2\sqrt{2}\Lambda_W^3)$ and $|Z_{\nu N}^R|= v^3/(4\sqrt{2}\Lambda_Z^3)$, respectively, and then to a scale for the $d = 7$ operator as $C_{Nl1} = 1/\Lambda^3$, with
\begin{align}
\Lambda^6 = \Lambda_W^6 \bigg(1 - \frac{\hat{\sigma}_{WZ}}{2(\hat{\sigma}_W - \hat{\sigma}_Z)}\bigg) + \Lambda_Z^6 \bigg(1 + \frac{2\hat{\sigma}_{WZ}}{\hat{\sigma}_W - \hat{\sigma}_Z}\bigg) \,.
\end{align}
Here, $\hat{\sigma}_{W}$ and $\hat{\sigma}_{Z}$ are the cross sections for $e^+e^- \to \nu N(\gamma)$ via $W^\pm$ and $Z$ exchange, respectively, while $\hat{\sigma}_{WZ}$ is the contribution from the interference. The hats again denote that the dependence on the WCs $W_N^L$ and $Z_{\nu N}^R$ is removed from the cross sections. For the DV sensitivities, we simply compute these from the $2\to 2$ cross section in Eq.~\eqref{eq:tot_Maj_cs}, while for the mono-$\gamma$ plus $\slashed{E}$ sensitivities, we insert Eq.~\eqref{eq:tot_Maj_cs} into Eq.~\eqref{eq:xsec_monophoton} and integrate over the relevant photon signal region. For LEP, this is given below Eq.~\eqref{eq:xsec_rescale}, while for FCC-ee we account for the cuts in Table~\ref{tab:universal_cuts}. Ultimately, we find that the sensitivities on the effective $W^\pm$ and $Z$ interactions dominate at $\sqrt{s} = 240$~GeV and $\sqrt{s} = 91.2$~GeV, respectively. The procedure is equivalent for the $d = 7$ operator $Q_{Sl1}$ which contributes to the WCs $W_N^L$ and $Z_{\nu N}^L$ at the EW scale. We note that the coefficient $C_{Nl1}$ is heavily constrained in the Majorana case by the non-observation of $0\nu\beta\beta$ decay~\cite{Dekens:2020ttz}.

To summarise, the mono-$\gamma$ plus $\slashed{E}$ (DV) sensitivities can be seen to reach $\Lambda\sim 1$--$2$~TeV ($\Lambda\sim 20$--$40$~TeV) for the $d = 6$ operators and $\Lambda\sim 600$--$900$~GeV ($\Lambda\sim 5$--$10$~TeV) for the $d = 7$ operators. The reach of the DV search is greater compared to Fig.~\ref{fig:two-heavy-barchart} because the $\delta = 1$ sensitivities in Figs.~\ref{fig:ff_DV_constraints} and~\ref{fig:gauge_DV_constraints} are used.

%%%%%%%%%%%%%%%%%%%%%%%%%%%%%%%%%%%%%%%%
\subsection{SMEFT Operators at $d\leq 7$}
%%%%%%%%%%%%%%%%%%%%%%%%%%%%%%%%%%%%%%%%

\begin{figure}[t!]
\centering
\includegraphics[width=0.8\textwidth]{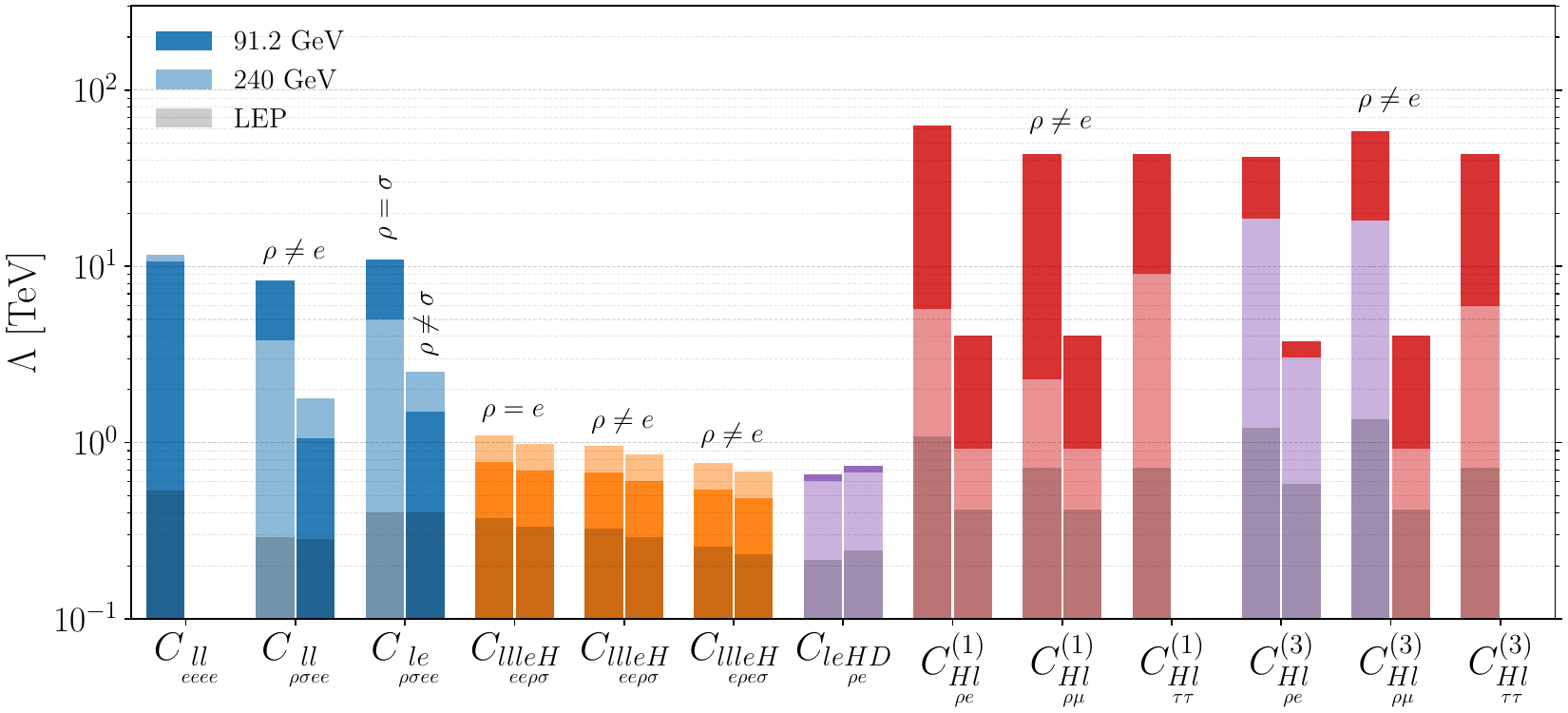}
\caption{Maximum reach to the scale of new physics $\Lambda$ for the $d = 6$ and $d = 7$ $\nu$SMEFT operators involving the active neutrinos, inducing $e^+ e^- \to \nu \nu (\gamma)$. For each operator, the FCC-ee sensitivities from the mono-$\gamma$ plus $\slashed{E}$ search are shown for $\rho = \sigma$ (left) and $\rho \neq \sigma$ (right). The mono-$\gamma$ plus $\slashed{E}$ constraints from LEP are shown as black striped bars.}
\label{fig:two-light-barchart}
\end{figure}

Finally, in Fig.~\ref{fig:two-light-barchart}, we show the maximum reach of FCC-ee to the $d = 6$ and $d = 7$ SMEFT operators generating effective interactions at the EW scale involving no HNLs. The only distinction between the Majorana and Dirac HNL scenarios is that the $\Delta L =\pm 2$ operators $Q_{llleH}$ and $Q_{leHD}$ are assumed to vanish in the latter scenario. The results shown for these operators only apply in the Majorana case. The translation of the FCC-ee mono-$\gamma$ plus $\slashed{E}$ bounds in Figs.~\ref{fig:diag-constraints},~\ref{fig:off-diag-constraints} and~\ref{fig:off-diag-constraints-2} to the SMEFT WCs is now slightly more involved, because the WCs can interfere with the SM contribution to $e^+e^- \to \nu \nu (\gamma)$. 

For the WCs that do not interfere with the SM, the appropriate matching condition can be used to rescale the constraint on $|C_i|$ in the $m_{N_2} \to 0$ limit, as for the operators in Figs.~\ref{fig:two-heavy-barchart} and~\ref{fig:one-light-one-heavy-barchart}. This is the case for the $d = 7$ SMEFT operators,
\begin{gather}
C_{\underset{\rho\sigma ee}{\nu e}}^{S,LL} = -\sqrt{2}v\bigg(C_{\underset{ee\{\rho\sigma\}}{llleH}} + \frac{1}{2}C_{\underset{e\{\rho e\sigma\}}{llleH}}\bigg)\,, \quad C_{\underset{\rho\sigma ee}{\nu e}}^{T,LL} = \frac{v}{4\sqrt{2}}C_{\underset{e[\rho e\sigma]}{llleH}} \,, \nonumber \\
\frac{2}{v^2}[W_{\nu}^R]_{\rho e} = -\frac{v}{\sqrt{2}}C_{\underset{\rho e}{leHD}}\,,
\label{eq:no-interference}
\end{gather}
where the curly (square) brackets denote the (anti-)symmetrisation of the flavour indices. As can be seen in Fig.~\ref{fig:two-light-barchart}, the constraints on $C_{llleH}$ depend on $\rho,\sigma = e,\mu,\tau$; in each case, the left and right bars show $\rho = \sigma$ and $\rho\neq\sigma$, respectively. The tensor operators relevant to $e^+e^-\to \nu \nu(\gamma)$ are only generated when the first and third flavour indices are of electron flavour and the second and fourth indices are different. When the coefficient for the tensor operator is non-zero, the maximum reach for the SMEFT operator is determined as $C_{llleH} = 1/\Lambda^3$, with $\Lambda^6 = \Lambda_S^6 + \Lambda_T^6$, where $\Lambda_S$ and $\Lambda_T$ are found separately from Eq.~\eqref{eq:no-interference}. For the SMEFT coefficient $C_{leHD}$, we show the maximum reach for $\rho = e$ (left) and $\rho \neq e$ (right) as purple bars. This coefficient is also constrained considerably by the non-detection of $0\nu\beta\beta$ decay~\cite{Cirigliano:2018yza}.

We next examine the maximum reach for the remaining $d = 6$ SMEFT operators, which can interfere with the SM depending on the flavour indices of the coefficient. Firstly, the operators $Q_{ll}$ and $Q_{le}$ contribute to the vector four-fermion operators as
\begin{gather}
C_{\underset{\rho\sigma ee}{\nu e}}^{V,LL} = C_{\underset{\rho\sigma ee}{ll}} + C_{\underset{ee\rho\sigma}{ll}}\,, \quad C_{\underset{\rho\sigma ee}{\nu e}}^{V,LR} = C_{\underset{\rho\sigma ee}{le}} \,.
\end{gather}
Using either Eq.~\eqref{eq:tot_Dirac_cs} or Eq.~\eqref{eq:tot_Maj_cs} in the $m_i, m_j\to 0$ limit, the contribution of these WCs to the total cross section for $e^+e^-\to \sum \nu \bar{\nu}$ is
\begin{align}
\sigma &= \sigma\big|_{\text{SM}}+\sum_{\rho,\sigma}\bigg[\frac{s}{48\pi}\Big(\big|C_{\underset{\rho\sigma ee}{\nu e}}^{V,LL}\big|^2 + \big|C_{\underset{\rho\sigma ee}{\nu e}}^{V,LR}\big|^2\Big) + \frac{G_F M_Z^2}{6\sqrt{2}\pi}\delta_{\rho\sigma}\chi_2\text{Re}\Big[g_L^e C_{\underset{\rho\sigma ee}{\nu e}}^{V,LL} + g_R^e C_{\underset{\rho\sigma ee}{\nu e}}^{V,LR}\Big]\nonumber\\
&\hspace{6.5em} + \frac{G_F M_W^2}{4\sqrt{2}\pi}\delta_{\rho e}\delta_{\sigma e}\text{Re}\Big[C_{\underset{\rho\sigma ee}{\nu e}}^{V,LL}\Big]\bigg(3 + 2\omega - 2(1+\omega)^2\log\bigg(\frac{1+\omega}{\omega}\bigg)\bigg)\bigg]\,,
\end{align}
with $\omega \equiv M_W^2/s$ and $\chi_2$ is given in Eq.~\eqref{eq:chi12}. The second and third terms in the summation are the interference of the WCs with the SM $Z$ and $W^\pm$ exchange diagrams, respectively. The WCs $C_{ll}$ and $C_{le}$ with $\rho = \sigma$ interfere with the SM $Z$ diagram away from the $Z$ pole, where the interference vanishes, while $C_{ll}$ with $\rho = \sigma = e$ interferes with the SM $W^\pm$ diagram. For $\rho \neq \sigma$, there is no interference with the SM.

The maximum reach of mono-$\gamma$ plus $\slashed{E}$ searches FCC-ee can be found as in Eq.~\eqref{eq:xsec_rescale}, but now including the interference terms if present,
\begin{align}
\label{eq:xsec_rescale-2}
|C_i|^2\hat{\sigma}_i = \text{Re}\big[C_j\big]\hat{\kappa}_j + |C_j|^2\hat{\sigma}_j + \ldots\,.
\end{align}
On the LH side we write the cross section for $e^+e^-\to NN\gamma$ in the $m_{N_2} \to 0$ limit, with the coefficient $C_{Ne}^{V,RR}$ extracted. The RH side is the cross section for $e^+e^- \to \sum \nu \bar{\nu} \gamma$, split into interference and non-interference terms, with the coefficient $C_{\nu e}^{V,LL}$ or $C_{\nu e}^{V,LR}$ removed. The hatted cross sections are computed at leading order (LO) by inserting the relevant $2\to 2$ cross sections into Eq.~\eqref{eq:xsec_monophoton} and integrating over the photon energy and angle accordingly. For the LEP bounds, we integrate over $0.06 < x_\gamma < 1 - m_{N_2}^2(2-\delta)^2/s$ and $|c_\gamma| < 1/\sqrt{2}$ for $\sqrt{s} = 200$~GeV. For FCC-ee, we instead integrate over the phase space that is not removed by the cuts in Table~\ref{tab:universal_cuts} and $p_T^\gamma > 1$~GeV, for $\sqrt{s} = 91.2$~GeV and~$240$~GeV. Incorporating the $p_T^\gamma$ cut is easiest with the change of variables $(x_\gamma, c_\gamma)\to (p_T^\gamma,c_\gamma)$ in the differential cross section, as explained below Eq.~\eqref{eq:xsec_monophoton}. The $p_T^\gamma$ cut has the advantage of shifting where the interference between $C_{ll}$ and $C_{le}$ and the SM $Z$ diagram vanishes to a value of $\sqrt{s}$ sufficiently above the $Z$ pole. The cross section $\hat{\kappa}_j$ is therefore not as sensitive of the value of $\sqrt{s}$ in the vicinity of the $Z$ pole as for $e^+e^- \to \sum \nu \bar{\nu}$. This also reduces the importance of next-to-leading order (NLO) corrections to $e^+e^-\to \sum\nu\bar{\nu}\gamma$ for the evaluation of $\hat{\kappa}_j$. In the following, we do not take into account the NLO corrections to remain consistent with the analysis of Sec.~\ref{sec:monophoton}, where we simulate the signals and background at LO. The corrections decrease and increase the SM contribution to $e^+e^-\to \sum\nu\bar{\nu}\gamma$ at $\sqrt{s} = 91.2$~GeV and $240$~GeV, respectively, and decrease the interference contributions at both $\sqrt{s}$ values. We expect this to have a non-negligible but small impact on the obtained limits. With the hatted cross sections now evaluated, we set $C_i$ in Eq.~\eqref{eq:xsec_rescale-2} to the maximum reach of LEP or FCC-ee and rearrange to find the maximum reach for $C_j$. The resulting reaches for $C_{ll}$ and $C_{le}$ are shown as blue bars in Fig.~\ref{fig:two-light-barchart}. We comment here that the  $C_{ll}$ and $C_{le}$ sensitivities can be improved if a shape analysis is considered instead of the counting analysis performed here. This is particularly applicable to these operators due to the existence and dominance of the interference terms, which modifies the shape of the $p_T^\gamma$ distribution.

Next, we consider the $d = 6$ SMEFT operators $Q_{Hl}^{(1)}$ and $Q_{Hl}^{(3)}$, which contribute to the effective $W^\pm$ and $Z$ interactions involving light neutrinos. However, the situation is further complicated by contribution of $Q_{Hl}^{(3)}$ to the process $\mu^-\to e^-\bar{\nu}_e \nu_\mu$, which is used to determine the value of the input parameter $G_F$. As such, we make use of the $\{\hat{M}_W, \hat{M}_Z, \hat{G}_F\}$ input scheme~\cite{Brivio:2017bnu}, where the canonically normalised values of these quantities are shifted linearly by the $d = 6$ coefficients. While the masses $M_W$ and $M_Z$ are shifted by operators not considered further here, the operator $Q_{Hl}^{(3)}$ shifts the Fermi constant as,
\begin{align}
\delta G_F = \frac{1}{2\hat{G}_F}\Big(C_{\underset{ee}{Hl}}^{(3)} + C_{\underset{\mu\mu}{Hl}}^{(3)}\Big)\,,
\end{align}
where $\hat{G}_F = 1.1663787 \times 10^{-5}~\text{GeV}^{-2}$ is the measured value. This shift enters the deviations to the SM $W^\pm$ and $Z$ interactions in Eq.~\eqref{eq:SM_couplings}, along with the direct contributions from $Q_{Hl}^{(1)}$ and $Q_{Hl}^{(3)}$, as
\begin{align}
\label{eq:WZ_shifts}
[\delta W_\nu^L]_{\rho e} &= -\frac{\delta G_F}{\sqrt{2}} \delta_{\rho e} +\hat{v}^2C_{\underset{\rho e}{Hl}}^{(3)}\,, \quad [\delta Z_\nu^L]_{\rho \sigma} = -\frac{\delta G_F}{\sqrt{2}}  g_L^\nu \delta_{\rho\sigma} - \frac{\hat{v}^2}{2}\Big(C_{\underset{\rho \sigma}{Hl}}^{(1)} - C_{\underset{\rho \sigma}{Hl}}^{(3)}\Big)\,,
\end{align}
respectively, where $\hat{v}^2 = 1/(\sqrt{2}\hat{G}_F)$. To determine the cross section for $e^+e^-\to \sum \nu \bar{\nu}\gamma$, the shifts in Eq.~\eqref{eq:WZ_shifts} are added to the SM contributions $[W_\nu^L]_{\rho e}^{\text{SM}} = \delta_{\rho e}$ and $[Z_\nu^L]_{\rho\sigma}^{\text{SM}} = g_L^\nu \delta_{\rho\sigma}$ and inserted into Eq.~\eqref{eq:tot_Dirac_cs}. Here, one must also take into account the deviation of the $Z$ coupling to electrons, i.e.
\begin{align}
[\delta Z_e^R]_{\alpha\beta} &= -\frac{\delta G_F}{\sqrt{2}} g_R^e \delta_{\alpha\beta}\,, \quad 
[\delta Z_e^L]_{\alpha\beta} = -\frac{\delta G_F}{\sqrt{2}} g_L^e \delta_{\alpha\beta} - \frac{\hat{v}^2}{2}\Big(C_{\underset{\alpha\beta}{Hl}}^{(1)} + C_{\underset{\alpha\beta}{Hl}}^{(3)}\Big)\,,
\end{align}
with $\alpha = \beta = e$. Finally, Eq.~\eqref{eq:tot_Dirac_cs} is inserted into Eq.~\eqref{eq:xsec_monophoton}, where we account for the QED coupling constant $e$ also being shifted by $\delta e/\hat{e} = - \delta G_F/\sqrt{2}$. For particular values of $C_{Hl}^{(1)}$ and $C_{Hl}^{(3)}$, we can integrate over the photon energy and angle to obtain the total cross section. We cross-check the accuracy of the analytical results by using the UFO file \texttt{SMEFTsim\_general\_MwScheme} provided by the \texttt{SMEFTsim} package~\cite{Brivio:2017btx,Brivio:2020onw} to simulate the process $e^+e^- \to \sum \nu \bar{\nu}\gamma$ in \texttt{MadGraph5\_aMC@NLO} for $C_{Hl}^{(1)}, C_{Hl}^{(3)} \neq 0$.

The maximum reach of mono-$\gamma$ plus $\slashed{E}$ searches FCC-ee to $C_{Hl}^{(1)}$ and $C_{Hl}^{(3)}$ can now be found as in Eq.~\eqref{eq:xsec_rescale-2}. Integrating over the photon phase space with the relevant cuts for LEP and FCC-ee included, we obtain a function of the coefficient $C_{Hl}^{(1)}$ or $C_{Hl}^{(3)}$ on the RH side of Eq.~\eqref{eq:xsec_rescale-2}. As before, there is a term linear in the WCs if they interfere with the SM, while a quadratic term is always present. However, there are now additional higher-order terms, denoted as ellipsis in Eq.~\eqref{eq:xsec_rescale-2}. These terms, being further suppressed by $\Lambda$, have a negligible impact on the maximum reach, which we have verified numerically. The resulting reaches for $C_{Hl}^{(1)}$ and $C_{Hl}^{(3)}$ are shown in Fig.~\ref{fig:two-light-barchart}, with the colors indicating whether $W^\pm$ (purple) and $Z$ (red) exchange contributes.

Before concluding this section, we note that the $d = 6$ SMEFT operators considered here are constrained by other present and future observables, at low and high energies. However, given the vast number of additional $d = 6$ SMEFT WCs which can also contribute to these and other probes, previous works~\cite{Han:2004az,Berthier:2015oma,Falkowski:2015krw,Efrati:2015eaa, Falkowski:2017pss,Ellis:2018gqa,Ellis:2020unq,Aoude:2020dwv,Ethier:2021bye,Corbett:2021eux,deBlas:2022ofj,Bartocci:2023nvp,Allwicher:2023shc,Celada:2024mcf,Ge:2024pfn,Greljo:2024ytg} have opted to perform global fits to ensure sufficient model independence, usually limiting the number of operators by enforcing a flavour symmetry such as the minimal flavour violation hypothesis. A complete analysis of the bounds from mono-$\gamma$ plus $\slashed{E}$ searches at FCC-ee in the context of global constraints is beyond the scope of this work, but we nevertheless list the relevant observables below and conduct a naive comparison of our results with the upper bounds extracted from the global fits.

The WCs $C_{ll}$ and $C_{le}$ are constrained by lepton pair production $e^+e^- \to \ell^+\ell^-$, with data available for the differential Bhabha scattering cross section $d\sigma_e/d\cos\theta$, the total cross sections $\sigma_\ell$ and forward-backward asymmetries $A_{\text{FB}}^{\ell}$ for $\ell = \mu,\tau$, and the $\tau$ polarisation $\mathcal{P}_\tau$, measured at different $\sqrt{s}$ values by LEP, SLD and VENUS~\cite{ALEPH:2005ab,ALEPH:2013dgf,VENUS:1997cjg}. At low energies, the coefficient $C_{ll}$ is constrained by the parity violating asymmetry $A_{\text{PV}}$ in M{\o}ller scattering, $e^-e^-\to e^-e^-$, measured by the SLAC E158 experiment~\cite{SLACE158:2005uay}. Both $C_{ll}$ and $C_{le}$ can also be probed by neutrino-electron scattering, such as $\nu_e e^- \to \nu e^-$ and $\nu_\mu e^- \to \nu e^-$ at CHARM~\cite{CHARM:1988tlj} and CHARM~II~\cite{CHARM-II:1994dzw}. Using these measurements, the global fit of~\cite{Falkowski:2017pss} sets the lower limit at 90\%~CL of $\Lambda \gtrsim 2$--$4$~TeV for the diagonal WCs of $C_{ll}$ and $C_{le}$, depending on the flavour. The off-diagonal WCs of $C_{ll}$ and $C_{le}$ are subject to stringent bounds from charged lepton flavour violating (cLFV) processes, $\mu\to eee$ at SINDRUM~\cite{SINDRUM:1987nra} and $\tau\to eee$, $\tau\to \mu ee$ at Belle~\cite{Hayasaka:2010np}. The former sets $\Lambda > 207$~TeV and $\Lambda > 164$~TeV for the $\rho = e$ and $\sigma = \mu$ WCs of $C_{ll}$ and $C_{le}$, respectively, while the latter imposes $\Lambda \gtrsim 8$--$11$~TeV for the other off-diagonal WCs~\cite{Crivellin:2013hpa,Pruna:2014asa,Calibbi:2017uvl}. Evidently, the bounds on $C_{ll}$ and $C_{le}$ from mono-$\gamma$ plus $\slashed{E}$ searches at FCC-ee are more stringent than current bounds on the diagonal WCs, but are not competitive with those on the off-diagonal WCs. Future measurements at FCC-ee and CEPC of $e^+e^- \to \ell^+\ell^-$ in the vicinity and above the $Z$ pole are also expected to improve the bounds on the diagonal WCs~\cite{Celada:2024mcf,Ge:2024pfn,Greljo:2024ytg}.

The WCs $C_{Hl}^{(1)}$ and $C_{Hl}^{(3)}$ are constrained by all of the observables discussed above, in addition to the following probes. Firstly, at low energies, the WCs contribute to the neutrino scattering processes $\nu_e N \to \nu X$ at CHARM~\cite{CHARM:1986vuz}, $\nu_\mu N \to \nu X$ at CCFR~\cite{CCFR:1997zzq}, and CE$\nu$NS at COHERENT~\cite{Du:2021rdg,Coloma:2024ict}. Experiments measuring atomic parity violation~\cite{Wood:1997zq} and parity-violating electron-proton~\cite{Qweak:2013zxf} and nucleus~\cite{Wang:2014guo,Beise:2004py} scattering constrain the effective weak charge $Q_W$, which can be expressed in terms of the WCs $C_{Hl}^{(1)}$ and $C_{Hl}^{(3)}$. However, it is well known that EWPOs at the $Z$ pole, i.e., the total $Z$ width $\Gamma_Z$, peak hadronic cross section $\sigma_{\text{had}}^{\text{peak},0}$ and ratios of hadronic and leptonic partial widths, provide the dominant constraints. The global fit of~\cite{Celada:2024mcf} requires $\Lambda \gtrsim 6$--$10$~TeV at 90\%~CL for $C_{Hl}^{(1)}$ and $C_{Hl}^{(3)}$. The same work considers the future sensitivity FCC-ee and CEPC to EWPOs and additional processes such as $e^+e^- \to q\bar{q}$, $e^+e^- \to Zh$, $e^+e^- \to \nu\bar{\nu} h$ and $e^+e^- \to W^+W^-$ away from the $Z$ pole. The sensitivities for the diagonal WCs are extended up to the tens of TeV. Intriguingly, with the exception of $C_{Hl}^{(3)}$ for $\rho = \sigma = e$, we find that mono-$\gamma$ plus $\slashed{E}$ searches at FCC-ee can reach beyond these limits. The greatest potential improvement is seen in the third generation couplings $C_{Hl}^{(1)}$ and $C_{Hl}^{(3)}$ for $\rho = \sigma = \tau$. For the off-diagonal WCs of $C_{Hl}^{(1)}$ and $C_{Hl}^{(3)}$, we find that cLFV bounds already constrain the scale of new physics to be $\Lambda > 164$~TeV for $\rho = e$ and $\sigma = \mu$ and $\Lambda \gtrsim 8$~TeV for the other off-diagonal combinations, already more stringent than the forecasted mono-$\gamma$ plus $\slashed{E}$ FCC-ee bounds.

%%%%%%%%%%%%%%%%%%%%%%%%%%%%%%%%%%%%%%%%
\section{Conclusions}
\label{sec:conclusions}
%%%%%%%%%%%%%%%%%%%%%%%%%%%%%%%%%%%%%%%%

HNLs are a well-motivated extension of the SM particle content, which, depending on their masses, can provide a mechanism that naturally leads to the light neutrino masses, serve as potential dark matter candidates, and offer insights into the matter-antimatter asymmetry of the universe via leptogenesis. In general, Majorana or Dirac HNLs can couple to the SM via active-sterile mixing ($V_{\alpha N}$) and/or EFT operator WCs ($C_i$) generated by heavy dynamics at the high scale $\Lambda$. In this context, we analyse the production and decay of HNLs for two proposed centre-of-mass energies at FCC-ee, $\sqrt{s} = 91.2$~GeV ($\mathcal{L} = 100$ ab$^{-1}$) and $\sqrt{s} = 240$~GeV ($\mathcal{L} = 5$ ab$^{-1}$), and subsequent final states leading to distinct mono-$\gamma$ plus $\slashed{E}$ and DV signatures. Simple cut-based analyses are proposed to take advantage of these signatures. Firstly, in Sec.~\ref{sec:active_sterile_mixing}, the sensitivity of mono-$\gamma$ plus $\slashed{E}$ searches to $V_{\alpha N}$ for a single Dirac HNL ($N$) is investigated. In Sec.~\ref{sec:ff_monophoton}, the sensitivity of mono-$\gamma$ plus $\slashed{E}$ searches to vector, scalar and tensor four-fermion operators and effective $W^\pm$ and $Z$ interactions is examined for a pair of Majorana or Dirac HNLs ($N_i$, $i = 1,2$) in the limit of vanishing $V_{\alpha N}$. Sensitivities to diagonal ($i=j$) and off-diagonal ($i \neq j$) WCs are studied alongside the effects of different mass splittings between the HNL pair. Finally, in Sec.~\ref{sec:displacedvertex}, the sensitivity of DV searches is analysed for the same EFT operators and particle content.

The reach of mono-$\gamma$ plus $\slashed{E}$ searches at FCC-ee to the electron-flavour active-sterile mixing $V_{eN}$ (Fig.~\ref{fig:mixing_exclusions}) lies almost entirely within the region already excluded by existing searches. However, the EFT operators are generally less constrained by current data, with mono-$\gamma$ plus $\slashed{E}$ and DV searches probing unconstrained regions of the parameter space for $V_{\alpha N} = 0$. The impact of non-vanishing mixing ($V_{\alpha N} \neq 0$) on the projected sensitivities is discussed in Sec.~\ref{sec:EFT_results} (mono-$\gamma$ plus $\slashed{E}$ search) and Sec.~\ref{sec:DV_results} (DV search). In Sec.~\ref{sec:discussion}, we map the maximum reach of our proposed searches at FCC-ee to the basis of $\nu$SMEFT operators in Tables~\ref{tab:vSMEFT-operators} and~\ref{tab:vSMEFT-operators-2}, using the matching conditions in App.~\ref{app:matching}. The maximal sensitivities to the scale of new physics $\Lambda$, assuming one operator at a time, are summarised in Figs.~\ref{fig:two-heavy-barchart}, \ref{fig:one-light-one-heavy-barchart} and \ref{fig:two-light-barchart} for the processes $e^+e^- \rightarrow NN(\gamma)$, $\nu N (\gamma)$ and $\nu\nu (\gamma)$, respectively.

Of the $\nu$SMEFT operators contributing to $e^+e^- \to NN(\gamma)$, it can be seen from Fig.~\ref{fig:two-heavy-barchart} that FCC-ee is more sensitive to the less-suppressed $d = 6$ vector four-fermion operators $Q_{eN}$ and $Q_{lN}$ and the $d = 6$ bosonic current operator $Q_{HN}$ (in addition to $Q_{eS}$, $Q_{lS}$ and $Q_{HS}$ in the Dirac scenario), with $\Lambda \sim 1$--$2$~TeV probed by mono-$\gamma$ plus $\slashed{E}$ searches (an improvement by a factor of $\sim 5$ with respect to existing LEP bounds) and $\Lambda \sim 20$--$35$ TeV by DV searches. The DV search sets completely new sensitivities with no equivalent existing bounds. For the coefficient $C_{HN}$ of the $d = 6$ operator $Q_{HN} = (\bar{N}\gamma_\mu N)(H^{\dagger} i \overleftrightarrow{D}^\mu H)$, FCC-ee also improves considerably on bounds from invisible decays of $\Upsilon(1S)$ (BaBar) and $Z$ (LEP), shown in Fig.~\ref{fig:nuSMEFT_constraints_comparison}. The bounds on the $d = 7$ operators $Q_{elNH}$ and $Q_{lNeH}$ ($Q_{elSNH}$ and $Q_{lSNeH}$ in the Dirac scenario) are less stringent due to the additional suppression by $\Lambda$, but are still appreciable, especially from the DV search.

The situation is analogous for the $\nu$SMEFT operators contributing to $e^+e^- \to \nu N(\gamma)$, with the reaches for the $d = 6$ scalar four-fermion operator $Q_{lNle}$ and $d = 6$ bosonic operator $Q_{HNe}$ being the most optimistic in Fig.~\ref{fig:one-light-one-heavy-barchart}, probing $\Lambda \sim 1.5$--$2.5$ TeV (mono-$\gamma$ plus $\slashed{E}$ search) and $\Lambda \sim 30$--$40$ TeV (DV search). The mono-$\gamma$ plus $\slashed{E}$ reaches are again $\sim 4$--$5$ times stronger than existing LEP constraints. While the operator $Q_{HNe} = (\bar{N}\gamma_\mu e)(\tilde{H}^{\dagger} i D^\mu H)$ in the Dirac HNL scenario is heavily constrained by existing bounds from NA62, T2K, BEBC, CHARM, Belle, ATLAS, CMS and PMNS unitarity constraints (Fig.~\ref{fig:nuSMEFT_constraints_comparison}), the DV search at FCC-ee can still reach to currently untested parts of the parameter space. In the Majorana HNL scenario, the DV search at FCC-ee is expected to not be competitive with bounds from $0\nu\beta\beta$ decay on $Q_{HNe}$. The reaches for the $d = 7$ operators $Q_{lNlH}$, $Q_{eNlH}$ and $Q_{Nl1(2)}$ in the Majorana scenario ($Q_{lSlH}$, $Q_{eSlH}$ and $Q_{Sl1(2)}$ in the Dirac scenario) are also considerable.

Finally, there are $d\leq 7$ operators in the SMEFT which contribute to $e^+e^- \to \nu\nu(\gamma)$. For these, we take into account the interference with the SM when present, which increases the mono-$\gamma$ plus $\slashed{E}$ reach for the $d = 6$ operators $Q_{ll}$, $Q_{le}$, $Q_{Hl}^{(1)}$ and $Q_{Hl}^{(3)}$ considerably, as seen in Fig.~\ref{fig:two-light-barchart}. For example, the scale of new physics that can be probed for $C_{Hl}^{(1)}$ and $C_{Hl}^{(3)}$ is $\Lambda \sim 40$--$60$ TeV, depending on the flavours involved. This can be an order of magnitude improvement over previous LEP bounds. Furthermore, these sensitivities are comparable with those from future FCC-ee and CEPC bounds from EWPOs, with bounds on the third generation couplings from this analysis potentially being an improvement. While they do not benefit from interference, the scale of new physics $\Lambda\sim 0.6$--$1$~TeV can still be probed for the $d = 7$ operators $Q_{llleH}$ and $Q_{leHD}$. In conclusion, this work highlights the potential of FCC-ee to probe a wide range of extensions leading to operators in the ($\nu$)SMEFT at low energies. We have performed a comprehensive study of $d\leq 7$ operators that can be probed by mono-$\gamma$ plus $\slashed{E}$ and DV signatures.

\acknowledgments
P.~D.~B. is supported by the Slovenian Research Agency under the research core funding No. P1-0035 and in part by the research grants N1-0253 and J1-4389. P.~D.~B. has received support from the European Union's Horizon 2020 research and innovation programme under the Marie Sk\l{}odowska-Curie grant agreement No 860881-HIDDeN. S.~K. is supported by the FWF research group funding FG1 and FWF project number P 36947-N. C.~M. is supported by the Newton International
Fellowship (NIF) of the Royal Society, UK with grant number NIF$\backslash$R1$\backslash$221737, he also wants to thank Supriya
Senapati for useful discussions. W.~P. is supported by the Excellence Programme of the Hungarian Ministry of Culture and Innovation under contract TKP2021-NKTA-64. F.~F.~D. acknowledges support from the UK Science and Technology Facilities Council (STFC) via the Consolidated
Grant ST/X000613/1.

\appendix
\newpage

%%%%%%%%%%%%%%%%%%%%%%%%%%%%%%%%%%%%%%%%
\section{Operators Matching and UV Completions}
\label{app:matching}
%%%%%%%%%%%%%%%%%%%%%%%%%%%%%%%%%%%%%%%%

Here, we give the tree-level matching conditions between the WCs of the EFT operators at the EW scale $\mathcal{O}_i$ and the WCs of the $d\leq 7$ $\nu$SMEFT operators $Q_i$ in Tables~\ref{tab:vSMEFT-operators} and~\ref{tab:vSMEFT-operators-2}. Firstly, in Table~\ref{tab:matching_3}, we give the matching conditions for the vector, scalar and tensor four-fermion interactions in Eqs.~\eqref{eq:L_fourfermion} and~\eqref{eq:L_fourfermion_S}. In Table~\ref{tab:matching_2}, we instead give the matching conditions for the effective $W^\pm$ and $Z$ interactions in Eqs.~\eqref{eq:L_gauge} and~\eqref{eq:L_gauge_S}. For each operator, we give the total number of parameters (in addition to those which are CP-even) for $n_\nu$ active neutrinos $\nu$ and $n_s$ gauge-singlet fields $N$ and $S$.

\begin{table}[t]
\centering
\renewcommand{\arraystretch}{1.2}
\setlength\tabcolsep{1.2pt}
\begin{tabular}{c|c|c|c|c}
\hline
\multicolumn{5}{c}{Vector Four-Fermion Operators}\\ \hline
\multirow{2}{*}{Operator} & \multirow{2}{*}{Coefficient} & \multicolumn{2}{c|}{Parameters} & \multirow{2}{*}{Matching ($X = R,L$)} \\ \cline{3-4}
& & Total & CP-even & \\ \hline

$\mathcal{O}^{V,LX}_{\nu e}$ & $C_{\underset{prst}{\nu e}}^{V,LX}$ & $n_\nu^2 n_e^2$ & $\frac{1}{2} n_\nu n_e (n_\nu n_e + 1)$ & $C_{\underset{prst}{le}}$ $,$ $C_{\underset{prst}{ll}} + C_{\underset{stpr}{ll}}$  \\

$\mathcal{O}^{V,RX}_{\nu N e} + \text{h.c.}$ & $C_{\underset{prst}{\nu N e}}^{V,RX}$ & $2 n_\nu n_s n_e^2$ & $n_\nu n_s n_e^2$ & $-\frac{v}{\sqrt{2}}C_{\underset{strp}{eNlH}}$ $,$ $-\frac{v}{\sqrt{2}}C_{\underset{strp}{lNlH}}$  \\

$\mathcal{O}^{V,RX}_{Ne}$ & $C_{\underset{prst}{Ne}}^{V,RX}$ & $n_s^2 n_e^2$ & $\frac{1}{2} n_s n_e(n_s n_e + 1)$ & $C_{\underset{stpr}{eN}}$ $,$ $C_{\underset{stpr}{lN}}$  \\ \hline

$\mathcal{O}^{V,LX}_{\nu Se} + \text{h.c.}$ & $C_{\underset{prst}{\nu Se}}^{V,LX}$ & $2 n_\nu n_s n_e^2$ & $n_\nu n_s n_e^2$ & $\frac{v}{\sqrt{2}}C_{\underset{tsrp}{eSlH}}^*$ $,$ $\frac{v}{\sqrt{2}}C_{\underset{tsrp}{lSlH}}^*$ \\

$\mathcal{O}^{V,LX}_{Se}$ & $C_{\underset{prst}{Se}}^{V,LX}$ & $n_s^2 n_e^2$ & $\frac{1}{2}n_s n_e \big(n_s n_e + 1\big)$ & $C_{\underset{stpr}{eS}}$ $,$ $C_{\underset{stpr}{lS}}$ \\
\hline
\end{tabular} \\
\vspace{1em}
\begin{tabular}{c|c|c|c|c}
\hline
\multicolumn{5}{c}{Scalar Four-Fermion Operators (+ h.c.)}\\ \hline
\multirow{2}{*}{Operator} & \multirow{2}{*}{Coefficient} & \multicolumn{2}{c|}{Parameters} & \multirow{2}{*}{Matching ($X = R,L$)} \\ \cline{3-4}
& & Total & CP-even & \\ \hline

$\mathcal{O}^{S,LX}_{\nu e}$ & $\frac{1}{2}C_{\underset{prst}{\nu e}}^{S,LX}$ & $n_\nu (n_\nu + 1)n_e^2$ & $\frac{1}{2} n_\nu (n_\nu + 1) n_e^2$ & ~$0$ $,$ $-\frac{v}{\sqrt{2}}\big(C_{\underset{st\{pr\}}{llleH}} + \frac{1}{2}C_{\underset{s\{ptr\}}{llleH}}\big)$ \\

$\mathcal{O}^{S,RX}_{\nu N e}$ & $C_{\underset{prst}{\nu N e}}^{S,RX}$ & $2 n_\nu n_s n_e^2$ & $n_\nu n_s n_e^2$ & $C_{\underset{prst}{lNle}} + \frac{1}{2}C_{\underset{srpt}{lNle}}$ $,$ $0$ \\

$\mathcal{O}^{S,RX}_{Ne}$ & $\frac{1}{2}C_{\underset{prst}{Ne}}^{S,RX}$ & $n_s (n_s + 1) n_e^2$ & $\frac{1}{2} n_s (n_s + 1)n_e^2$ & $-\frac{v}{2\sqrt{2}}C_{\underset{s\{pr\}t}{lNeH}}$ $,$ $\frac{v}{\sqrt{2}} C_{\underset{stpr}{elNH}}$ \\ \hline

$\mathcal{O}^{S,RX}_{SNe}$ & $C_{\underset{prst}{SNe}}^{S,RX}$ & $2n_\nu n_s n_e^2$ & $n_\nu n_s n_e^2$ & $-\frac{v}{2\sqrt{2}}C_{\underset{sprt}{lSNeH}}$ $,$ $\frac{v}{\sqrt{2}} C_{\underset{stpr}{elSNH}}$ \\
\hline
\end{tabular}\\
\vspace{1em}
\begin{tabular}{c|c|c|c|c}
\hline
\multicolumn{5}{c}{Tensor Four-Fermion Operators (+ h.c.)}\\ \hline
\multirow{2}{*}{Operator} &\multirow{2}{*}{Coefficient}  & \multicolumn{2}{c|}{Parameters} & \multirow{2}{*}{Matching} \\ \cline{3-4}
& & Total & CP-even & \\ \hline

$\mathcal{O}^{T,LL}_{\nu e}$ & $\frac{1}{2}C_{\underset{prst}{\nu e}}^{T,LL}$ & $n_\nu (n_\nu - 1) n_e^2$ & $\frac{1}{2} n_\nu (n_\nu - 1) n_e^2$ & $\frac{v}{8\sqrt{2}}C_{\underset{s[ptr]}{llleH}}$ \\

$\mathcal{O}^{T,RR}_{\nu N e}$ & $C_{\underset{prst}{\nu N e}}^{T,RR}$ & $2 n_\nu n_s n_e^2$ & $n_\nu n_s n_e^2$ & $\frac{1}{8}C_{\underset{srpt}{lNle}}$\\

$\mathcal{O}^{T,RR}_{Ne}$ & $\frac{1}{2}C_{\underset{prst}{Ne}}^{T,RR}$ & $n_s (n_s - 1) n_e^2$ & $\frac{1}{2} n_s (n_s - 1) n_e^2$ & $\frac{v}{8\sqrt{2}}C_{\underset{s[pr]t}{lNeH}}$ \\ \hline

$\mathcal{O}^{T,RR}_{SNe}$ & $C_{\underset{prst}{SNe}}^{T,RR}$ & $2n_\nu n_s n_e^2$ & $n_\nu n_s n_e^2$ & $-\frac{v}{8\sqrt{2}}C_{\underset{sprt}{lSNeH}}$ \\
\hline
\end{tabular}
\caption{Matching between the vector (top), scalar (centre) and tensor (bottom) four-fermion interactions and the $d\leq 7$ $\nu$SMEFT operators.}
\label{tab:matching_3}
\end{table}
\begin{table}[t]
\centering
\renewcommand{\arraystretch}{1.2}
\setlength\tabcolsep{1.2pt}
\begin{tabular}{c|c|c|c|c}
\hline
\multicolumn{5}{c}{ Effective $W^\pm$ Interactions (+ h.c.)}\\ \hline
\multirow{2}{*}{Operator} & \multirow{2}{*}{Coefficient} & \multicolumn{2}{c|}{Parameters} & \multirow{2}{*}{Matching} \\ \cline{3-4}
& & Total & CP-even & \\ \hline

$\mathcal{O}_{\nu e W}^L$, $\mathcal{O}_{\nu e W}^R$ & $[W_{\nu}^L]_{pr}$, $[W_{\nu}^R]_{pr}$ & $2 n_\nu n_e$ & $n_\nu n_e$ &   $\delta_{pr} + v^2 C_{\underset{pr}{Hl}}^{(3)}$ $,$ $-\frac{v^3}{2\sqrt{2}}C_{\underset{pr}{leHD}}$ \\

$\mathcal{O}_{NeW}^R$, $\mathcal{O}_{NeW}^L$ & $[W_{N}^R]_{pr}$, $[W_{N}^L]_{pr}$ & $2 n_s n_e$ & $n_s n_e$ & $\frac{v^2}{2}C_{\underset{pr}{HNe}}$ $,$ $-\frac{v^3}{2\sqrt{2}}C_{\underset{pr}{Nl1}}$ \\ 
\hline

$\mathcal{O}_{SeW}^L$ & $[W_{S}^L]_{pr}$ & $2 n_s n_e$ & $n_s n_e$ & $-\frac{v^3}{2\sqrt{2}}C_{\underset{pr}{Sl1}}$ \\ \hline
\end{tabular}\\
\vspace{1em}
\begin{tabular}{c|c|c|c|c}
\hline
\multicolumn{5}{c}{ Effective $Z$ Interactions}\\ \hline
\multirow{2}{*}{Operator} & \multirow{2}{*}{Coefficient} & \multicolumn{2}{c|}{Parameters} & \multirow{2}{*}{Matching} \\ \cline{3-4}
& & Total & CP-even & \\ \hline
$\mathcal{O}_{\nu Z}^L$ & $[Z_{\nu}^L]_{pr}$ & $n_\nu^2$ & $\frac{1}{2}n_\nu(n_\nu + 1)$ & $g_{L}^\nu \delta_{pr} - \frac{v^2}{2}\big(C_{\underset{pr}{H l}}^{(1)} - C_{\underset{pr}{H l}}^{(3)}\big)$\\

$\mathcal{O}_{\nu N Z}^R + \text{h.c.}$ & $[Z_{\nu N}^R]_{pr}$ & $2 n_\nu n_s$ & $n_\nu n_s$ & $\frac{v^3}{4\sqrt{2}}\big(C_{
\underset{rp}{Nl1}} + 2C_{
\underset{rp}{Nl2}}\big)$ \\

$\mathcal{O}_{NZ}^R$ & $[Z_{N}^R]_{pr}$ & $n_s^2$ & $\frac{1}{2}n_s (n_s + 1)$ & $-\frac{v^2}{2}C_{
\underset{pr}{HN}}$ \\ \hline

$\mathcal{O}_{\nu SZ}^L$ & $[Z_{\nu S}^L]_{pr}$ & $2n_\nu n_s$ & $n_\nu n_s$ & $-\frac{v^3}{4\sqrt{2}}\big(C_{
\underset{rp}{Sl1}}^* + 2C_{
\underset{rp}{Sl2}}^*\big)$ \\

$\mathcal{O}_{SZ}^L$ & $[Z_{S}^L]_{pr}$ & $n_s^2$ & $\frac{1}{2}n_s \big(n_s + 1\big)$ & $-\frac{v^2}{2}C_{
\underset{pr}{HS}}$ \\
\hline
\end{tabular}
\caption{Matching between the effective $W^\pm$ (top) and $Z$ (bottom) interactions and the $d\leq 7$ $\nu$SMEFT operators.}
\label{tab:matching_2}
\end{table}

The WCs in the Majorana and Dirac HNL scenarios can be rotated to the mass basis according to Eqs.~\eqref{eq:mixing_Majorana} and~\eqref{eq:mixing_Dirac}, respectively, to obtain Eqs.~\eqref{eq:L_fourfermion_mass} and \eqref{eq:L_gauge_mass}. As the light neutrinos are further diagonalised as $\nu_\alpha' = U_{\alpha i}\nu_i'$ with $\alpha = e,\mu,\tau$ and $i = 1,2,3$, we can consider WCs for operators containing $\nu_\alpha'$ or $\nu_i'$. We use the former in this work.

In Eqs.~\eqref{eq:L_fourfermion_mass} and~\eqref{eq:L_gauge_mass}, the WCs satisfy
\begin{align}
C_{\underset{\alpha\beta\rho\sigma}{\nu e}}^{V,XY} &= C_{\underset{\beta\alpha\sigma\rho}{\nu e}}^{V,XY*} \,, \quad 
C_{\underset{\alpha j\rho\sigma}{\nu Ne}}^{V,XY} = C_{\underset{j\alpha\sigma\rho}{N\nu e}}^{V,XY*} \,, \quad
C_{\underset{ij\rho\sigma}{Ne}}^{V,XY} = C_{\underset{ji\sigma\rho}{Ne}}^{V,XY*} \,, \nonumber \\
C_{\underset{\alpha\beta\rho\sigma}{\nu e}}^{S,XY} & = C_{\underset{\beta\alpha\sigma\rho}{\nu e}}^{S,YX*} \,, \quad C_{\underset{\alpha j\rho\sigma}{\nu Ne}}^{S,XY} = C_{\underset{j\alpha\sigma\rho}{N\nu e}}^{S,YX*} \,, \quad C_{\underset{ij\rho\sigma}{Ne}}^{S,XY} = C_{\underset{ji\sigma\rho}{Ne}}^{S,YX*} \,, \nonumber \\
C_{\underset{\alpha\beta\rho\sigma}{\nu e}}^{T,XX} &= C_{\underset{\beta\alpha\sigma\rho}{\nu e}}^{T,YY*} \,,\quad C_{\underset{\alpha j\rho\sigma}{\nu Ne}}^{T,XY} = C_{\underset{j\alpha\sigma\rho}{N\nu e}}^{T,YX*} \,, \quad C_{\underset{ij\rho\sigma}{Ne}}^{T,XX} = C_{\underset{ji\sigma\rho}{Ne}}^{T,YY*} \,, \nonumber \\
[Z_\nu^X]_{\alpha\beta} &= [Z_\nu^X]_{\beta\alpha}^* \,, \quad [Z_{\nu N}^X]_{\alpha j} = [Z_{N\nu}^X]_{j\alpha}^* \,, \quad [Z_N^X]_{ij} = [Z_{N}^X]_{ji}^* \,,
\end{align}
in both the Majorana and Dirac cases. In the Majorana scenario, the following relations also apply,
\begin{align}
C_{\underset{\alpha\beta\rho\sigma}{\nu e}}^{V,XY} &= - C_{\underset{\beta\alpha\rho\sigma}{\nu e}}^{V,YY} \,, \quad 
C_{\underset{\alpha j\rho\sigma}{\nu Ne}}^{V,XY} = - C_{\underset{j\alpha\rho\sigma}{N\nu e}}^{V,YY} \,, \quad
C_{\underset{ij\rho\sigma}{Ne}}^{V,XY} = - C_{\underset{ji\rho\sigma}{Ne}}^{V,YY} \,, \nonumber \\
C_{\underset{\alpha\beta\rho\sigma}{\nu e}}^{S,XY} & = C_{\underset{\beta\alpha\rho\sigma}{\nu e}}^{S,XY} \,, \quad C_{\underset{\alpha j\rho\sigma}{\nu Ne}}^{S,XY} = C_{\underset{j\alpha\rho\sigma}{N\nu e}}^{S,XY} \,, \quad C_{\underset{ij\rho\sigma}{Ne}}^{S,XY} = C_{\underset{ji\rho\sigma}{Ne}}^{S,XY} \,, \nonumber \\
C_{\underset{\alpha\beta\rho\sigma}{\nu e}}^{T,XX} &= - C_{\underset{\beta\alpha\rho\sigma}{\nu e}}^{T,XX} \,,\quad C_{\underset{\alpha j\rho\sigma}{\nu Ne}}^{T,XY} = - C_{\underset{j\alpha\rho\sigma}{N\nu e}}^{T,XY} \,, \quad C_{\underset{ij\rho\sigma}{Ne}}^{T,XX} = - C_{\underset{ji\rho\sigma}{Ne}}^{T,XX} \,, \nonumber \\
[Z_\nu^X]_{\alpha\beta} &= - [Z_\nu^Y]_{\beta\alpha} \,, \quad [Z_{\nu N}^X]_{\alpha j} = - [Z_{N\nu}^Y]_{j\alpha} \,, \quad [Z_N^X]_{ij} = - [Z_{N}^Y]_{ji} \,.
\end{align}

In this analysis we do not include the effects of RG running. The anomalous dimension matrix from gauge and Yukawa couplings for the $d = 6$ $\nu$SMEFT operators in Table~\ref{tab:vSMEFT-operators} has been calculated in~\cite{Datta:2020ocb,Datta:2021akg,Ardu:2024tzb} and implemented numerically in~\cite{Aebischer:2024csk}. For us, the running due to the Yukawa couplings vanishes as we consider the $Y_\nu = 0$ limit in this work. Meanwhile, the running and mixing proportional to the $U(1)_Y$ gauge coupling $g'$ is negligible for $\Lambda \sim 1$~TeV. The same argument can be made for the $d = 7$ operators. The bounds on the WCs in this work are nevertheless valid at the EW scale.

\begin{table}[t!]
\centering
\renewcommand{\arraystretch}{1.2}
\setlength\tabcolsep{4pt}
\begin{tabular}{c|ccccc}
\hline
Scalar  &  $\mathcal{S}$ &  $\mathcal{S}_1$ &  $\varphi$ &  $\Xi$ &  $\Xi_1$ \\\hline
Irrep.  &$(1,1)_0$ & $(1,1)_1$ & $(1,2)_{\frac{1}{2}}$ & $(1,3)_0$ & $(1,3)_1$ \\ \hline
\end{tabular}\\
\vspace{1em}
\begin{tabular}{c|cccccc}
\hline
Fermion  &  $\mathcal{N}$ &  $E$ &  $\Delta_1$ &  $\Delta_3$ &  $\Sigma$ &  $\Sigma_1$ \\\hline
Irrep.  &$(1,1)_0$ & $(1,1)_{-1}$ & $(1,2)_{-\frac{1}{2}}$ & $(1,2)_{-\frac{3}{2}}$ & $(1,3)_0$ & $(1,3)_{-1}$ \\ \hline
\end{tabular}\\
\vspace{1em}
\begin{tabular}{c|cccccc}
\hline
Vector  &  $\mathcal{B}$ &  $\mathcal{B}_1$ &  $\mathcal{W}$ &  $\mathcal{W}_1$ &  $\mathcal{L}_1$ &  $\mathcal{L}_3$ \\\hline
Irrep.  &$(1,1)_0$ & $(1,1)_{1}$ & $(1,3)_{0}$ & $(1,3)_{1}$ & $(1,2)_{\frac{1}{2}}$ & $(1,2)_{-\frac{3}{2}}$ \\\hline
\end{tabular}
\caption{New scalar (top), fermion (middle) and vector (bottom) fields with tree-level matching to the $\nu$SMEFT operators in Tables~\ref{tab:vSMEFT-operators} and~\ref{tab:vSMEFT-operators-2}, following the naming convention of~\cite{deBlas:2017xtg}. The irreducible representations under the SM gauge group are indicated.}
\label{tab:UV_fields}
\end{table}

The $\nu$SMEFT operators of dimension $d \leq 7$ in Tables~\ref{tab:vSMEFT-operators} and~\ref{tab:vSMEFT-operators-2} can be generated by heavy new degrees of freedom at the scale $\Lambda \gg v$. Given the large number of possible representations of heavy new fields under the SM gauge group, which may contribute to the operators at tree-level, one-loop or a higher number of loops, it is practical to consider only a subset of UV complete models. We therefore examine only the general extensions of~\cite{deBlas:2017xtg}: the lowest irreducible representations of scalar boson, vector-like fermion and vector boson fields under the SM, shown in Table~\ref{tab:UV_fields}, which generate the operators at tree-level.

The tree-level matching of these fields to $d = 6$ and $d = 7$ SMEFT operators has been systematically performed in \cite{deBlas:2017xtg,Li:2023cwy}, while the matching to $d = 7$ in the $\nu$SMEFT has been studied in~\cite{Beltran:2023ymm}. In Table~\ref{tab:UV_completions}, we show the single- and two-particle UV completions of each $\nu$SMEFT operator in Table~\ref{tab:vSMEFT-operators}. These tree-level UV completions also apply to the operators in Table~\ref{tab:vSMEFT-operators-2}; as $N$ and $S$ are SM gauge singlets, any diagram inducing an operator containing $N$ will also generate an operator with $N\to S^c$.

\begin{table}[t!]
\footnotesize
\centering
\renewcommand{\arraystretch}{1.2}
\setlength\tabcolsep{4pt}
\begin{tabular}{c|c}
\hline
\multicolumn{2}{c}{$\psi^2 H^2$} \\ \hline

$Q_{5}$ & $\Xi_1$, $\mathcal{N}$, $\Sigma$ \\
$Q_{N}$ & $\mathcal{S}$, $\Delta_1$ \\
\hline
\end{tabular}
\hspace{1em}
\begin{tabular}{c|c|c|c}
\hline 

\multicolumn{2}{c|}{$\psi^2 H^3$} & \multicolumn{2}{c}{$\psi^4$} \\\hline

\multirow{2}{*}{$Q_{lNH}$} & $\varphi$, $E$, $\Delta_1$, $\Sigma$, $\Sigma_1$, & $Q_{ll}$ & $\mathcal{S}_1$, $\Xi_1$, $\mathcal{B}$, $\mathcal{W}$ \\ 

& $(\mathcal{S}, \mathcal{N})$  &  $Q_{le}$ & $\varphi$, $\mathcal{B}$, $\mathcal{L}_1$, $\mathcal{L}_3$  \\ \cline{1-2}

\multicolumn{2}{c|}{$\psi^2 H^2 D$} &  $Q_{lNle}$ & $\mathcal{S}_1$, $\varphi$  \\ \cline{1-2}

$Q^{(1), (3)}_{Hl}$ & $\mathcal{N}$, $E$, $\Sigma$, $\Sigma_1$ & $Q_{lN}$ & $\varphi$, $\mathcal{B}$, $\mathcal{L}_1$ \\

$Q_{HN}$ & $\Delta_1$, $\mathcal{B}$ & $Q_{eN}$ & $\mathcal{S}_1$, $\mathcal{B}$, $\mathcal{B}_1$ \\

$Q_{HNe}$ & $\Delta_1$, $\mathcal{B}_1$ & & \\
\hline
\end{tabular} \\
\vspace{1em}
\begin{tabular}{c|c|c|c}
\hline 
\multicolumn{2}{c|}{$\psi^2 H^4$} & \multicolumn{2}{c}{$\psi^4 H$} \\\hline

$Q_{lH}$ & $\Xi_1$, $\mathcal{N}$, $\Sigma$ & \multirow{2}{*}{$Q_{llleH}$} & $\mathcal{N}$, $\Sigma$, \\ \cline{2-2}

\multirow{3}{*}{$Q_{NH}$} & $\mathcal{S}$, &  &  $(\mathcal{S}_1, \varphi)$, $(\mathcal{S}_1, \Delta_3)$, $(\varphi, \Xi_1)$, $(\Xi_1, \Delta_3)$  \\ \cline{4-4}

&  $(\varphi, \Delta_1)$, $(\Xi, \Delta_1)$, $(\Xi, \Sigma)$, $(\Xi_1, \Delta_1)$, & \multirow{4}{*}{$Q_{lNlH}$} &  $(\mathcal{S}_1, \varphi)$, $(\mathcal{S}_1, E)$, $(\mathcal{S}_1, \Delta_1)$, $(\varphi, \Xi_1)$,  \\

& $(\Xi_1, \Sigma_1)$, $(\mathcal{N}, \Delta_1)$, $(\Delta_1, \Sigma)$, $(\Delta_1, \Sigma_1)$ &  &  $(\varphi, \mathcal{N})$, $(\varphi, \Sigma)$, $(\Xi_1, \Delta_1)$, $(\Xi_1, \Sigma_1)$, \\ \cline{1-2}

\multicolumn{2}{c|}{$\psi^2 H^3 D$} & &  $(\mathcal{N}, \mathcal{B})$, $(\mathcal{N}, \mathcal{L}_1)$, $(E, \mathcal{L}_1)$, $(\Delta_1, \mathcal{B})$, \\ \cline{1-2}

\multirow{5}{*}{$Q_{Nl1(2)}$} &  $(\mathcal{S}, \mathcal{N})$, $(\mathcal{S}, \Delta_1)$, $(\mathcal{S}, \mathcal{L}_1)$, $(\Xi, \Delta_1)$, &  &  $(\Delta_1, \mathcal{W})$, $(\Sigma, \mathcal{W})$, $(\Sigma, \mathcal{L}_1)$, $(\Sigma_1, \mathcal{L}_1)$ \\ \cline{4-4}

&  $(\Xi, \Sigma)$, $(\Xi, \mathcal{L}_1)$, $(\Xi_1, \Delta_1)$, $(\Xi_1, \Sigma_1)$, & \multirow{3}{*}{$Q_{eNlH}$} &  $(\mathcal{S}_1, \varphi)$, $(\mathcal{S}_1, \mathcal{N})$, $(\mathcal{S}_1, \Delta_3)$, $(\varphi, \Delta_1)$, \\ 

&  $(\Xi_1, \mathcal{L}_1)$, $(\mathcal{N}, \Delta_1)$, $(\mathcal{N}, \mathcal{B})$, $(\Delta_1, \Sigma)$, &  &  $(\mathcal{N}, \mathcal{B})$, $(\mathcal{N}, \mathcal{B}_1)$, $(\Delta_1, \mathcal{B})$, $(\Delta_1, \mathcal{B}_1)$, \\ 

&  $(\Delta_1, \Sigma_1)$, $(\Delta_1, \mathcal{B})$, $(\Delta_1, \mathcal{B}_1)$, $(\Delta_1, \mathcal{W})$, &  &  $(\Delta_1, \mathcal{L}_1)$, $(\Delta_1, \mathcal{L}_3)$, $(\Delta_3, \mathcal{L}_1)$, $(\Delta_3, \mathcal{L}_3)$ \\ \cline{4-4}

&  $(\Delta_1, \mathcal{W}_1)$, $(\Sigma, \mathcal{W})$, $(\Sigma_1, \mathcal{W}_1)$ & \multirow{2}{*}{$Q_{lNeH}$} &  $(\mathcal{S}, \varphi)$, $(\mathcal{S}, E)$,  $(\mathcal{S}, \Delta_1)$, $(\mathcal{S}_1, \varphi)$, \\ \cline{2-2}

\multirow{2}{*}{$Q_{leHD}$} & $\mathcal{N}$, $\Sigma$, &  &  $(\mathcal{S}_1, E)$, $(\mathcal{S}_1, \Delta_1)$, $(\varphi, \Delta_1)$ \\ \cline{4-4}
  
&  $(\Xi_1, \Delta_1)$, $(\Delta_1, \mathcal{B}_1)$, $(\mathcal{B}_1, \mathcal{L}_3)$ & \multirow{2}{*}{$Q_{elNH}$} &  $(\mathcal{S}, \varphi)$, $(\mathcal{S}, \Delta_1)$, $(\varphi, \Delta_1)$, $(\mathcal{S}_1, E)$, \\

& &  &  $(E, \mathcal{B}_1)$, $(\Delta_1, \mathcal{B}_1)$, $(\Delta_1, \mathcal{L}_1)$\\
\hline
\end{tabular}
\caption{One- and two-particle tree-level UV completions of the $\nu$SMEFT operators which can be probed by FCC-ee. Two-particle UV completions are gathered in parentheses and are only shown if the fields cannot induce the operator individually.}
\label{tab:UV_completions}
\end{table}

\begin{figure}[t!]
\centering
\includegraphics[width=0.3\textwidth]{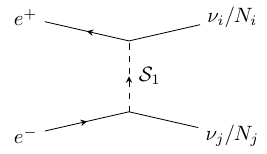}
\hspace{1.5em}
\includegraphics[width=0.3\textwidth]{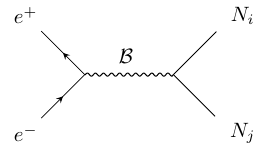}
\caption{Two UV complete scenarios generating at tree-level the processes $e^+e^-\to \nu\nu$, $e^+e^-\to \nu N$ and $e^+e^-\to NN$; a singly-charged scalar $\mathcal{S}_1$ (left) and gauge-singlet vector boson $\mathcal{B}$ (right).}
\label{fig:UV_completions}
\end{figure}

We now consider two particularly simple UV scenarios. If one adds the singly-charged scalar $\mathcal{S}_1$ to the SM field content, it is possible to write the renormalisable terms,
\begin{align}
\mathcal{L} \supset -  [y_{\mathcal{S}_1}^{ll}]_{pr}\epsilon_{ij}\bar{L}_{p}^{ic} L_r^j \mathcal{S}_1 - [y_{\mathcal{S}_1}^{Ne}]_{pr} \bar{N}_{Rp}^c e_{Rr} \mathcal{S}_1 + \text{h.c.}\,,
\end{align}
with $[y_{\mathcal{S}_1}^{ll}]_{pr} = -[y_{\mathcal{S}_1}^{ll}]_{rp}$. Integrating out $\mathcal{S}_1$ yields the matching relations
\begin{align}
C_{\underset{prst}{ll}} = \frac{[y_{\mathcal{S}_1}^{ll}]_{ps}^*[y_{\mathcal{S}_1}^{ll}]_{rt}}{M_{\mathcal{S}_1}^2}\,,\quad C_{\underset{prst}{lNle}} = - \frac{2[y_{\mathcal{S}_1}^{ll}]_{ps}^*[y_{\mathcal{S}_1}^{Ne}]_{rt}}{M_{\mathcal{S}_1}^2} \,, \quad C_{\underset{prst}{eN}} = \frac{[y_{\mathcal{S}_1}^{Ne}]_{sp}^* [y_{\mathcal{S}_1}^{Ne}]_{tr}}{2M_{\mathcal{S}_1}^2}\,.
\end{align}
Thus, the heavy singly-charged scalar $\mathcal{S}_1$ can modify the SM process $e^+e^-\to \nu\nu(\gamma)$ via $C_{ll}$ and result in the $e^+e^-\to \nu N(\gamma)$ and $e^+e^-\to N N(\gamma)$ processes via $C_{lNle}$ and $C_{eN}$, respectively, shown in Fig.~\ref{fig:UV_completions} (left). We note that, at one-loop, integrating out $\mathcal{S}_1$ also generates the $d = 5$ dipole operator $Q_{NNB}$. Thus, stringent $Z$ pole bounds on the dipole operator~\cite{Chun:2024mus} can also constrain this model.

Alternatively, a TeV-scale gauge-singlet vector boson $\mathcal{B}$ may be present. Such a field naturally arises as the gauge boson of an additional $U(1)_X$ gauge group, such as the $Z'$ of a gauged $U(1)_{B-L}$. The generic scenario allows to write the terms
\begin{align}
\mathcal{L} \supset  - [g_\mathcal{B}^f]_{pr}(\bar{f}_p\gamma_\mu f_r)\mathcal{B}^\mu - g_\mathcal{B}^H (H^\dagger i\overleftrightarrow{D}_\mu H)\mathcal{B}^\mu\,,
\end{align}
with $f = Q,u,d,L,e,N$ and $[g_\mathcal{B}^f]_{pr} = [g_\mathcal{B}^f]_{rp}^*$. If $\mathcal{B}$ is associated with an extra gauge symmetry, the couplings $[g_\mathcal{B}^f]$ and $[g_\mathcal{B}^H]$ correspond to the charges of the fields under $U(1)_X$. For $U(1)_{B-L}$, the couplings are $[g_\mathcal{B}^{f}] = g_{B-L} Y_{B-L}^{f}$, with $g_{B-L}$ the new gauge coupling and $Y_{B-L}^f = -1~(1/3)$ for leptons (quarks). One can consider $U(1)_{B-L}$ to be an unbroken symmetry, requiring the neutrinos to be Dirac fermions and for $Z'$ to obtain its mass via the St\"{u}ckelberg mechanism~\cite{Stueckelberg:1938hvi,Heeck:2014zfa}. Alternatively, $U(1)_{B-L}$ can be broken, with the RH neutrino mass $M_R$ proportional to the symmetry breaking scale~\cite{Deppisch:2019kvs, Liu:2022kid, Liu:2024fey} and the neutrinos being Majorana fermions. Regardless of this distinction, integrating out $Z'$ induces a large number of effective operators, leading to stringent constraints from collider experiments~\cite{Carena:2004xs}. A scenario where mono-$\gamma$ or DV searches at FCC-ee could provide competitive constraints would be if $\mathcal{B}$ couples only to $N$ and $e$. It is straightforward to find the following matching conditions after integrating out $\mathcal{B}$,
\begin{gather}
C_{\underset{prst}{ee}} = -\frac{[g_\mathcal{B}^e]_{pr}^*[g_\mathcal{B}^e]_{st}}{2M_{\mathcal{B}}^2}\,,\quad
C_{\underset{prst}{eN}} = -\frac{[g_\mathcal{B}^e]_{pr}^*[g_\mathcal{B}^N]_{st}}{M_{\mathcal{B}}^2} \,, \quad C_{\underset{prst}{NN}} = -\frac{[g_\mathcal{B}^N]_{pr}^*[g_\mathcal{B}^N]_{st}}{2M_{\mathcal{B}}^2}\,,
\end{gather}
with $Q_{ee} = (\bar{e} \gamma_\mu e)(\bar{e} \gamma^\mu e)$ and $Q_{NN} = (\bar{N} \gamma_\mu N)(\bar{N} \gamma^\mu N)$. The operator $Q_{eN}$ induces the $e^+e^-\to N N(\gamma)$ process, shown in Fig.~\ref{fig:UV_completions} (right).

%

%%%%%%%%%%%%%%%%%%%%%%%%%%%%%%%%%%%%%%%%
\section{HNL Production Cross Sections}
\label{app:2to2_xsec}
%%%%%%%%%%%%%%%%%%%%%%%%%%%%%%%%%%%%%%%%

In this appendix we provide analytical expressions for the cross sections of $e^+ e^- \to \nu\nu(\gamma)$, $e^+ e^- \to \nu N(\gamma)$ and $e^+ e^- \to NN(\gamma)$, which can be induced by the $\nu$SMEFT operators in Tables~\ref{tab:vSMEFT-operators} and~\ref{tab:vSMEFT-operators-2}. For complete generality, we derive the leading order (LO) cross sections for the scattering processes,
\begin{align}
\label{eq:2to2_processes}
\ell_\alpha^+\ell_\beta^- \to \sum_{i\leq j}\mathcal{N}_i \mathcal{N}_j (\gamma) \quad (\text{Majorana})\,,\quad \ell_\alpha^+\ell_\beta^- \to \sum_{i,j}\mathcal{N}_i \bar{\mathcal{N}}_j (\gamma)\quad (\text{Dirac})\,,
\end{align}
with the four-momenta $p_\alpha + p_\beta = p_i + p_j$. In the Majorana case, $\mathcal{N}_i = \nu_i$ for $i = 1,2,3$ and $\mathcal{N}_{i+3} = N_i$ for $i = 1,\ldots n_s$, where both $\nu_i$ and $N_i$ are Majorana fermions. In the Dirac case, $\mathcal{N}_i = \nu_i$ for $i = 1,2,3$ and $\mathcal{N}_{i+3} = N_{i}$ for $i = 1,\ldots n_s$, where $\nu_i$ are massless Weyl fermions and $N_{i}$ are Dirac fermions. To derive the following expressions, we work in the Feynman gauge and use the Feynman rule prescription for Majorana fermions in Ref.~\cite{Denner:1992vza}.

The differential cross section for the processes in Eq.~\eqref{eq:2to2_processes}, without a final-state photon and neglecting the masses of the initial-state charged leptons, can be written at LO as
\begin{align}
\label{eq:tot_Dirac_cs_theta}
\frac{d\sigma}{dc_\theta} &= \bigg(\frac{1}{2}\bigg)\frac{1}{128\pi}\sum_{i, j}^{3+n_s}\lambda^{\frac{1}{2}}(s, m_i^2, m_j^2) \nonumber \\
&\hspace{1em} \times\bigg[\bigg(1 - \frac{(m_i^2 - m_j^2)^2}{s^2} + \frac{\lambda(s,m_i^2,m_j^2)}{s^2}c_\theta^2\bigg)\bigg(\big|L_{\underset{ij\alpha\beta}{\mathcal{N}e}}^{V,RR}\big|^2 + \big|L_{\underset{ij\alpha\beta}{\mathcal{N}e}}^{V,RL}\big|^2\bigg) \nonumber \\
&\hspace{3em} + \bigg(1 - \frac{m_i^2 + m_j^2}{s}\bigg)\Big(\big|L_{\underset{ij\alpha\beta}{\mathcal{N}e}}^{S,RR}\big|^2 + \big|L_{\underset{ij\alpha\beta}{\mathcal{N}e}}^{S,RL}\big|^2\Big)  \nonumber \\
&\hspace{3em} + 16\bigg(\frac{m_i^2 + m_j^2}{s} - \frac{(m_i^2 - m_j^2)^2}{s^2} + \frac{\lambda(s,m_i^2,m_j^2)}{s^2}c_\theta^2\bigg)\big|L_{\underset{ij\alpha\beta}{\mathcal{N}e}}^{T,RR}\big|^2 \nonumber \\
&\hspace{3em} + \frac{4 m_i m_j}{s}\text{Re}\Big[2\,L_{\underset{ij\alpha\beta}{\mathcal{N}e}}^{V,LR}\,L_{\underset{ij\alpha\beta}{\mathcal{N}e}}^{V,RR *} - L_{\underset{ij\alpha\beta}{\mathcal{N}e}}^{S,LR}\,L_{\underset{ij\alpha\beta}{\mathcal{N}e}}^{S,RR *}\Big] 
\nonumber \\
&\hspace{3em} + \frac{2\lambda^{\frac{1}{2}}(s,m_i^2,m_j^2)}{s}c_\theta\Big(\big|L_{\underset{ij\alpha\beta}{\mathcal{N}e}}^{V,RR}\big|^2 -\big|L_{\underset{ij\alpha\beta}{\mathcal{N}e}}^{V,RL}\big|^2 - 4 \,\text{Re}\Big[L_{\underset{ij\alpha\beta}{\mathcal{N}e}}^{S,RR}\,L_{\underset{ij\alpha\beta}{\mathcal{N}e}}^{T,RR*}\Big]\Big)\bigg] \nonumber \\
&\hspace{3em}+ (L\leftrightarrow R) \,,
\end{align}
where $s = (p_\alpha + p_\beta)^2 = (p_i + p_j)^2$ is the centre-of-mass energy, $\lambda(x,y,z) = (x - y - z)^2 - 4 y z$ is the K\"{a}ll\'en function and $c_\theta = \cos \theta$, with $\theta$ the angle between the incoming charged lepton $\ell_\beta^-$ and the outgoing $\mathcal{N}_i$. In the Dirac case, the coefficients in Eq.~\eqref{eq:tot_Dirac_cs_theta} are given by
\begin{align}
\label{eq:Dirac-coefficients}
L^{V,XX}_{\underset{ij\alpha\beta}{\mathcal{N}e}}
\Big|_{\text{Dirac}} &\equiv C^{V,XX}_{\underset{ij\alpha\beta}{\mathcal{N}e}} + \chi_Z [Z_{\mathcal{N}}^X]_{ij} [Z_e^X]_{\alpha\beta} + \chi_W^\alpha [W_{\mathcal{N}}^{XX}]_{ij\alpha\beta} \,, \nonumber \\
L^{V,XY}_{\underset{ij\alpha\beta}{\mathcal{N}e}}\Big|_{\text{Dirac}} &\equiv C^{V,XY}_{\underset{ij\alpha\beta}{\mathcal{N}e}} + \chi_Z [Z_{\mathcal{N}}^X]_{ij} [Z_e^Y]_{\alpha\beta} + \chi_W^\alpha \frac{m_i m_j}{2M_W^2} [W_{\mathcal{N}}^{YY}]_{ij\alpha\beta} \,,\nonumber \\
L^{S,XX}_{\underset{ij\alpha\beta}{\mathcal{N}e}}\Big|_{\text{Dirac}} &\equiv C^{S,XX}_{\underset{ij\alpha\beta}{\mathcal{N}e}} +  \chi_W^\alpha \frac{m_i m_j}{2M_W^2} [W_{\mathcal{N}}^{XY}]_{ij\alpha\beta} \,,\nonumber \\
L^{S,XY}_{\underset{ij\alpha\beta}{\mathcal{N}e}}\Big|_{\text{Dirac}} &\equiv C^{S,XY}_{\underset{ij\alpha\beta}{\mathcal{N}e}} - 2\chi_W^\alpha [W_{\mathcal{N}}^{YX}]_{ij\alpha\beta} \,,\nonumber \\
L^{T,XX}_{\underset{ij\alpha\beta}{\mathcal{N}e}}\Big|_{\text{Dirac}} &\equiv C^{T,XX}_{\underset{ij\alpha\beta}{\mathcal{N}e}} +  \chi_W^\alpha \frac{m_i m_j}{8M_W^2} [W_{\mathcal{N}}^{XY}]_{ij\alpha\beta} \,, 
\end{align}
for $X \neq Y = R,L$. The final line in Eq.~\eqref{eq:tot_Dirac_cs_theta} indicates that, for each term, a term with $L\leftrightarrow R$ should be added. We define for convenience,
\begin{gather}
[W_{\mathcal{N}}^{XY}]_{ij\alpha\beta} \equiv [W_{{\mathcal{N}}}^X]_{i\beta} [W_{{\mathcal{N}}}^Y]_{j\alpha}^{*} \,,
\end{gather}
and use the charged lepton neutral-current couplings,
\begin{gather}
[Z_{e}^R]_{\alpha\beta} = g_R^e \delta_{\alpha\beta} \,,\quad [Z_{e}^L]_{\alpha\beta} = g_L^e \delta_{\alpha\beta} - \frac{v^2}{2}\big(C_{\underset{\alpha\beta}{Hl}}^{(1)} + C_{\underset{\alpha\beta}{Hl}}^{(3)}\big) \,,
\end{gather}
where $g_R^e = s_w^2$ and $g_L^e = -1/2 + s_w^2$, with $s_w = \sin\theta_w$. The propagator factors
\begin{gather}
\chi_W^\alpha = \frac{g^2}{2}\frac{1}{t_\alpha^2 - M_W^2 + i \Gamma_W M_W} \,, \quad \chi_Z = \frac{g^2}{c_w^2}\frac{1}{s - M_Z^2 + i \Gamma_Z M_Z}\,\,,
\label{eq:prop_factors}
\end{gather}
account for the $t$-channel exchange of $W^{\pm}$ and $s$-channel exchange of $Z$, respectively. The $t$-channel combination of four-momenta appearing in $\chi_W^{\alpha}$ is
$t_{\alpha} = (p_j - p_{\alpha})^2$.

In the Majorana case, the coefficients in Eq.~\eqref{eq:tot_Dirac_cs_theta} are instead given by
\begin{align}
\label{eq:Maj-coefficients}
L^{V,XX}_{\underset{ij\alpha\beta}{\mathcal{N}e}}\Big|_{\text{Maj}} &\equiv L^{V,XX}_{\underset{ij\alpha\beta}{\mathcal{N}e}}\Big|_{\text{Dirac}} - \chi_W^\beta \frac{m_i m_j}{2M_W^2} [W_{\mathcal{N}}^{XX}]_{ji\alpha\beta} \,, \nonumber \\
L^{V,XY}_{\underset{ij\alpha\beta}{\mathcal{N}e}}\Big|_{\text{Maj}} &\equiv L^{V,XY}_{\underset{ij\alpha\beta}{\mathcal{N}e}}\Big|_{\text{Dirac}} - \chi_W^\beta [W_{\mathcal{N}}^{YY}]_{ji\alpha\beta} \,,\nonumber \\
L^{S,XX}_{\underset{ij\alpha\beta}{\mathcal{N}e}}\Big|_{\text{Maj}} &\equiv L^{S,XX}_{\underset{ij\alpha\beta}{\mathcal{N}e}}\Big|_{\text{Dirac}} + \chi_W^\beta \frac{m_i m_j}{2M_W^2} [W_{\mathcal{N}}^{XY}]_{ji\alpha\beta}
\,,\nonumber \\
L^{S,XY}_{\underset{ij\alpha\beta}{\mathcal{N}e}}\Big|_{\text{Maj}} &\equiv L^{S,XY}_{\underset{ij\alpha\beta}{\mathcal{N}e}}\Big|_{\text{Dirac}} - 2 \chi_W^\beta  [W_{\mathcal{N}}^{YX}]_{ji\alpha\beta} \,,\nonumber \\
L^{T,XX}_{\underset{ij\alpha\beta}{\mathcal{N}e}}\Big|_{\text{Maj}} &\equiv L^{T,XX}_{\underset{ij\alpha\beta}{\mathcal{N}e}}\Big|_{\text{Dirac}} - \chi_W^\beta \frac{m_i m_j}{8M_W^2} [W_{\mathcal{N}}^{XY}]_{ji\alpha\beta}\,,
\end{align}
for $X \neq Y = R,L$. The additional terms with respect to Eq.~\eqref{eq:Dirac-coefficients} involving the effective $W^\pm$ couplings arise from diagrams where $\mathcal{N}_i$ and $\mathcal{N}_j$ are interchanged in the final state. The Majorana cross section should also be multiplied by the factor of $1/2$ in parenthesis, which takes into account that the sum over $i,j$ in Eq.~\eqref{eq:tot_Dirac_cs_theta} double counts the process $\ell_\alpha^+ \ell_\beta^- \to \mathcal{N}_i \mathcal{N}_j$ for $i \neq j$. Furthermore, the factor of $1/2$ provides the necessary symmetry factor for identical final state particles, i.e. $\ell_\alpha^+ \ell_\beta^- \to \mathcal{N}_i \mathcal{N}_j$ for $i = j$.

To compute the total cross sections as a function of the centre-of-mass energy, the differential cross section $d\sigma/dc_\theta$ in Eq.~\eqref{eq:tot_Dirac_cs_theta} is integrated over the $c_\theta$ range $[-1,1]$\footnote{We take the limit $\Gamma_W \ll M_W$ so that the terms originating from the $t$-channel $W^\pm$ exchange can be performed analytically.}. In the Dirac case, the LO cross section for $\ell_{\alpha}^+\ell_\beta^- \to \sum_{i,j}\mathcal{N}_i\bar{\mathcal{N}}_j$, including all interference terms, is
\begin{align}
\label{eq:tot_Dirac_cs}
\sigma(s)\big|_{\text{Dirac}} &= \frac{1}{192\pi}\sum_{i, j}\lambda^{\frac{1}{2}}(s,m_i^2,m_j^2) \nonumber \\
&\hspace{1.5em}\times\bigg[\frac{192 G_F^2 M_W^4}{s^2} \bigg(F_W^{ij} \big|[W_{\mathcal{N}}^{RR}]_{ij\alpha\beta}\big|^2 + G_W^{ij} \big|[W_{\mathcal{N}}^{RL}]_{ij\alpha\beta}\big|^2\bigg) \nonumber \\
&\hspace{3em} + 4\bigg(1 - \frac{m_i^2 + m_j^2}{2s} - \frac{(m_i^2 - m_j^2)^2}{2s^2}\bigg)\Big(\big|L_{\underset{ij\alpha\beta}{\mathcal{N}e}}^{V,RR}\big|^2 + \big|L_{\underset{ij\alpha\beta}{\mathcal{N}e}}^{V,RL}\big|^2\Big) \nonumber \\
&\hspace{3em} + 3\bigg(1 - \frac{m_i^2 + m_j^2}{s}\bigg)\Big(\big|L_{\underset{ij\alpha\beta}{\mathcal{N}e}}^{S,RR}\big|^2 + \big|L_{\underset{ij\alpha\beta}{\mathcal{N}e}}^{S,RL}\big|^2\Big) \nonumber \\
&\hspace{3em} + 16\bigg(1 + \frac{m_i^2 + m_j^2}{s} - \frac{2(m_i^2 - m_j^2)^2}{s^2}\bigg)\big|L_{\underset{ij\alpha\beta}{\mathcal{N}e}}^{T,RR}\big|^2 \nonumber \\
&\hspace{3em} + \frac{12m_i m_j}{s}\text{Re}\Big[2 \, L_{\underset{ij\alpha\beta}{\mathcal{N}e}}^{V,LR} \, L_{\underset{ij\alpha\beta}{\mathcal{N}e}}^{V,RR *} - L_{\underset{ij\alpha\beta}{\mathcal{N}e}}^{S,LR} \, L_{\underset{ij\alpha\beta}{\mathcal{N}e}}^{S,RR *}\Big] \nonumber \\
& \hspace{3em} + \frac{48 G_F M_W^2}{\sqrt{2}s}\text{Re}\bigg[\bigg(F_{VW}^{ij}L_{\underset{ij\alpha\beta}{\mathcal{N}e}}^{V,RR} + G_{VW}^{ij}L_{\underset{ij\alpha\beta}{\mathcal{N}e}}^{V,LR}\bigg)[W_{\mathcal{N}}^{RR}]_{ij\alpha\beta}^* \nonumber \\
&\hspace{10.5em} + \bigg(F_{SW}^{ij} L_{\underset{ij\alpha\beta}{\mathcal{N}e}}^{S,RR} + G_{SW}^{ij} L_{\underset{ij\alpha\beta}{\mathcal{N}e}}^{S,LR} + F_{TW}^{ij} L_{\underset{ij\alpha\beta}{\mathcal{N}e}}^{T,RR}\Bigg) [W_{\mathcal{N}}^{RL}]_{ij\alpha\beta}^*\bigg]\bigg] \nonumber \\
& \hspace{3em} + (L \leftrightarrow R)\,,
\end{align}
where $G_F = 1/(\sqrt{2}v^2)$ is the Fermi constant and the final line again indicates that additional terms with $L\leftrightarrow R$ are required. In the Majorana case the cross section given by,
\begin{align}
\label{eq:tot_Maj_cs}
\sigma(s)\big|_{\text{Maj}} &= \frac{1}{192\pi}\sum_{i \leq j} \bigg(1 - \frac{\delta_{ij}}{2}\bigg)\lambda^{\frac{1}{2}}(s,m_i^2,m_j^2) \nonumber \\
&\hspace{1.5em}\times\bigg[\frac{192 G_F^2 M_W^4}{s^2} \bigg(F_W^{ij} \Big(\big|[W_{\mathcal{N}}^{RR}]_{ij\alpha\beta}\big|^2 + \big|[W_{\mathcal{N}}^{LL}]_{ij\alpha\beta}\big|^2\Big) \nonumber \\
&\hspace{9em}+ G_W^{ij} \Big(\big|[W_{\mathcal{N}}^{RL}]_{ij\alpha\beta}\big|^2 + \big|[W_{\mathcal{N}}^{LR}]_{ij\alpha\beta}\big|^2\Big) \nonumber \\
&\hspace{9em} + F_{WW}^{ij}\,\text{Re}\Big[[W_{\mathcal{N}}^{RR}]_{ij\alpha\beta} [W_{\mathcal{N}}^{RR}]_{ij\beta\alpha} + [W_{\mathcal{N}}^{LL}]_{ij\alpha\beta} [W_{\mathcal{N}}^{LL}]_{ij\beta\alpha} \Big]\nonumber \\
&\hspace{9em} + G_{WW}^{ij}\,\text{Re}\Big[ [W_{\mathcal{N}}^{LR}]_{ij\alpha\beta} [W_{\mathcal{N}}^{RL}]_{ij\beta\alpha}\Big]\bigg)  \nonumber \\
&\hspace{3em} + 4\bigg(1 - \frac{m_i^2 + m_j^2}{2s} - \frac{(m_i^2 - m_j^2)^2}{2s^2}\bigg) \Big(\big|L_{\underset{ij\alpha\beta}{\mathcal{N}e}}^{V,RR}\big|^2 + \big|L_{\underset{ij\alpha\beta}{\mathcal{N}e}}^{V,RL}\big|^2\Big) \nonumber \\
&\hspace{3em} + 3\bigg(1 - \frac{m_i^2 + m_j^2}{s}\bigg)\Big(\big|L_{\underset{ij\alpha\beta}{\mathcal{N}e}}^{S,RR}\big|^2 + \big|L_{\underset{ij\alpha\beta}{\mathcal{N}e}}^{S,RL}\big|^2\Big) \nonumber \\
&\hspace{3em} + 16\bigg(1 + \frac{m_i^2 + m_j^2}{s} - \frac{2(m_i^2 - m_j^2)^2}{s^2}\bigg)\big|L_{\underset{ij\alpha\beta}{\mathcal{N}e}}^{T,RR}\big|^2 \nonumber \\
&\hspace{3em} - \frac{12m_i m_j}{s}\text{Re}\Big[L_{\underset{ij\alpha\beta}{\mathcal{N}e}}^{V,RR} \, L_{\underset{ij\beta\alpha}{\mathcal{N}e}}^{V,RR} + L_{\underset{ij\alpha\beta}{\mathcal{N}e}}^{V,RL} \, L_{\underset{ij\beta\alpha}{\mathcal{N}e}}^{V,RL} + L_{\underset{ij\alpha\beta}{\mathcal{N}e}}^{S,RL} \, L_{\underset{ij\beta\alpha}{\mathcal{N}e}}^{S,RR}\Big] \nonumber \\
& \hspace{3em} + \frac{48 G_F M_W^2}{\sqrt{2} s}\text{Re}\bigg[ 
L_{\underset{ij\alpha\beta}{\mathcal{N}e}}^{V,RR}\bigg(F_{VW}^{ij}[W_{\mathcal{N}}^{RR}]_{ij
\alpha\beta}^* - G_{VW}^{ij}[W_{\mathcal{N}}^{RR}]_{ji\alpha\beta}^*\bigg) \nonumber\\
&\hspace{10em} + L_{\underset{ij\alpha\beta}{\mathcal{N}e}}^{V,RL}\bigg(G_{VW}^{ij}[W_{\mathcal{N}}^{LL}]_{ij
\alpha\beta}^* - F_{VW}^{ij}[W_{\mathcal{N}}^{LL}]_{ji
\alpha\beta}^*\bigg) \nonumber \\ 
&\hspace{10em} + F_{SW}^{ij} L_{\underset{ij\alpha\beta}{\mathcal{N}e}}^{S,RR}[W_{\mathcal{N}}^{RL}]_{\{ij\}
\alpha\beta}^* + G_{SW}^{ij} L_{\underset{ij\alpha\beta}{\mathcal{N}e}}^{S,RL}[W_{\mathcal{N}}^{LR}]_{\{ij\}
\alpha\beta}^* \nonumber \\
&\hspace{10em} + F_{TW}^{ij} L_{\underset{ij\alpha\beta}{\mathcal{N}e}}^{T,RR}[W_{\mathcal{N}}^{RL}]_{[ij]
\alpha\beta}^*\bigg]\bigg] \nonumber \\
& \hspace{3em} + (\alpha \leftrightarrow \beta, ~\chi \leftrightarrow \chi^*)\,,
\end{align}
where $[W_{\mathcal{N}}^{XY}]_{\{ij\}
\alpha\beta} \equiv [W_{\mathcal{N}}^{XY}]_{ij
\alpha\beta} + [W_{\mathcal{N}}^{XY}]_{ji
\alpha\beta}$ and $[W_{\mathcal{N}}^{XY}]_{[ij]
\alpha\beta} \equiv [W_{\mathcal{N}}^{XY}]_{ij
\alpha\beta} - [W_{\mathcal{N}}^{XY}]_{ji
\alpha\beta}$. In both Eqs.~\eqref{eq:tot_Dirac_cs} and~\eqref{eq:tot_Maj_cs}, the coefficients are given by
\begin{align}
L^{V,XY}_{\underset{ij\alpha\beta}{\mathcal{N}e}} &\equiv C^{V,XY}_{\underset{ij\alpha\beta}{\mathcal{N}e}} + \chi_Z [Z_{\mathcal{N}}^X]_{ij} [Z_e^Y]_{\alpha\beta} \,, \nonumber \\
L^{S,XY}_{\underset{ij\alpha\beta}{\mathcal{N}e}} &\equiv C^{S,XY}_{\underset{ij\alpha\beta}{\mathcal{N}e}} \,, \nonumber \\
L^{T,XX}_{\underset{ij\alpha\beta}{\mathcal{N}e}} &\equiv C^{T,XX}_{\underset{ij\alpha\beta}{\mathcal{N}e}}\,.
\label{eq:total_cs_coefficients}
\end{align}
The final line in Eq.~\eqref{eq:tot_Maj_cs} indicates that for each term, an additional term must be included with $\alpha \leftrightarrow \beta$ and $\chi \leftrightarrow \chi^*$ in the coefficients in Eq.~\eqref{eq:total_cs_coefficients}. The factors $F_X^{ij}$ and $G_X^{ij}$ are given by,
\begin{align}
F_{W}^{ij} &= 1 + \frac{1}{2\omega} - (1 + \omega_{ij})L_{ij} + A_{ij}^2\bigg(1 - \frac{2\omega}{(1 + 2\omega_{ij})^2 - \lambda_{ij}}  - \omega_{ij}L_{ij}\bigg) \,, \nonumber \\
G_{W}^{ij} &= \frac{1}{2\omega} - \frac{2\omega}{(1 + 2\omega_{ij})^2 - \lambda_{ij}} + A_{ij}^2 \bigg(1 - \frac{2\omega}{(1 + 2\omega_{ij})^2 - \lambda_{ij}} - \omega_{ij}L_{ij}\bigg)\,, \nonumber \\
F_{WW}^{ij} &= A_{ij}\bigg(1 - \frac{2\omega}{1+2\omega_{ij}}\bigg(\frac{1}{2\omega} + \frac{\omega_{ij}}{\omega} + \frac{(1+2\omega_{ij})^2 - \lambda_{ij}}{4\omega}- A_{ij}^2\bigg)L_{ij}\bigg)\,,\nonumber\\
G_{WW}^{ij} &= 2\bigg(1 - \frac{2\omega}{1+2\omega_{ij}}\bigg)L_{ij} + A_{ij}^2\bigg(1 + \frac{\omega}{1+2\omega_{ij}}\bigg(8 - \frac{(1+2\omega_{ij})^2 - \lambda_{ij}}{2\omega}\bigg)L_{ij}\bigg)\,, \nonumber \\
F_{VW}^{ij} &= 3 + 2\omega_{ij} - 2\Big((1 + \omega)(1 + 2\omega_{ij} - \omega) + 2\omega(1 + 2\omega)A_{ij}^2\Big)L_{ij} \,, \nonumber \\
G_{VW}^{ij} &= -A_{ij}\bigg(1 - 2\omega_{ij} + 2\omega\bigg(1 + \frac{(1 + 2\omega_{ij})^2 - \lambda_{ij}}{4\omega}\bigg)L_{ij}\bigg) \,, \nonumber \\
F_{SW}^{ij} &= - A_{ij}\big(1 + 3\omega L_{ij}\big)\,, \nonumber \\
G_{SW}^{ij} &= \Big(1 + 2\omega_{ij} - 2\omega\big(1 - A_{ij}^2\big)\Big)L_{ij}\,,\nonumber \\
F_{TW}^{ij} &= 8A_{ij}\bigg(\omega_{ij} + \frac{\omega}{2}\bigg(1 - \frac{(1 + 2\omega_{ij})^2 - \lambda_{ij}}{2\omega}\bigg)L_{ij}\bigg)\,,
\end{align}
where we define,
\begin{gather}
\lambda_{ij} \equiv \lambda\bigg(1, \frac{m_i^2}{s}, \frac{m_j^2}{s}\bigg)\,, \quad \omega_{ij} \equiv \omega - \frac{m_i^2 + m_j^2}{2s}\,, \quad A_{ij} \equiv \frac{m_i m_j}{2M_W^2}\,,
\end{gather}
and,
\begin{gather}
L_{ij} \equiv \frac{1}{\sqrt{\lambda_{ij}}}\log\bigg(\frac{1 + 2\omega_{ij} + \sqrt{\lambda_{ij}}}{1 + 2\omega_{ij} - \sqrt{\lambda_{ij}}}\bigg)\,,
\end{gather}
for convenience.

Using Eq.~\eqref{eq:tot_Dirac_cs}, we can calculate the cross section for the SM process $\ell_\alpha^+ \ell_\beta^- \to \sum \nu \bar{\nu}$, where the neutrinos are massless Weyl fermions with $\nu_{Lp} = P_L\delta_{pi}\nu_i$. The SM charged- and neutral-current couplings in Eq.~\eqref{eq:SM_couplings} are inserted into Eq.~\eqref{eq:tot_Dirac_cs} and the limit $m_i, m_j \to0$ taken. This yields
\begin{align}
\label{eq:tot_cs_SM}
\sigma(s)\big|_{\text{SM}} &= \frac{G_F^2 M_Z^4}{6\pi s}\bigg[\chi_1 N_\nu \big((g_R^e)^2 + (g_L^e)^2\big)\delta_{\alpha\beta} + 6 c_w^4\bigg(1 + \frac{1}{2\omega} - (1 + \omega)\log\bigg(\frac{1 + \omega}{\omega}\bigg)\bigg) \nonumber \\
& \hspace{5em} + 3 \chi_2 g_L^e c_w^2 \delta_{\alpha\beta}  \bigg(3 + 2\omega - 2(1 + \omega)^2\log\bigg(\frac{1 + \omega}{\omega}\bigg)\bigg)\bigg]\,,
\end{align}
with $\omega \equiv M_W^2/s$ and
\begin{align}
\label{eq:chi12}
\chi_1 \equiv \frac{s^2}{(s - M_Z^2)^2 + (\Gamma_Z M_Z)^2}\,,\quad \chi_2 \equiv \frac{s (s - M_Z^2)}{(s - M_Z^2)^2 + (\Gamma_Z M_Z)^2} \,.
\end{align}
The first term in Eq.~\eqref{eq:tot_cs_SM} corresponds to $s$-channel $Z$ exchange, the second to $t$-channel $W^\pm$ exchange, and the last to the interference between the two contributions. The cross section in Eq.~\eqref{eq:tot_cs_SM} is also applicable for light Majorana or Dirac neutrinos with non-zero active-sterile mixing, as long as the PMNS mixing matrix is approximately unitary. For non-zero EFT interactions, Eqs.~\eqref{eq:tot_Dirac_cs} and~\eqref{eq:tot_Maj_cs} provide the leading interference effects between the SM and heavy new physics contributions to the process $\ell_\alpha^+ \ell_\beta^- \to \sum \nu \bar{\nu}$, which we make use of in Sec.~\ref{sec:discussion}. 

The expressions in Eqs.~\eqref{eq:tot_Dirac_cs} and~\eqref{eq:tot_Maj_cs} can also be used to compute the cross section for the processes $\ell_\alpha^+ \ell_\beta^- \to \sum_{ij}\nu_{i} N_j$ (Majorana) and $\ell_\alpha^+ \ell_\beta^- \to \sum_{ij}\nu_i \bar{N}_{j} + \bar{\nu}_i N_j$ (Dirac) induced by the active-sterile mixing $V_{\alpha N_j}$.
In both cases, the total cross section is
\begin{align}
\label{eq:tot_Maj_cs_mixing}
\sigma(s)\big|_{\text{mix}} &= \frac{G_F^2 M_Z^4}{6\pi s} \sum_{\rho,j}|V_{\rho N_j}|^2 (1 - y_j) \nonumber \\
&\hspace{1em}\times \bigg[(1 - y_j)(2 + y_j)\chi_1\big((g_R^e)^2 + (g_L^e)^2\big)\delta_{\alpha\beta}  \nonumber \\
& \hspace{2.5em} + 6 c_w^4 (\delta_{\alpha\rho} + \delta_{\beta\rho})\bigg(1 + \frac{1}{2\omega} - \bigg(1 + \omega - \frac{y_j}{2}\bigg)L_{ij}\bigg) \nonumber \\
& \hspace{2.5em} + 6 \chi_2 g_L^e c_w^2 \delta_{\alpha\beta} \delta_{\alpha\rho} \Big(3 + 2\omega - y_j - 2(1 + \omega)(1 + \omega - y_j)L_{ij}\Big)\bigg]\,,
\end{align}
where $L_{ij} \equiv \log[(1 + \omega - y_j)/\omega] / (1-y_j)$, and $\chi_1$, $\chi_2$ are given in Eq.~\eqref{eq:chi12}. 
As in Eq.~\eqref{eq:tot_cs_SM}, the first and second terms in parenthesis in Eq.~\eqref{eq:tot_Maj_cs_mixing} correspond to $Z$ and $W^\pm$ exchange, respectively, and the last term to $Z-W^\pm$ interference.

\begin{figure}[t!]
\centering
\includegraphics[width=0.45\textwidth]{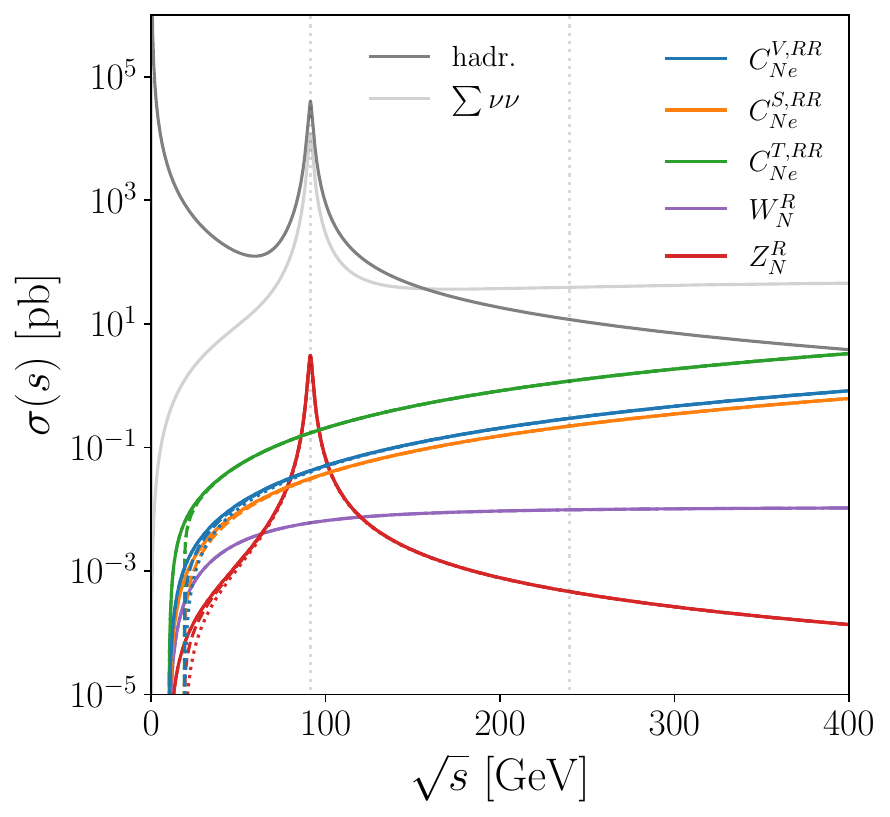}
\includegraphics[width=0.45\textwidth]{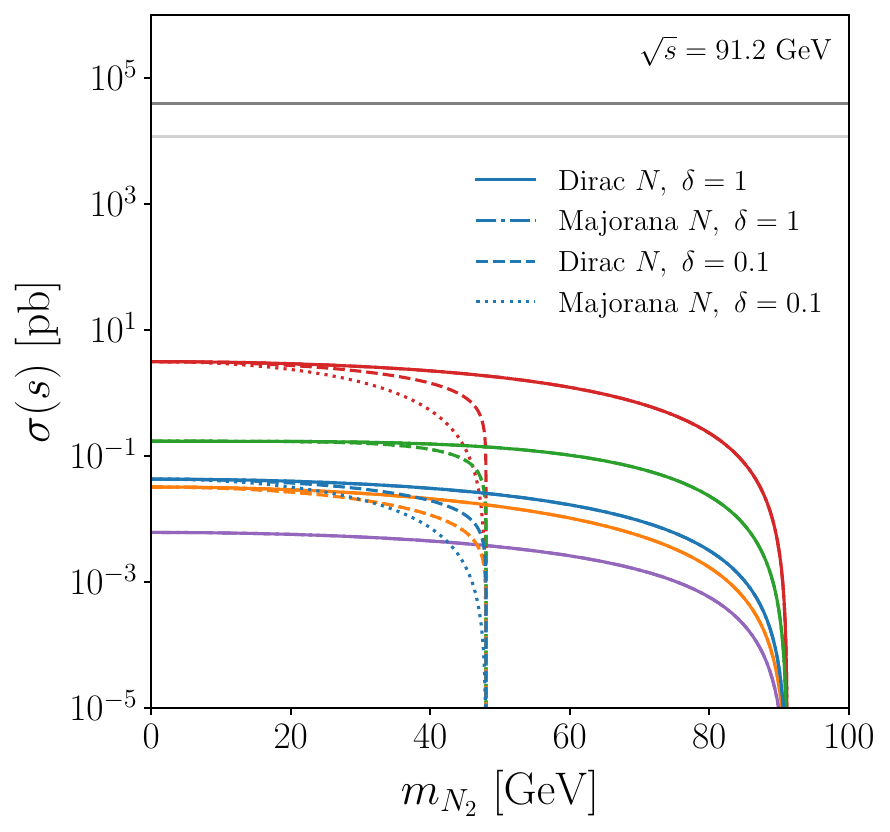}
\caption{(Left) LO total cross sections $\sigma(s)$ as a function of $\sqrt{s}$ for $e^+ e^- \to N_1 N_2$, induced by the off-diagonal four-fermion and effective $Z$ interactions, and $e^+ e^- \to \nu_e N_2$, induced by the effective $W^{\pm}$ interaction. (Right) The same cross sections, but as a function of $m_{N_2}$ for $\sqrt{s} = 91.2$~GeV. The cross sections are plotted for Dirac and Majorana $N$ and two values of the mass splitting ratio $\delta$.}
\label{fig:xsec_plot}
\end{figure}

Using Eqs.~\eqref{eq:tot_Dirac_cs} and~\eqref{eq:tot_Maj_cs}, we now explore the contributions of EFT operators to the single and pair production of HNLs. In Fig.~\ref{fig:xsec_plot} (left) we plot the LO cross section for the process $e^+ e^- \to N_1 N_2$ as a function of $\sqrt{s}$, turning on one off-diagonal four-fermion and effective $Z$ interaction at a time. In the same plot, we show the cross section for $e^+ e^- \to \nu_e N_2$ induced by the effective $W^{\pm}$ coupling. The benchmark values of the WCs are
\begin{align}
C_{\underset{12ee}{Ne}}^{V,RR}\,,~C_{\underset{12ee}{Ne}}^{S,RR}\,,~C_{\underset{12ee}{Ne}}^{T,RR} = 1 ~ \text{TeV}^{-2}\,, \
\quad [W_{N}^{R}]_{2e}\,,~[Z_{N}^{R}]_{12} = 10^{-2}\,,
\end{align}
and two mass splitting ratios $\delta$ are chosen for Majorana or  Dirac HNLs; $\delta = 1$ (or equivalently, $m_{N_1} = 0$) and $\delta = 0.1$. For comparison, we also plot the SM prediction for $e^+ e^- \to \sum \nu \bar{\nu}$ in Eq.~\eqref{eq:tot_cs_SM} and the total hadronic cross section $e^+e^- \to q \bar{q}$, neglecting quark masses and QCD corrections.

As expected, the impact of the effective $Z$ interaction is maximised near the $Z$ pole, while the cross sections for the four-fermion and effective $W^\pm$ interactions grow linearly and logarithmically with $s$, respectively. In Fig.~\ref{fig:xsec_plot} (right), we plot the same cross sections as a function of $m_{N_2}$ for a fixed value of the centre-of-mass energy, $\sqrt{s} = 91.2$~GeV. In both plots, the impact of the mass splitting $\delta$ is apparent. In the left plot, the minimum value of $\sqrt{s}$ required to produce $N_1$ and $N_2$, $\sqrt{s} > m_{N_1} + m_{N_2} = m_{N_2}(2 - \delta)$, can be seen. In the right plot, this is equivalent to the upper limit $m_{N_2} < \sqrt{s}/(2 - \delta)$ for fixed $\sqrt{s}$ and $\delta$.

In Fig.~\ref{fig:xsec_plot}, the cross sections for the production of Dirac (solid and dashed lines) and Majorana HNLs (dot-dashed and dotted lines) are compared. With the normalisation of the Lagrangians in Eqs.~\eqref{eq:L_fourfermion_mass} and~\eqref{eq:L_gauge_mass} (in particular, the prefactor of 1/2 for the Majorana four-fermion and effective $Z$ operators), the cross sections for $\ell_\alpha^+ \ell_\beta^- \to \mathcal{N}_i \mathcal{N}_j$ (Majorana) and $\ell_\alpha^+ \ell_\beta^- \to \mathcal{N}_i \bar{\mathcal{N}}_j + \mathcal{N}_j \bar{\mathcal{N}}_i$ (Dirac) with $i\neq j$, coincide\footnote{There is an interesting exception for the cross sections $\ell_\alpha^+ \ell_\beta^- \to \nu_i \nu_j$ (Majorana) and $\ell_\alpha^+ \ell_\beta^- \to \nu_i \bar{\nu}_j + \nu_j \bar{\nu}_i$ (Dirac). If the coefficients $[W_\nu^R]_{i\alpha}$ are non-zero for general $i$ and $\alpha$, $t$-channel diagrams with the SM charged-current at one vertex and $W^R_\nu$ at the other give different overall cross sections in the Majorana and Dirac cases, even neglecting the neutrino masses. This is the result of the additional term in Eq.~\eqref{eq:tot_Maj_cs} with respect to Eq.~\eqref{eq:tot_Dirac_cs} that is proportional to $G_{WW}^{ij}$, which does not vanish in the $m_i, m_j \to 0$ limit.} in the limit $m_{i}, m_{j} \to 0$ for the same values of $C_{\mathcal{N}e}^{V,RR}$, $C_{\mathcal{N}e}^{S,RR}$, $C_{\mathcal{N}e}^{T,RR}$, $W_{\mathcal{N}}^{R}$ and $Z_{\mathcal{N}}^{R}$. The differences in the cross sections now arise solely from the Dirac versus Majorana nature of the HNLs. For $\delta = 1$, the Dirac and Majorana cross sections are identical, while for $\delta = 0.1$, the vector four-fermion and effective $Z$ interaction induced cross sections fall off faster in the Majorana case compared to the Dirac case. This is because, in the Majorana case, there are additional interference terms proportional to $\sim m_i m_j \text{Re}\big[(C_{Ne}^{V,RR})^2\big]$ and $\sim m_i m_j \text{Re}\big[(Z_{N}^{R})^2\big]$, which vanish when one of the outgoing states is massless ($\delta = 1$), but result in a reduced cross section for $m_i \sim m_j$. Such terms are not present for single scalar or tensor coefficients.

\begin{figure}[t!]
\centering
\includegraphics[width=0.45\textwidth]{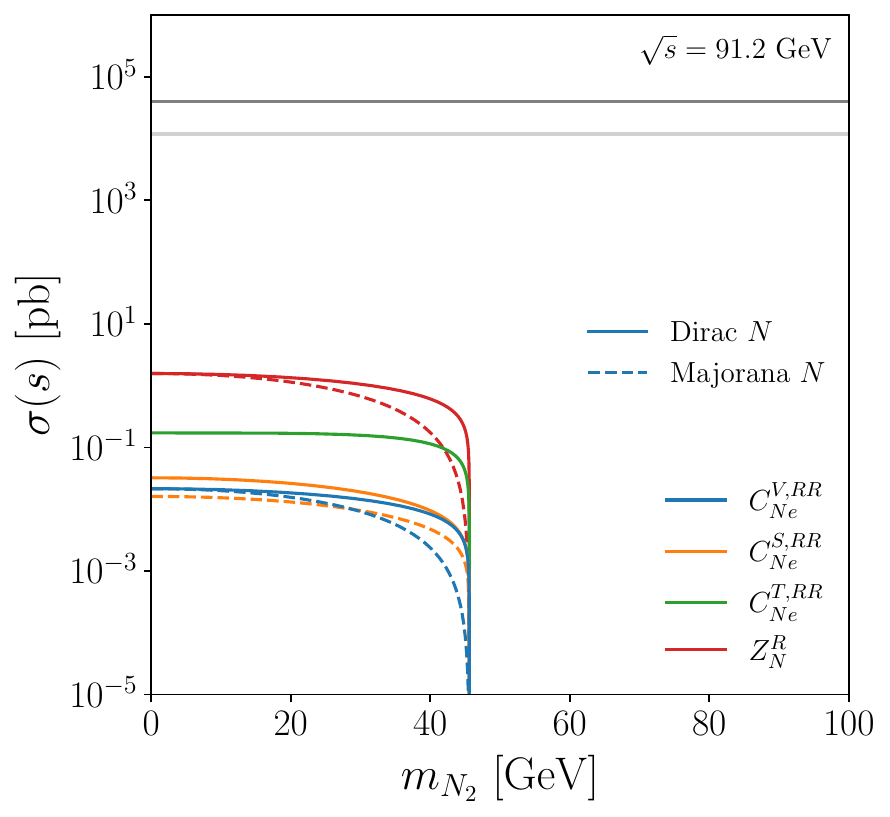}
\caption{LO cross sections for $e^+ e^- \to N_2 N_2$, induced by the diagonal four-fermion and effective $Z$ interactions, shown as a function of $m_{N_2}$ for $\sqrt{s} = 91.2$~GeV and the Dirac and Majorana scenarios.}
\label{fig:xsec_plot_2}
\end{figure}

In Fig.~\ref{fig:xsec_plot_2}, we plot the cross sections for the processes $e^+e^- \to N_2 N_2$ (Majorana) and $e^+e^- \to N_2 \bar{N}_2$ (Dirac) induced by the diagonal four-fermion and effective $Z$ interactions, taking the benchmark WC values
\begin{align}
C_{\underset{22ee}{Ne}}^{V,RR}\,,~C_{\underset{22ee}{Ne}}^{S,RR},~C_{\underset{22ee}{Ne}}^{T,RR} = 1 ~ \text{TeV}^{-2}\,, \
\quad [Z_{N}^{R}]_{22} = 10^{-2}\,.
\end{align}
With the choice of normalisation of the Majorana and Dirac Lagrangians, it can be seen that the vector four-fermion and effective $Z$ cross sections again coincide in the massless limit. However, this is no longer true for the scalar and tensor operators. In the Majorana scenario, the diagonal tensor coupling $C_{Ne}^{T,RR}$ and associated cross section vanishes, while the cross section via $C_{Ne}^{S,RR}$ is a factor of two smaller in the Majorana case compared to the Dirac case. The Majorana cross sections via $C_{Ne}^{V,RR}$ and $Z_{N}^R$ can again be seen to fall off faster as a function of $m_{N_2}$ compared to the Dirac cross sections.

In this work, the $2\to 2$ cross sections above are used to estimate the sensitivity of DV signatures at FCC-ee to the $\nu$SMEFT operators in Table~\ref{tab:vSMEFT-operators}. However, for the complementary mono-$\gamma$ signature, the cross sections in Eq.~\eqref{eq:2to2_processes} with a photon in the final state are required. For these $2\to 3$ processes, diagrams such as those in Fig.~\ref{fig:feynman-diagrams} contribute. Additional diagrams with the photon attached to the other incoming charged lepton should be included. Finally, for the $t$-channel diagram induced by the effective $W^\pm$ interactions, a diagram must be added with the photon attached to the intermediate $W^\pm$. In previous analyses, the contribution from this diagram has been neglected due to the suppression from two $W^\pm$ propagators. Following this approximation and taking only the leading contributions from initial state radiation, the mono-$\gamma$ cross sections can be obtained from $\sigma(s)$ in Eqs.~\eqref{eq:tot_Dirac_cs} and~\eqref{eq:tot_Maj_cs} with
\begin{align}
\label{eq:xsec_monophoton}
\frac{d^2\sigma}{dx_\gamma dc_\gamma} = \sigma(s(1-x_\gamma)) \frac{\alpha}{\pi} \frac{1+(1-x_\gamma)^2}{x_\gamma} \frac{1}{1-\beta_e^2 c_\gamma^2} \,,
\end{align}
where $\alpha = e^2/(4\pi)$ is the QED fine-structure constant, $x_\gamma = 2E_\gamma/\sqrt{s}$ is the fraction of the beam energy carried away by the photon and $c_\gamma = \cos\theta_\gamma$, with $\theta_\gamma$ the angle of the photon with respect to the beam axis. The differential cross section in Eq.~\eqref{eq:xsec_monophoton} follows the usual dependence on $x_\gamma$ and $c_\gamma$. In all cases, the cross section peaks at $x_\gamma \sim 0$ and $c_\gamma \sim \pm 1$. For the four-fermion operators, the cross section decreases monotonically with $x_\gamma$, while for the $s$-channel diagram induced by the effective $Z$ interaction, there is an additional peak at $x_\gamma \sim 1 - M_Z^2/s$. In Sec.~\ref{sec:discussion}, it is convenient to perform the change of variables $(x_\gamma, c_\gamma)\to (p_T^\gamma,c_\gamma)$ to implement the $p_T^\gamma > 1$~GeV cut. This can be done with the replacement $E_\gamma = p_T^\gamma/\sqrt{1-c_\gamma^2}$ and multiplying Eq.~\eqref{eq:xsec_monophoton} by the appropriate Jacobian factor.

For the full analysis of the mono-$\gamma$ plus $\slashed{E}$ signature, it is necessary to apply cuts on the outgoing photon energy $E_\gamma$, transverse momentum $p_T^\gamma$, and angle $\theta_\gamma$. In order to do this, we simulate the $e^+e^- \to \mathcal{N}_i \mathcal{N}_j \gamma$ process in \texttt{MadGraph5\_aMC@NLO}~\cite{Alwall:2014hca}. We implemented the Lagrangians in Eqs.~\eqref{eq:L_fourfermion_mass} and~\eqref{eq:L_gauge_mass} in \texttt{Feynrules}~\cite{Alloul:2013bka} model files, which were then converted to UFO files for input to MadGraph. In the Dirac scenario, all of the effective operators in Eq.~\eqref{eq:L_fourfermion_mass} and~\eqref{eq:L_gauge_mass} can be simulated. However, MadGraph does not support multi-fermion interactions with the violation of fermion-flow~\cite{Alwall:2014hca}. One way to circumvent this limitation is to add heavy fields to the model file which, when integrated out, produce the four-fermion operators in the Majorana case~\cite{Zapata:2022qwo}. 

However, we use a simpler method to reproduce the Majorana cross sections for the four-fermion operators. In the \texttt{Feynrules} model file, the Majorana Lagrangian is written identically to the Dirac Lagrangian, and \texttt{SelfConjugate -> False} is taken for the states $\mathcal{N}_i$ and $\mathcal{N}_j$. However, one can make the replacements
\begin{gather}
C^{V,LX}_{\underset{ij\alpha\beta}{\mathcal{N}e}} \to -C^{V,RX*}_{\underset{ij\beta\alpha}{\mathcal{N}e}}\,, \quad C^{S,LX}_{\underset{ij\alpha\beta}{\mathcal{N}e}} \to C^{S,RY*}_{\underset{ij\beta\alpha}{\mathcal{N}e}} \,, \quad C^{T,LL}_{\underset{ij\alpha\beta}{\mathcal{N}e}} \to - C^{T,RR*}_{\underset{ij\beta\alpha}{\mathcal{N}e}}\,, \nonumber \\
[Z_{\mathcal{N}}^L]_{ij} \to -[Z_{\mathcal{N}}^R]_{ij}^* \,,
\label{eq:DiractoMaj_replacements}
\end{gather}
for the four-fermion operators containing $P_L$ in the neutrino bilinear $(\bar{\mathcal{N}}_i \Gamma \mathcal{N}_j)$. Then, the Majorana cross section for $\ell_\alpha^+\ell_\beta^- \to \mathcal{N}_i \mathcal{N}_j(\gamma)$ can be correctly reproduced for $i = j$ and $i\neq j$ by simulating $\ell_\alpha^+\ell_\beta^- \to \mathcal{N}_i \bar{\mathcal{N}}_i(\gamma)$ and $\ell_\alpha^+\ell_\beta^- \to \mathcal{N}_i \bar{\mathcal{N}}_j(\gamma) + \bar{\mathcal{N}}_i \mathcal{N}_j (\gamma)$, respectively. To finalise the procedure, all four-fermion couplings should be multiplied by $1/\sqrt{2}$; at the cross section level, the resulting factor of $1/2$ provides the necessary symmetry factor for $i = j$ and takes into account the double counting inherent in the simulated process above for $i \neq j$. For the effective $Z$ and $W^\pm$ interactions, we can simply set \texttt{SelfConjugate -> True}.

Before concluding this appendix, we briefly discuss the applicability of the EFT formalism. As discussed in App.~\ref{app:matching}, the $\nu$SMEFT operators can be the result of UV physics at tree-level, one-loop or multiple loops. The simplest scenarios are UV completions with $s$- and $t$-channel contributions to $\ell_\alpha^+\ell_\beta^- \to \mathcal{N}_i \mathcal{N}_j(\gamma)$, such as those shown in Fig.~\ref{fig:UV_completions}. In the case of the heavy vector boson mediator $\mathcal{B}$, the propagator of the full scattering cross section can be expanded for $\sqrt{s} \ll M_\mathcal{B}$ as 
\begin{align}
\frac{1}{s - M_\mathcal{B}^2 + i \Gamma_\mathcal{B}M_\mathcal{B}}  = -\frac{1}{M_\mathcal{B}^2} + \mathcal{O}\bigg(\frac{s}{M_\mathcal{B}^2}\bigg)\,,
\end{align}
which gives the same result as integrating $\mathcal{B}$ out of the full theory, matching to the $\nu$SMEFT coefficient $C_{eN}$ and then to the rotated $\nu$SMEFT coefficient $C_{Ne}^{V,RR}$, which is inserted into Eq.~\eqref{eq:tot_Dirac_cs} or~\eqref{eq:tot_Maj_cs}. The same expansion can be performed for $t$-channel and one-loop cross sections in the UV theory. The EFT is then only valid for $\sqrt{s}$ sufficiently smaller than $M$, where $M$ is the generic heavy mediator mass; $\sqrt{s}\lesssim M/3$ is sufficient not to observe, for example, the resonant and logarithmic scaling of $s$- and $t$-channel cross sections, respectively. The naive scale of new physics $\Lambda$ is related to $M$ via UV couplings and $1/(16\pi^2)$ loop factors. However, decreasing the couplings and increasing the number of loop factors only increases the size of $\Lambda$ with respect to $M$. Therefore, the condition $\Lambda \gtrsim 3\sqrt{s}$ also applies for the EFT to be valid.

%%%%%%%%%%%%%%%%%%%%%%%%%%%%%%%%%%%%%%%%
\section{HNL Decay Rates}
\label{app:decays}
%%%%%%%%%%%%%%%%%%%%%%%%%%%%%%%%%%%%%%%%

The $\nu$SMEFT operators in Tables~\ref{tab:vSMEFT-operators} and~\ref{tab:vSMEFT-operators-2} and resulting effective Lagrangians in Eqs.~\eqref{eq:L_fourfermion_mass} and~\eqref{eq:L_gauge_mass}, which are relevant for the production of HNLs via the processes in Eq.~\eqref{eq:2to2_processes}, can also lead to their decay, as depicted in Fig.~\ref{fig:feynman-diagrams-decay}. Here, we review the possible decay modes, which are summarised in Table~\ref{tab:HNL_decays}, and give approximate formulae for the decay rates.

\begin{table}[t!]
\centering
\renewcommand{\arraystretch}{1.3}
\setlength\tabcolsep{3.2pt}
\begin{tabular}{c|c|c|c}
\hline
\multicolumn{4}{c}{ $\mathcal{N}_j$ Decays}\\ \hline
Operator & $(q^2)^{\text{max}} \lesssim 1~\text{GeV}^2$ & $(q^2)^{\text{max}} \gtrsim 1~\text{GeV}^2$ & $(q^2)^{\text{max}} \gtrsim M_W^2, M_Z^2, M_h^2$ \\ \hline
$C_{\mathcal{N}\mathcal{N}}^{V,XY}$  & $\mathcal{N}_i \mathcal{N}_k \mathcal{N}_l$ & $\mathcal{N}_i \mathcal{N}_k \mathcal{N}_l$ & $\mathcal{N}_i \mathcal{N}_k \mathcal{N}_l$ \\ \hline

$C_{\mathcal{N}e}^{V,XY}$, $C_{\mathcal{N}e}^{S,XY}$, $C_{\mathcal{N}e}^{T,XX}$  & $\mathcal{N}_i \ell_\alpha^-\ell_\beta^+$ & $\mathcal{N}_i \ell_\alpha^-\ell_\beta^+$ & $\mathcal{N}_i \ell_\alpha^-\ell_\beta^+$ \\ \hline

\multirow{2}{*}{$[W_{\mathcal{N}}^{X}]_{j\alpha}$} & $\mathcal{N}_i \ell_\alpha^-\ell_\beta^+$ & $\mathcal{N}_i \ell_\alpha^-\ell_\beta^+$ & \multirow{2}{*}{$\ell^{\mp}_\alpha W^\pm$}\\

& $\ell^{\mp}_\alpha P^\pm/V^\pm$ & $\ell^{-}_\alpha u\bar{d}$ ($\ell^{+}_\alpha \bar{u} d$) & \\ \hline
\multirow{3}{*}{$[Z_{\mathcal{N}}^{X}]_{ij}$} & $\mathcal{N}_i \mathcal{N}_k \mathcal{N}_l$ & $\mathcal{N}_i \mathcal{N}_k \mathcal{N}_l$ & \multirow{3}{*}{$\mathcal{N}_i Z$} \\

& $\mathcal{N}_i \ell_\alpha^-\ell_\alpha^+$ & $\mathcal{N}_i \ell_\alpha^-\ell_\alpha^+$ & \\
& $\mathcal{N}_i P^0/V^0$ & $\mathcal{N}_i q\bar{q}$ & \\ \hline
\end{tabular}
\caption{Majorana HNL decays induced by the $\nu$SMEFT operators in Tables~\ref{tab:vSMEFT-operators}, for three different $(q^2)^{\text{max}}$ regimes.}
\label{tab:HNL_decays}
\end{table}

All of the operators in Eq.~\eqref{eq:L_fourfermion_mass} induce the decay $\mathcal{N}_j \to \mathcal{N}_i e^-e^+$. For the four-fermion operators, we take all WCs except $\alpha = \beta = e$ to vanish, so this is the only available decay channel. However, the effective $Z$ interaction can also mediate $\mathcal{N}_j \to \mathcal{N}_i \ell_\alpha^-\ell_\alpha^+$ for $\alpha = \mu,\tau$, while the effective $W^\pm$ interactions can lead to $\mathcal{N}_j \to \mathcal{N}_i \ell_\alpha^-\ell_\beta^+$ with $\alpha = e$ and $\beta = \mu,\tau$. In the Majorana case, $\mathcal{N}_j \to \mathcal{N}_i \ell_\alpha^-\ell_\beta^+$ with $\alpha = \mu,\tau$ and $\beta = e$ is also possible. Assuming that the HNL mass is well below the EW scale, the decay rates for these processes are given by\footnote{In Eq.~\eqref{eq:decay-approx}, we also neglect interference terms which become important near the kinematic threshold $m_j \sim m_i + m_\alpha + m_\beta$.}
\begin{align}
\label{eq:decay-approx}
\Gamma(\mathcal{N}_j \to \mathcal{N}_i f_\alpha \bar{f}_\beta) &= \frac{N_cm_{j}^5}{1536\pi^3}\bigg[I_1^{i\alpha\beta}\bigg(\big|
L_{\underset{ij\alpha\beta}{\mathcal{N}f}}^{V,RR}\big|^2 + \big|
L_{\underset{ij\alpha\beta}{\mathcal{N}f}}^{V,RL}\big|^2
\nonumber \\
&\hspace{8em}+ \frac{1}{4}\Big(\big|
L_{\underset{ij\alpha\beta}{\mathcal{N}f}}^{S,RR}\big|^2 + \big|
L_{\underset{ij\alpha\beta}{\mathcal{N}f}}^{S,RL}\big|^2\Big)
+ 12\big|L_{\underset{ij\alpha\beta}{\mathcal{N}f}}^{T,RR}\big|^2\bigg) \nonumber  \\
&\hspace{5.5em} - I_3^{\alpha\beta i}\,\text{Re}\Big[L_{\underset{ij\alpha\beta}{\mathcal{N}f}}^{V,LR} \,L_{\underset{ij\alpha\beta}{\mathcal{N}f}}^{V,RR*} - \frac{1}{2}\,L_{\underset{ij\alpha\beta}{\mathcal{N}f}}^{S,LR} \,L_{\underset{ij\alpha\beta}{\mathcal{N}f}}^{S,RR*}\Big]\bigg] \nonumber \\
&\hspace{1.5em} + (L\leftrightarrow R)\,,
\end{align}
for $f = e = \ell$ and the number of colors $N_c = 1$. The coefficients are given in the Dirac and Majorana cases by Eqs.~\eqref{eq:Dirac-coefficients} and~\eqref{eq:Maj-coefficients}, respectively, with the following limit taken for the propagator factors in Eq.~\eqref{eq:prop_factors},
\begin{align}
\chi^\alpha_W \to -\frac{g^2}{2M_W^2}\,, \quad \chi_Z \to -\frac{g^2}{c_w^2M_Z^2}\,.
\end{align}
This is equivalent to matching the $\nu$SMEFT to $\nu$LEFT operators (for which $W^\pm$ and $Z$ are integrated out) at the EW scale and neglecting QED corrections~\cite{Jenkins:2017dyc}. In Eq.~\eqref{eq:decay-approx}, we use the shorthands $I_1^{i\alpha\beta} = I_1(y_i, y_\alpha, y_\beta)$ and $I_3^{\alpha\beta i} = I_3(y_\alpha, y_\beta, y_i)$, where $y_X \equiv m_X/m_j$ and the functions $I_1(x,y,z)$ and $I_3(x,y,z)$ are given by,
\begin{align}
I_1(x,y,z) &= 12\int_{(x+y)^2}^{(1-z)^2}\frac{ds}{s}\,(1 + z^2 - s)(s - x^2 - y^2)\lambda^{\frac{1}{2}}(1,s,z^2)\lambda^{\frac{1}{2}}(s,x^2,y^2)\,, \nonumber \\
I_3(x,y,z) &= 24z\int_{(x + y)^2}^{(1-z)^2}\frac{ds}{s}\,(s - x^2 - y^2)\lambda^{\frac{1}{2}}(1,s,z^2)\lambda^{\frac{1}{2}}(s,x^2,y^2)\,.
\end{align}
While Eq.~\eqref{eq:decay-approx} is a good approximation for HNLs in the GeV range, for masses closer to the $W^\pm$ and $Z$ masses, we implement a more accurate estimate of $\Gamma(\mathcal{N}_j \to \mathcal{N}_i \ell_\alpha^- \ell_\beta^+)$ retaining full dependence on the propagators, discussed further below Eq.~\eqref{eq:2body-WZh}.

The effective $W^\pm$ and $Z$ interactions also lead to the decays of HNLs to quarks. For maximum momentum transfers to the $q\bar{q}$ system above the QCD scale $\sim 1$~GeV, the quarks do not hadronise and appear as jets in the final state. Firstly, the effective $W^\pm$ interactions induce the process $\mathcal{N}_j \to \ell_\alpha^- u_\rho \bar{d}_\sigma$ with
\begin{align}
\label{eq:decay_rate_Wud}
\Gamma(\mathcal{N}_j \to \ell_\alpha^- u_\rho \bar{d}_\sigma) = \frac{N_c m_{j}^5}{1536\pi^3} \bigg[I_1^{\alpha\rho\sigma}\Big(\big|
L_{\underset{\alpha j\rho\sigma}{e\mathcal{N}ud}}^{V,RL}\big|^2 + \big|
L_{\underset{\alpha j\rho\sigma}{e\mathcal{N}ud}}^{V,LL}\big|^2\Big) - I_3^{\rho\sigma\alpha}\,\text{Re}\Big[L_{\underset{\alpha j\rho\sigma}{e\mathcal{N}ud}}^{V,RL}\,L_{\underset{\alpha j\rho\sigma}{e\mathcal{N}ud}}^{V,LL*}\Big] \bigg] \,,
\end{align}
for $N_c = 3$ and,
\begin{align}
L^{V,XL}_{\underset{\alpha j \rho\sigma}{e\mathcal{N}ud}} &\equiv -\frac{g^2}{2M_W^2} [W_\mathcal{N}^X]_{j\alpha}^* V_{\rho\sigma}  \,,
\end{align}
where $V$ is the CKM matrix, defined to rotate the LH down-type quark fields to the mass basis as $d_{Lr} = P_L V_{r\sigma} d_\sigma$. For Majorana HNLs, the decay $\mathcal{N}_j \to \ell_\alpha^+ \bar{u}_\rho d_\sigma$ is also possible, with a decay rate equal to Eq.~\eqref{eq:decay_rate_Wud}. Likewise, the effective $Z$ interaction induces the process $\mathcal{N}_j \to \mathcal{N}_i q_\alpha \bar{q}_\beta$ with the decay rate given by Eq.~\eqref{eq:decay-approx} with $f = q$ and $N_c = 3$, where $\alpha,\beta = u,c$ for $q = u$ and $\alpha,\beta = d,s,b$ for $q = d$. In the Dirac scenario, the coefficients are
\begin{align}
\label{eq:effective-quark-couplings}
L^{V,XY}_{\underset{ij\alpha\beta}{\mathcal{N}q}} \equiv -\frac{g^2}{c_w^2 M_Z^2} [Z_\mathcal{N}^X]_{ij} [Z_{q}^Y]_{\alpha\beta} \,, \quad
L^{S,XY}_{\underset{ij\alpha\beta}{\mathcal{N}q}} \equiv 0\,, \quad L^{T,RR}_{\underset{ij\alpha\beta}{\mathcal{N}q}} \equiv 0 \,, 
\end{align}
where $[Z_{q}^R]_{\alpha\beta} = g_R^q \delta_{\alpha\beta}$ and $[Z_{q}^L]_{\alpha\beta} = g_L^q \delta_{\alpha\beta}$ with $g_R^u = - 2s_w^2/3$, $g_L^u = 1/2 - 2s_w^2/3$, $g_R^u = s_w^2/3$ and $g_L^u = - 1/2 + s_w^2/3$. In the Majorana scenario, the coefficients are equal to Eq.~\eqref{eq:effective-quark-couplings} with the replacement $[Z_{\mathcal{N}}^L]_{ij} \to -[Z_{\mathcal{N}}^R]_{ij}^*$. Following the approach of \cite{Bondarenko:2018ptm}, large QCD corrections to decays with outgoing quarks are taken into account by multiplying Eqs.~\eqref{eq:decay-approx} and~\eqref{eq:decay_rate_Wud} by the factor~\cite{Gorishnii:1990vf}
\begin{align}
\label{eq:QCD_correction}
1 + \Delta_{\text{QCD}} \equiv \frac{\Gamma(\tau\to\nu_\tau + \text{hadr.})}{\sum_{q}\Gamma(\tau\to\nu_\tau + \bar{u}q)\big|_{\text{tree}}} = 1+\frac{\alpha_s}{\pi} + 5.2 \frac{\alpha_s^2}{\pi^2} + 26.4 \frac{\alpha_s^3}{\pi^3} \,.
\end{align}
where the strong coupling constant $\alpha_s$ is evaluated at the maximum momentum transfer to the outgoing quarks; $(q^2)^{\text{max}} = (m_j - m_\alpha)^2$ for $\mathcal{N}_j \to \ell_\alpha^- u_\rho \bar{d}_\sigma$ and $(q^2)^{\text{max}} = (m_j - m_i)^2$ for $\mathcal{N}_j \to \mathcal{N}_i q_\alpha \bar{q}_\beta$.

Finally, we comment that some of the $\nu$SMEFT operators in Tables~\ref{tab:vSMEFT-operators} and~\ref{tab:vSMEFT-operators-2} generate four-neutrino operators in the broken phase. These operators do not contribute to HNL production, but can induce their decay. For example, in the Majorana scenario, $Q_{lN}$ and $Q_{lNlH}$ induce the operators $\mathcal{O}_{N \nu}^{V,RL} = (\bar{N} \gamma_\mu N)(\bar{\nu}_L \gamma^\mu \nu_L)$ and $\mathcal{O}_{\nu N \nu}^{V,RL} = (\bar{\nu}_L^c \gamma_\mu N)(\bar{\nu}_L \gamma^\mu \nu_L)$. The operators $Q_{lS}$ and $Q_{lSlH}$ do the same in the Dirac case. When the active-sterile mixing is negligible, it is sufficient to estimate these decay rates as $\mathcal{N}_j \to \mathcal{N}_i \nu_\alpha \bar{\nu}_\beta$, using Eq.~\eqref{eq:decay-approx} with $f = \nu$, $N_c = 1$ and $y_\alpha = y_\beta = 0$. Only the coefficients
\begin{align}
\label{eq:eff_NNvv}
L_{\underset{ij\alpha\beta}{\mathcal{N}\nu}}^{V,XL} = C_{\underset{ij\alpha\beta}{\mathcal{N}\nu}}^{V,XL} - \frac{g^2}{c_w^2 M_Z^2}[Z_\mathcal{N}^X]_{ij} [Z_{\nu}^L]_{\alpha\beta}\,,
\end{align}
are non-zero, with $[Z^L_{\nu}] = g_L^\nu \delta_{\alpha\beta}$. The HNL decay width from $C_{lN}$, for example, is twice as large as that from $C_{eN}$ for $m_j \gg m_i + 2m_e$, because the decay $N_j \to N_i \nu_e\bar{\nu}_e$ is open with an equal rate to $N_j \to N_i e^-e^+$.

We now summarise the two-body HNL decays induced by the effective $W^\pm$ and $Z$ interactions. If the maximum momentum transfer $(q^2)^{\text{max}}$ of the process is instead below the QCD scale, the final-state quarks hadronise. Thus, the effective $W^\pm$ interactions lead to the decay of HNLs to charged pseudoscalar and vector mesons, $\mathcal{N}_j \to \ell_\alpha^- P^+$ and $\mathcal{N}_j \to \ell_\alpha^- V^+$, respectively, with the lightest states being $P^\pm = \{\pi^\pm,K^\pm,D^\pm,D_s^\pm\}$ and $V^\pm = \{\rho^\pm,K^{*\pm}\}$. The decay rates for these processes are
\begin{align}
\label{eq:2body_charged}
\Gamma(\mathcal{N}_j \to \ell_\alpha^- P^+) &= \frac{G_F^2 f_P^2 m_{j}^3 |V_{q\bar{q}}|^2}{16\pi} \lambda^{\frac{1}{2}}(1, y_\alpha^2, y_P^2) \Big[F(y_\alpha, y_P)\Big(\big|[W_{\mathcal{N}}^R]_{j\alpha}\big|^2 + \big|[W_{\mathcal{N}}^L]_{j\alpha}\big|^2\Big) \nonumber \\
& \hspace{13.5em} + 4 y_\alpha y_P^2\text{Re}\Big[[W_{\mathcal{N}}^R]_{j\alpha} [W_{\mathcal{N}}^L]_{j\alpha}^*\Big]\Big]\,, \\
\Gamma(\mathcal{N}_j \to \ell_\alpha^- V^+) &= \frac{G_F^2 f_V^2 m_{j}^3 |V_{q\bar{q}}|^2}{16\pi} \lambda^{\frac{1}{2}}(1, y_\alpha^2, y_V^2) \Big[G(y_\alpha, y_V)\Big(\big|[W_{\mathcal{N}}^R]_{j\alpha}\big|^2 + [W_{\mathcal{N}}^L]_{j\alpha}\big|^2\Big) \nonumber \\
&\hspace{13.5em} - 12 y_\alpha y_V^2 \text{Re}\Big[[W_{\mathcal{N}}^R]_{j\alpha} [W_{\mathcal{N}}^L]_{j\alpha}^*\Big]\Big] \,,
\end{align}
where $F(x,y) = 1 - y^2 - x^2(2 - x^2 + y^2)$, $G(x,y) = (1 - y^2)(1 + 2y^2) + x^2(x^2 + y^2 - 2)$, and the pseudoscalar and vector form factors are defined via $\bra{0}\bar{q} \gamma_\mu \gamma_5 q\ket{P} = i f_P p_\mu$ and $\bra{0}\bar{q} \gamma_\mu q\ket{V} = i f_V m_V \epsilon_\mu$, respectively. The values taken for these form factors are given in Table~IV of~\cite{Feng:2024zfe}. The appropriate CKM matrix element $V_{q\bar{q}}$ should be used depending on the quark content of the meson. For Majorana HNLs, the effective $W^\pm$ interactions also induce the decays $\mathcal{N}_j \to \ell_\alpha^+ P^-$ and $\mathcal{N}_j \to \ell_\alpha^+ V^-$ with the same rates as Eq.~\eqref{eq:2body_charged}.

The effective $Z$ interactions also lead to the decay of HNLs to neutral pseudoscalar and vector mesons, $\mathcal{N}_j \to \mathcal{N}_i P^0$ and $\mathcal{N}_j \to \mathcal{N}_i V^0$, respectively, with $P^0 = \{\pi^0,\eta,\eta'\}$ and $V^0 = \{\rho,\omega,\phi\}$. These have the decay rates
\begin{align}
\Gamma(\mathcal{N}_j \to \mathcal{N}_i P^0) &= \frac{G_F^2 f_P^2 m_{j}^3}{8\pi}  \lambda^{\frac{1}{2}}(1, y_i^2, y_P^2) \Big[F(y_i, y_P)\Big(\big|[Z_{\mathcal{N}}^R]_{ij}\big|^2 + \big|[Z_{\mathcal{N}}^L]_{ij}\big|^2\Big) \nonumber \\
& \hspace{11.5em} + 4 y_i y_P^2\text{Re}\Big[[Z_{\mathcal{N}}^R]_{ij} [Z_{\mathcal{N}}^L]_{ij}^*\Big]\Big]\,,  \\
\Gamma(\mathcal{N}_j \to \mathcal{N}_i V^0) &= \frac{G_F^2 f_V^2 \kappa_V^2 m_{j}^3}{8\pi} \lambda^{\frac{1}{2}}(1, y_i^2, y_V^2) \Big[G(y_i, y_V)\Big(\big|[Z_{\mathcal{N}}^R]_{ij}\big|^2 + [Z_{\mathcal{N}}^L]_{ij}\big|^2 \Big) \nonumber \\
&\hspace{12.5em} - 12 y_i y_V^2 \text{Re}\Big[[Z_{\mathcal{N}}^R]_{ij} [Z_{\mathcal{N}}^L]_{ij}^*\Big]\Big]\,,
\end{align}
where the $\kappa_V$ factors arise from the light quark composition of the vector mesons and take the values $\kappa_\rho = 1 - 2s_w^2$, $\kappa_\omega = - 2s_w^2/3$ and $\kappa_\phi = -\sqrt{2}(1/2 - 2 s_w^2/3)$~\cite{Coloma:2020lgy}.

\begin{figure}[t!]
\centering
\includegraphics[width=0.495\textwidth]{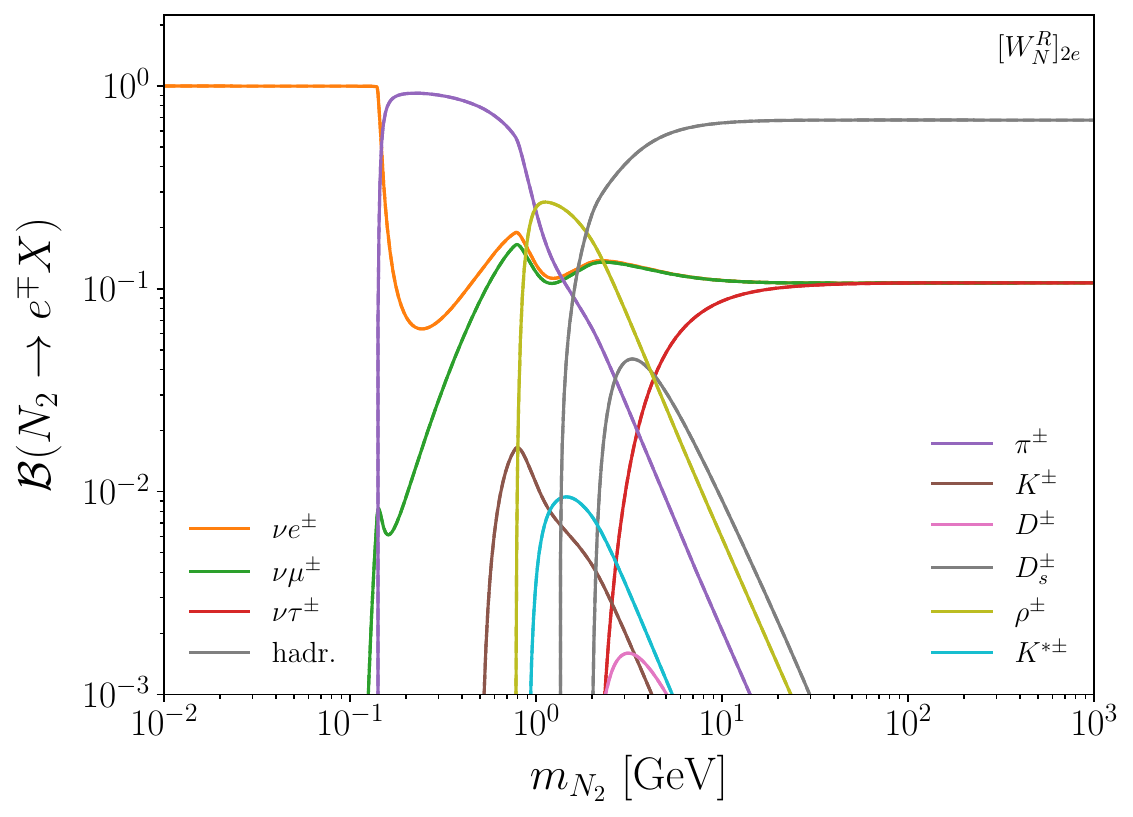}\\
\includegraphics[width=0.495\textwidth]{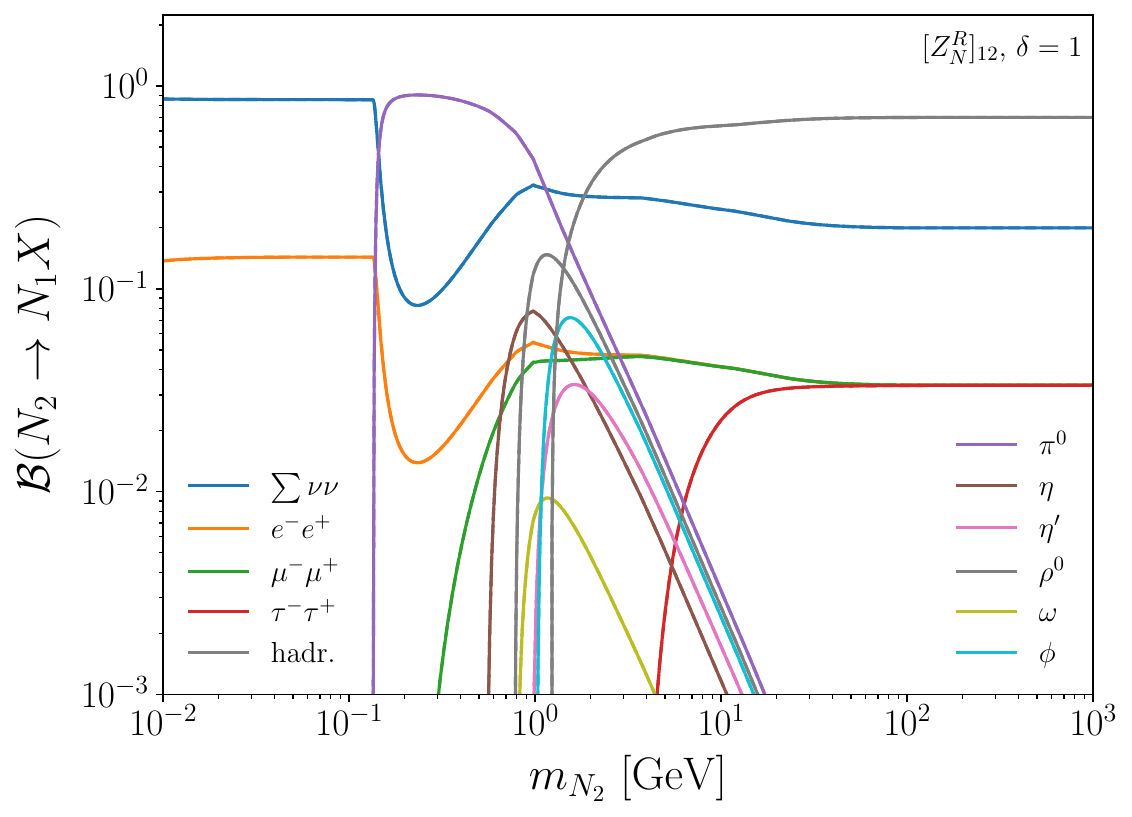}
\includegraphics[width=0.495\textwidth]{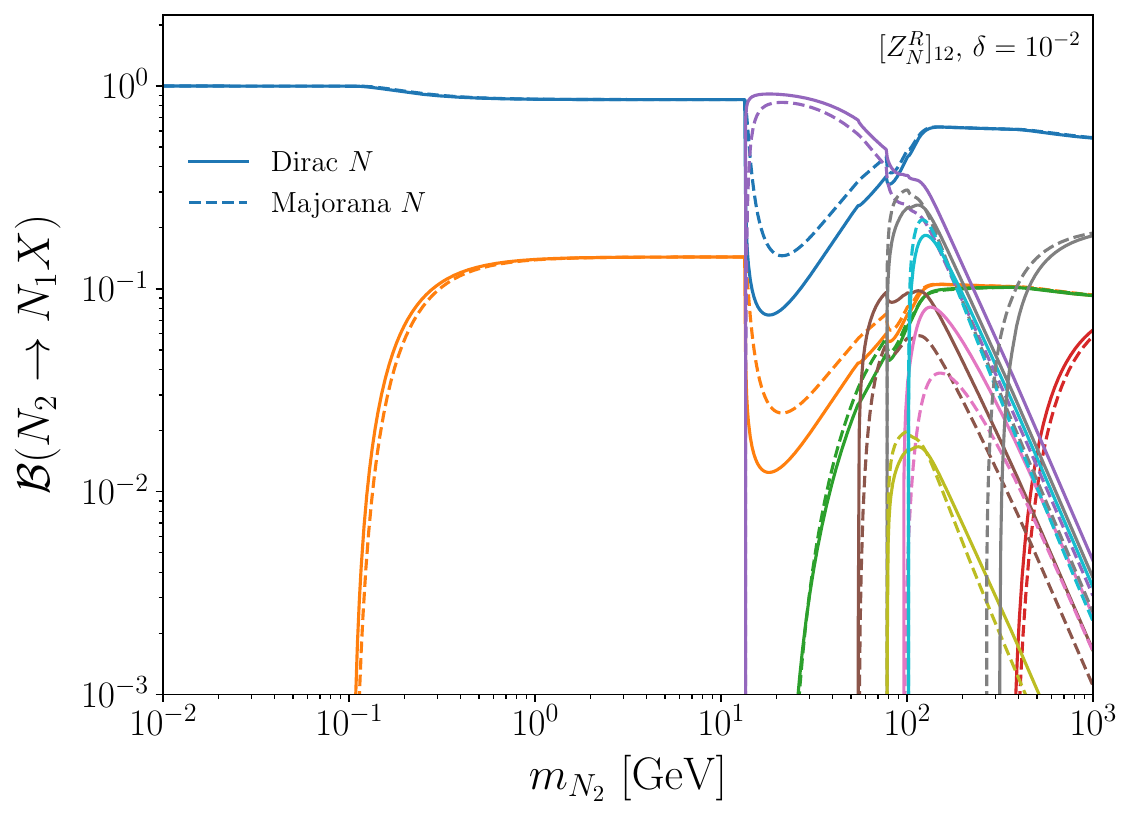}
\caption{(Above) Branching ratios of the decays $N_2 \to e^\mp X$ for non-zero values of the coefficient $[W_{N}^R]_{2e}$ and for Majorana (dashed) and Dirac (solid) HNLs. (Below) Branching ratios of the decays $N_2 \to N_1 X$ for non-zero values of $[Z_{N}^R]_{12}$, for the mass splitting ratios $\delta = 1$ (left) and $\delta = 10^{-2}$ (right).}
\label{fig:BR_plots}
\end{figure}

Finally, for $\mathcal{O}(100)$~GeV HNL masses, the effective $W^\pm$ and $Z$ interactions enable the two-body decays $\mathcal{N}_j \to \ell_\alpha^- W^+$, $\mathcal{N}_j \to \mathcal{N}_i Z$ and $\mathcal{N}_j \to \mathcal{N}_i h$, when the momentum transfer $q^2$ exceeds the on-shell production thresholds of $W^\pm$ and $Z$. The decay rates are
\begin{align}
\label{eq:2body-WZh}
\Gamma(\mathcal{N}_j \to \ell_{\alpha}^- W^{+}) &= \frac{G_F m_j^3}{8\sqrt{2}\pi}\lambda^{\frac{1}{2}}(1, y_\alpha^2,  y_W^2)\Big[G(y_\alpha, y_W)\Big(\big|[W_{\mathcal{N}}^R]_{j\alpha}\big|^2 + \big|[W_{\mathcal{N}}^L]_{j\alpha}\big|^2\Big) \nonumber \\ 
&\hspace{10.5em} - 12 y_\alpha y_W^2 \text{Re}\Big[[W_{\mathcal{N}}^R]_{j\alpha} [W_{\mathcal{N}}^L]_{j\alpha}^*\Big]\Big]\,,\nonumber\\
\Gamma(\mathcal{N}_j \to \mathcal{N}_i Z) &= \frac{G_F m_j^3}{4\sqrt{2}\pi} \lambda^{\frac{1}{2}}(1, y_i^2, y_Z^2) \Big[G(y_\alpha, y_Z)\Big(\big|[Z_{\mathcal{N}}^R]_{ij}\big|^2 + [Z_{\mathcal{N}}^L]_{ij}\big|^2\Big) \nonumber \\
&\hspace{10.5em} - 12 y_i y_Z^2 \text{Re}\Big[[Z_{\mathcal{N}}^R]_{ij} [Z_{\mathcal{N}}^L]_{ij}^*\Big]\Big]\,, 
\end{align}
The three-body decays $\mathcal{N}_j\to \mathcal{N}_i f_\alpha \bar{f}_\beta$ (for $f = \ell, u, d$), $\mathcal{N}_j\to \ell_\alpha^- u_\rho \bar{d}_\sigma$ and $\mathcal{N}_j\to \mathcal{N}_i \mathcal{N}_k \bar{\mathcal{N}}_l$ are implemented with the full propagator structure of the intermediate states $W^\pm$ and $Z$. For momentum transfers larger than $M_W$ and $M_Z$, these expressions already account for the two-body decay rates above; for $(q^2)^{\text{max}} > M_W^2, M_Z^2$, the integration over phase space is dominated by the propagator poles. In this regime, the narrow width approximation (NWA) can be used. For $(q^2)^{\text{max}} < M_W^2, M_Z^2$, the total HNL width is given in the Majorana case by
\begin{align}
\Gamma_{\mathcal{N}_j}\big|_{\text{Maj}} &= \sum_{i\leq k \leq l}\Gamma_{\mathcal{N}_i\mathcal{N}_k\mathcal{N}_l} + \sum_{i, \alpha,\beta}\Gamma_{\mathcal{N}_i\ell_\alpha^-\ell_\beta^+}\nonumber \\
&\hspace{1.5em} + \sum_{i}\bigg[\Theta_{ij}\sum_{ P^0}\Gamma_{\mathcal{N}_i P^0} + \Theta_{ij}\sum_{ V^0}\Gamma_{\mathcal{N}_i V^0} + (1 - \Theta_{ij})\sum_{\alpha, \beta}\Gamma_{\mathcal{N}_i q_\alpha \bar{q}_\beta}\bigg] \nonumber \\
&\hspace{1.5em} + 2\sum_{\alpha}\bigg[\Theta_{\alpha j}\sum_{P^+}\Gamma_{\ell_\alpha^{-} P^+} + \Theta_{\alpha j}\sum_{V^+}\Gamma_{\ell_\alpha^{-} V^+} + (1 - \Theta_{\alpha j})\sum_{\rho, \sigma}\Gamma_{\ell_\alpha^- u_\rho \bar{d}_\sigma}\bigg]\,,
\label{eq:tot_decay_rate_Maj}
\end{align}
and in the Dirac case by,
\begin{align}
\Gamma_{\mathcal{N}_j}\big|_{\text{Dirac}} &= \sum_{i\leq k,  l}\Gamma_{\mathcal{N}_i\mathcal{N}_k\bar{\mathcal{N}}_l} + \sum_{i, \alpha,\beta}\Gamma_{\mathcal{N}_i\ell_\alpha^-\ell_\beta^+} \nonumber \\
&\hspace{1.5em} + \sum_{i}\bigg[\Theta_{ij}\sum_{ P^0}\Gamma_{\mathcal{N}_i P^0} + \Theta_{ij}\sum_{ V^0}\Gamma_{\mathcal{N}_i V^0} + (1 - \Theta_{ij})\sum_{\alpha, \beta}\Gamma_{\mathcal{N}_i q_\alpha \bar{q}_\beta}\bigg] \nonumber \\
&\hspace{1.5em} + \sum_{\alpha}\bigg[\Theta_{\alpha j}\sum_{P^+}\Gamma_{\ell_\alpha^{-} P^+} + \Theta_{\alpha j}\sum_{V^+}\Gamma_{\ell_\alpha^{-} V^+} + (1 - \Theta_{\alpha j})\sum_{\rho, \sigma}\Gamma_{\ell_\alpha^- u_\rho \bar{d}_\sigma}\bigg]\,,
\label{eq:tot_decay_rate_Dirac}
\end{align}
where we introduce the shorthand $\Gamma_{X} \equiv \Gamma(\mathcal{N}_j \to X)$ and
\begin{align}
\Theta_{ab} \equiv \Theta\bigg(1 - \frac{(m_b - m_a)^2}{1~\text{GeV}^2}\bigg) \,.
\end{align}
When $(q^2)^{\text{max}}$ exceeds $M_W^2$ and $M_Z^2$, the contributions of $W^\pm$ and $Z$ can be replaced by the NWA expression, respectively. For $(q^2)^{\text{max}} > M_W^2, M_Z^2$, this can be written as
\begin{align}
\label{eq:NWA}
\Gamma_{\mathcal{N}_j}^{\text{NWA}} &= \Gamma_{\mathcal{N}_j}^{\psi^4} + (2)\Gamma_{\ell^- W^+}\sum_X\text{BR}_{W^+ \to X} + \Gamma_{\mathcal{N}_i Z}\sum_X\text{BR}_{Z\to X}\,,
\end{align}
where $\Gamma_{\mathcal{N}_j}^{\psi^4}$ is the contribution of only the four-fermion operators to Eq.~\eqref{eq:tot_decay_rate_Maj} or~\eqref{eq:tot_decay_rate_Dirac}. The factor of two in parentheses is necessary for the Majorana HNL decay rate. In Eq.~\eqref{eq:NWA}, the decays of $W^\pm$ and $Z$ also include the channels $W^\pm \to \mathcal{N}_i \ell_\beta^\pm$ and $Z \to \mathcal{N}_i \mathcal{N}_j$, respectively, with rates given by
\begin{align}
\Gamma(W^{+} \to \mathcal{N}_i \ell_{\beta}^+) &= \frac{G_F M_W^3}{12\sqrt{2}\pi}\bigg[\bigg(2 - \frac{m_i^2 + m_\beta^2}{M_W^2} - \frac{(m_i^2 - m_\beta^2)^2}{M_W^4}\bigg)\Big(\big|[W_{\mathcal{N}}^R]_{i\beta}\big|^2 + [W_{\mathcal{N}}^L]_{i\beta}\big|^2\Big) \nonumber \\ 
&\hspace{5.3em} + \frac{12m_i m_\beta}{M_W^2} \text{Re}\Big[[W_{\mathcal{N}}^L]_{i\beta} [W_{\mathcal{N}}^R]_{i\beta}^*\Big]\bigg]\lambda^{\frac{1}{2}}\bigg(1, \frac{m_i^2}{M_W^2}, \frac{m_\beta^2}{M_W^2}\bigg) \,,\nonumber\\
\Gamma(Z\to \mathcal{N}_i \mathcal{N}_j) &= \bigg(1 - \frac{\delta_{ij}}{2}\bigg)\frac{G_F M_Z^3}{6\sqrt{2}\pi}\bigg[\bigg(2 - \frac{m_i^2 + m_j^2}{M_Z^2} - \frac{(m_i^2 - m_j^2)^2}{M_Z^4}\Big)\Big(\big|[Z_{\mathcal{N}}^R]_{ij}\big|^2 + \big|[Z_{\mathcal{N}}^L]_{ij}\big|^2\Big) \nonumber \\ 
&\hspace{9.8em} + \frac{12 m_i m_j}{M_Z^2} \text{Re}\Big[[Z_{\mathcal{N}}^L]_{ij} [Z_{\mathcal{N}}^R]_{ij}^*\Big]\bigg]\lambda^{\frac{1}{2}}\bigg(1, \frac{m_i^2}{M_Z^2}, \frac{m_j^2}{M_Z^2}\bigg) \,,
\end{align}
where the prefactors in parenthesis are present in the Majorana case. Including all possible final states, the branching ratios in Eq.~\eqref{eq:NWA} sum to unity and the total HNL decay width is just the sum of three-body decays, induced by the four-fermion operators, and the two-body decays in Eq.~\eqref{eq:2body-WZh}.

We now briefly explore the branching ratios of HNL decays via the operators of interest in this work. In particular, the decay mode with $e^-e^+$ in the final state, which is considered as the displaced vertex signature in Sec.~\ref{sec:displacedvertex}. For the majority of the four-fermion operators, this is the only decay mode present, giving $\text{BR}(N_2 \to N_1 e^-e^+) = 1$. However, as discussed below Eq.~\eqref{eq:eff_NNvv}, some of the $\nu$SMEFT operators induce the decay $N_2 \to N_1 \nu_e \bar{\nu}_e$, resulting in $\text{BR}(N_2 \to N_1 e^-e^+) = 1/2$. Finally, we show in Fig.~\ref{fig:BR_plots} the branching ratios resulting from the effective $W^\pm$ (above) and $Z$ (below) interactions. Clearly, the presence of more decay modes lead the branching ratios to depend on the mass of $N_2$, with $\text{BR}(N_2 \to \nu e^-e^+)$ and $\text{BR}(N_2 \to N_1 e^-e^+)$ decreasing for HNL masses above the production thresholds for hadronic final states. For large $m_{N_2}$, the branching ratios plateau at the values $\text{BR}(N_2 \to \nu e^-e^+) \approx 1/9$ and $\text{BR}(N_2 \to N_1 e^-e^+) \approx 3.4\times 10^{-2}$. In the case of the effective $Z$ interaction, the branching fractions also depend on the mass splitting ratio. For small $\delta$ (right), the kinematic thresholds are pushed to larger values of $m_{N_2}$, while a difference in the branching ratios of Dirac (solid) and Majorana (dashed) $N_2$ becomes evident. 

\bibliography{bibliography.bib}
\bibliographystyle{JHEP}

\end{document}